\documentclass[a4paper,11pt]{article}

\usepackage{jinstpub} 
\usepackage{siunitx}
\DeclareSIUnit[number-unit-product = ]\percent{\char`\%}
\usepackage{booktabs}
\usepackage{subcaption}
\usepackage{graphicx}
\usepackage[symbol]{footmisc}

\usepackage{enumitem}
\setlist[itemize]{itemsep=0.1em}
\setlist[enumerate]{itemsep=0.1em}
\setlist[description]{itemsep=0.1em}

\usepackage{lineno}
\begin{document}

\title{\boldmath SND@LHC: The Scattering and Neutrino Detector at the LHC}

\author{The SND@LHC Collaboration\footnote{See on the back for the author list}}
%\institute{See on the  back for the author list \label{addr1}}

\abstract{SND@LHC is a compact and stand-alone experiment designed to perform measurements with neutrinos produced at the LHC in the %unexplored
pseudo-rapidity region of ${\num{7.2} < \eta < \num{8.4}}$.
%, complementary to all the other experiments at the LHC.
The experiment is located \SI{480}{m} downstream of the ATLAS interaction point, in the TI18 tunnel.
The detector is composed of a hybrid system based on an \SI{830}{kg} target made of tungsten plates, interleaved with emulsion and electronic trackers, also acting as an electromagnetic calorimeter, and followed by a hadronic calorimeter and a muon identification system. 
%downstream by a calorimeter and a muon system.
%The configuration allows for efficiently distinguishing 
The detector is able to distinguish interactions of all three neutrino flavours, which allows probing the physics of heavy flavour production at the LHC in the very forward region.
%the region that is not accessible to other LHC experiments.
This region is of particular interest for future circular colliders and for very high energy astrophysical neutrino experiments.
%The detector is also able to search for Feebly Interacting Particles that scatter or decay within it.
The detector is also able to search for the scattering of Feebly Interacting Particles.
In its first phase, the detector is ready to operate throughout LHC Run 3 and collect a total of \SI{250}{fb^{-1}}.
}

\maketitle

\flushbottom

\newpage

\section{Introduction}
\label{sec:introduction}
\subsection{Overview}
SND@LHC is a compact experiment proposed to exploit the high flux of energetic neutrinos of all  flavours from the LHC~\cite{Beni:2019gxv,snd_technical_proposal}.
It is located slightly off-axis, covering the unexplored pseudo-rapidity ($\eta$) range from \num{7.2} to \num{8.4}, in which a large fraction of neutrinos originate from charmed-hadrons decays.
%SND@LHC is a compact experiment proposed to make measurements with neutrinos of all three neutrino flavours from the LHC  in the pseudo-rapidity  range of $7.2 < \eta < 8.4$. This range of pseudo-rapidity is currently unexplored~\cite{Aaij:2015bpa}, and a large fraction of the corresponding neutrinos originate from charmed-hadron decays. 
Thus, neutrinos can probe heavy-flavour production in a region that is not accessible to other large LHC experiments, which are designed to study high-$p_T$ physics at $\eta$<5.
%$: ATLAS covers a pseudorapidity of $\eta < 4.5$, CMS of $\eta < 2.5$, ALICE up to $\eta < XXX$, while LHCb is sensitive in the range $2 < \eta < 5$.

Together with the FASER$\nu$~\cite{Abreu:2019yak} experiment, SND@LHC will make the first observations of neutrinos produced by a collider, in an energy range which was inaccessible  at accelerators so far.
%FASER$\nu$ will explore the pseudo-rapidity range $\eta > 8.9$, where the compositions of the various sources of neutrinos are different.
%In addition to distinct detector concepts, SND@LHC and FASER$\nu$ will explore different pseudo-rapidity ranges in which the relative compositions of the various sources of neutrinos are different. 
%Hence, the neutrino physics programmes of the two experiments are complementary. 
SND@LHC is also sensitive to Feebly Interacting Particles (FIPs) through scattering off nuclei and electrons in the detector target.
The direct-search strategy gives the experiment sensitivity in a region of the FIP mass-coupling parameter space that is complementary to other indirect searches~\cite{sym12101648}.

%in a region of the parameter space which extends current bounds in a way which is complementary to other searches. 

In order to shield the detector from most of the charged particles produced in the LHC collisions, SND@LHC is located in the TI18 tunnel,
%unused since the decommissioning of LEP,
about \SI{480}{m} downstream of the ATLAS interaction point (IP1).
Charged-lepton identification and the measurement of the neutrino energy are essential to distinguish among the three flavours in neutrino charged-current interactions and to identify and study the corresponding neutrino production process. 
These features were the main drivers in the design of the SND@LHC apparatus, that had also to account for geometrical constraints of the selected location.
The detector was installed in TI18 in 2021 during the Long Shutdown~2 and it has started to collect data since the beginning of the LHC Run~3 in April 2022. The SND@LHC experiment will run throughout the whole Run~3 and it is expected to collect \SI{250}{fb^{-1}} of data in 2022--25, corresponding to two thousands high-energy neutrino interactions of all flavours in the detector target.

%is expected to collect \SI{250}{fb^{-1}} of data in 2022--25, 
%during Run~3 of the LHC.

\begin{figure}[htbp]
\centering
\includegraphics[width=1.0\columnwidth]{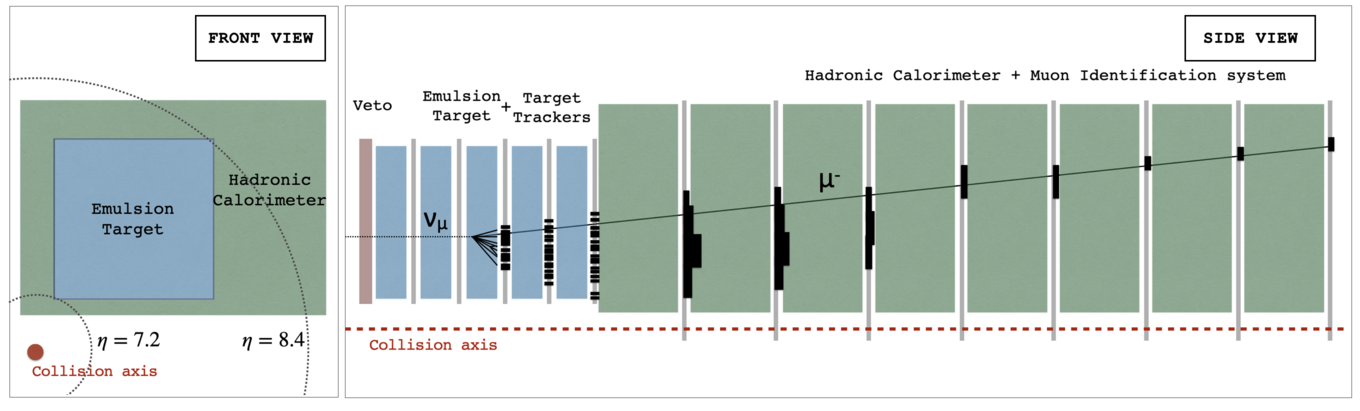}
\caption{Schematic layout of the SND@LHC detector.}
\label{fig:layout_schema}
\end{figure}

The SND@LHC detector is composed of several parts (see Figure~\ref{fig:layout_schema}).
The veto system tags events with charged particles entering the detector from the front.
It is followed by the emulsion target, which acts as a vertex detector, and the target trackers that provide the timestamp to the events reconstructed in the emulsions.
The combination of the emulsion target and the target tracker also acts as an electromagnetic calorimeter.
A shielding surrounding the target has been put in place to absorb low-energy neutrons and as a thermal insulation chamber.
The target system is followed by a hadronic calorimeter and a muon identification system.

The detector concept and the physics goals of the SND@LHC experiment have been described in the Technical Proposal~\cite{snd_technical_proposal}.
This document details the detector layout, construction and installation phases. 
%This section gives an overview of the physics and the experiment. 
%We then describe the detector environment in Section~\ref{sec:environment}. 
Sections~\ref{sec:veto} to~\ref{sec:muon} describe the sub-systems of the detector.
Section~\ref{sec:daq} describes the data acquisition and online systems, while Section~\ref{sec:offline} discusses the offline software and simulation framework.
Sections~\ref{sec:commissioning} and \ref{sec:installation} give details about the commissioning and installation of the detector.
Finally, in Section~\ref{sec:outlook} we give some ideas about a possible upgrade of the detector.

\subsection{The physics case}
\label{subsec:physics}
Neutrinos allow for precise tests of the Standard Model (SM)~\cite{Brock:1993sz,Conrad:1997ne,Formaggio:2013kya,Lellis:2004yn}. They are a probe for new physics~\cite{Marfatia:2015hva,Arguelles:2019xgp} and provide a unique view of the Universe~\cite{Ackermann:2019ows}. 
The neutrino-nucleon cross section region between 350 GeV and 10 TeV is currently unexplored~\cite{pdg,PhysRevLett.122.041101}.
%Neutrino interactions have been measured in the energy regime below \SI{350}{GeV}.
%An overview of all available $\nu$ cross-section measurements is given in~\cite{pdg}.
%The IceCube collaboration reported a few tens of events in the region \SI{10}{TeV}--\SI{1}{PeV}~\cite{Aartsen:2017kpd}.
Indeed, measurements of neutrino interactions in the last decades were mainly performed at low energies for neutrino oscillation studies. 
%, where neutrino oscillations over the available baselines are enhanced. 

Neutrinos in $pp$ interactions at the CERN LHC arise promptly from leptonic $W$ and $Z$ decays, and $b$ and $c$ decays. They are subsequently also produced in the decays of pions and kaons.
%LHC neutrinos offer the unique possibility of observing their interactions with  matter in the largely unexplored range from a few hundred GeV to a few TeV in a laboratory.
%Furthermore, the contribution of the $\tau$ flavour to the LHC neutrino flux is sizeable.

The use of LHC as a neutrino factory was first envisaged about 30 years ago~\cite{DeRujula:1984pg,DeRujula:1992sn,Vannucci:253670}, in particular for the then undiscovered $\nu_{\tau}$~\cite{Jarlskog:215298}.
The idea suggested a detector intercepting the very forward flux (${\eta > 7}$) of neutrinos (about \SI{5}{\percent} have $\tau$ flavour) from $b$ and $c$ decays.
Recently, it was pointed out~\cite{Buontempo:2018gta} that at larger angles (${4 <  \eta < 5}$) leptonic $W$ and $Z$ decays also provide an additional contribution to the neutrino flux, of which one third has $\tau$ flavour. The role of an off-axis setup has been emphasised in a recent paper~\cite{Beni:2020yfy}. 
%In these events, the charged lepton could be detected in coincidence at the collision point by the existing collider detector, thus providing an independent determination of the neutrino flavour.

Today, two factors make it possible and particularly interesting to add a compact neutrino detector at the LHC.
The high intensity of $pp$ collisions achieved by the machine turns into a large expected neutrino flux in the forward direction~\cite{Beni:2019gxv}, and the high neutrino energies imply relatively large neutrino cross-sections.
As a result, even a detector with a relatively modest size to fit into one of the existing underground areas close to the LHC tunnel has significant physics potential.
Machine-induced backgrounds decrease rapidly while moving along and away from the beam line.

A detailed study of a possible underground location for a neutrino detector  was conducted in 2018~\cite{Beni:2019gxv}, during the LHC Run~2.
Four locations were considered for hosting a possible neutrino detector: the CMS quadrupole region (\SI{25}{m} from the CMS Interaction Point (IP5)), UJ53 and UJ57 ($\sim$\SI{90} and \SI{120}{m} from IP5), RR53 and RR57 ($\sim$\SI{240}{m} from IP5), TI18 ($\sim$\SI{480}{m} from IP1).
The potential sites were studied on the basis of expected neutrino rates, flavour composition and energy spectrum, predicted backgrounds, and in-situ measurements performed with a nuclear emulsion detector and radiation monitors.
TI18 emerged as the most favourable location.
Assuming a luminosity of \SI{250}{fb^{-1}} in the LHC Run~3, a detector with a mass of \SI{830}{kg} located in TI18 can observe and study about two thousand high-energy neutrino interactions of all flavours.
%~\cite{Beni:2019gxv}.

%The FASER collaboration~\cite{Ariga:2019ufm} in 2019 proposed to extend its physics case and also measure neutrinos with a detector, FASER$\nu$~\cite{Abreu:2019yak}, located in the TI12 tunnel, on the opposite side of IP1. 
%reThe location is on-axis at ${\eta > 9}$. 

The main physics goals of the SND@LHC experiment are summarised in the following sections.

\begin{figure}[htbp]
\centering
\includegraphics[width=0.6\columnwidth]{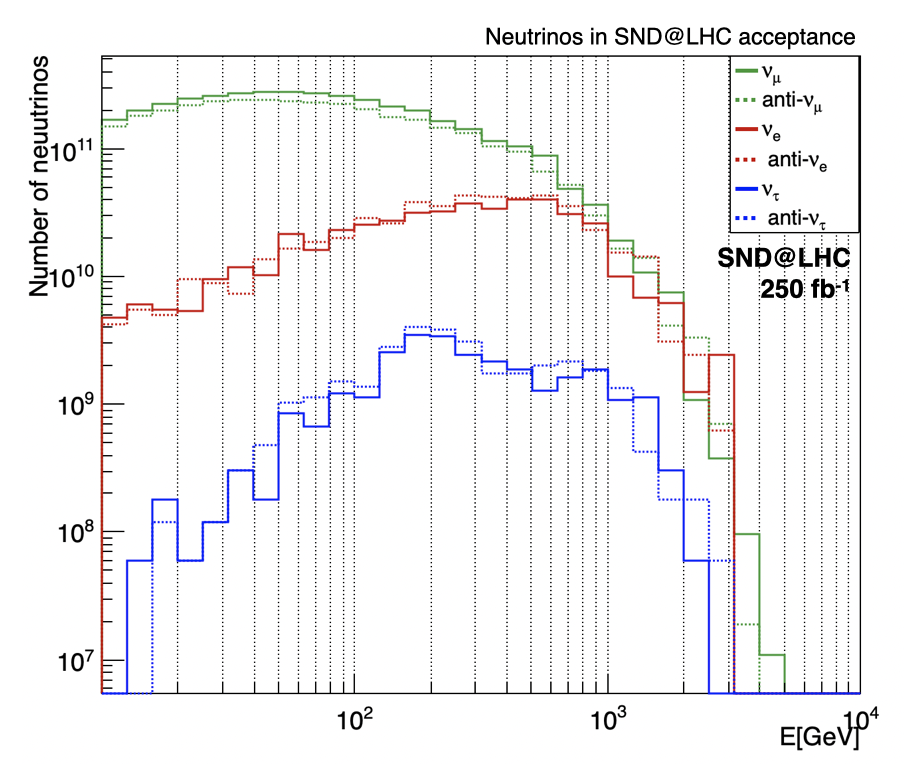}
\caption{Energy spectrum of the different types of incoming  neutrinos and anti-neutrinos as predicted by simulation, normalised to \SI{250}{fb^{-1}}.}
%the DPMJET~\cite{Roesler_2001,DPMJET}/FLUKA~\cite{Fluka2,Fluka3} simulation. 
%The result of the simulation has been normalised to produce neutrino spectra for \SI{250}{fb^{-1}}.}
\label{fig:neutrinos_snd}
\end{figure}

\subsection*{Neutrino physics}
Figure~\ref{fig:neutrinos_snd} shows the energy spectrum of incoming neutrinos and anti-neutrinos in the pseudo-rapidity range covered by the SND@LHC detector, ${7.2 < \eta < 8.4}$, normalised to \SI{250}{fb^{-1}}.
Neutrino production in proton-proton collisions at the LHC is simulated with the FLUKA Monte Carlo code~\cite{Fluka2,Fluka3}. DPMJET3 (Dual Parton Model, including charm)~\cite{Roesler_2001,DPMJET}  is used for the event generation, and FLUKA performs the particle propagation towards the SND@LHC detector with the help of the FLUKA model of the LHC accelerator~\cite{Boccone:2014hxd}. 

FLUKA also takes care of simulating the production of neutrinos from decays of long-lived products of the $pp$ collisions and of particles produced in re-interactions with the surrounding material. \textsc{Genie}~\cite{cite:GENIE} is then used to simulate neutrino interactions with the SND@LHC detector material.
About 1700 charged-current (CC) and 550 neutral current (NC) neutrino interactions are expected in the target volume, mainly from muon neutrinos (72\%) and electron neutrinos (23\%).

In the explored $\eta$ range, electron neutrinos and anti-neutrinos are predominantly produced by charmed-hadron decays. 
%As a result, SND@LHC is capable of measuring charmed-hadron production indirectly through the observation of electron neutrinos and anti-neutrinos.  
Therefore, if one assumes that the deep-inelastic charged-current cross-section of the electron neutrino follows the SM prediction, as also supported by the HERA results in their SM interpretation~\cite{Chekanov:2003vw,Ahmed:1994fa}, electron neutrinos can be used as a probe of the production of charm. 
%after unfolding the instrumental effects and subtracting the contribution from kaon decays~\cite{Chekanov:2003vw}.  
Taking into account uncertainties in the correlation between the yield of charmed hadrons in a given $\eta$ region with the neutrinos in the measured $\eta$ region, it was evaluated that the measurement of the charmed-hadron production in $pp$ collision can be done with a statistical uncertainty of about \SI{5}{\percent}, while the leading contribution to the uncertainty is the systematic error of \SI{35}{\percent}~\cite{snd_technical_proposal}. 

Furthermore, the measurement of the charmed hadrons can be translated into a measurement of the corresponding open charm production in the same rapidity window, given the linear correlation between the parent charm quark and the hadron. 
The dominant partonic process for associated charm production at the LHC is the scattering of two gluons producing a $c\overline{c}$ pair~\cite{Gleisberg:2003xi}. 
The average lowest momentum fraction ($x$) of interacting gluons probed by SND@LHC is around \num{e-6}.
The extraction of the gluon PDF at such low values of $x$, where it is completely unknown, could provide valuable information for future experiments probing the same low $x$ range, such as FCC~\cite{FCC:2018byv}.
It can also reduce uncertainties on the flux of very-high-energy (PeV scale) atmospheric neutrinos produced in charm decays, essential for the evidence of neutrinos from astrophysical sources~\cite{Bhattacharya:2016jce,Jeong:2021vqp}.

Since the three neutrino flavours can be identified, the lepton flavour universality can be tested in the neutrino sector by measuring the ratio of $\nu_e$/$\nu_\tau$ and  $\nu_e$/$\nu_\mu$ interactions.
Both $\nu_e$ and $\nu_\tau$ are mainly produced by semi-leptonic and fully leptonic decays of charmed hadrons. 
Unlike $\nu_\tau$s that are produced almost only in $D_s$ decays, $\nu_e$s are produced in the decay of all ground-state charmed hadrons, essentially $D^0$, $D$, $D_s$ and $\Lambda_c$. 
Therefore, the $\nu_e/\nu_\tau$ ratio depends only on the charm hadronisation fractions and decay branching ratios. The systematic uncertainties due to the charm-quark production mechanism cancel out, and the ratio becomes sensitive to the $\nu$-nucleon interaction cross-section ratio of the two neutrino species, which is affected by the uncertainty on hadronisation processes.
The estimate of the branching ratios has a systematic uncertainty of about 22\% while the statistical uncertainty is dominated by the low statistics of the $\nu_\tau$ sample, which corresponds to a 30\% accuracy.

The situation is rather different for $\nu_e$s when compared to $\nu_\mu$s. 
The $\nu_\mu$s are much more abundant but heavily contaminated by  $\pi$ and $K$ decays, and therefore the production mechanism cannot be considered the same as in the case of $\nu_e$. 
However, this contamination is mostly concentrating at low energies. 
Above \SI{600}{GeV}, the contamination is predicted to be reduced to about 35\%, and stable with the energy. Moreover, charmed-hadrons decays have practically equal branching ratios into electron and muon neutrinos. 
Therefore the $\nu_e/\nu_\mu$ ratio is not affected by the systematic uncertainties in the weighted branching fractions, but rather by uncertainties due to $\pi$ and $K$ production in this $\eta$ range and to their propagation through  the machine elements along the beamline, that can be assessed thanks to the available measurements used to constrain the simulation.   
The $\nu_e/\nu_\mu$ ratio provides a test of the lepton flavour universality with an  uncertainty of 15\%, with an equal 10\% statistical and systematic contribution.

SND@LHC plans to measure the ratio between charged-current (CC) and neutral-current (NC) interactions as an internal consistency test. 
Indeed, by summing over neutrinos and anti-neutrinos, the ratio between NC and CC deep-inelastic interaction cross-sections at a given energy can be written as a simple function of the  Weinberg angle, with a correction factor accounting for the non-isoscalarity of the target~\cite{Alekhin:2015byh}. 
In the approximation that the differential $\nu$ and $\bar{\nu}$ fluxes, as a function of their energy, are equal, the same formula also applies  to the observed interactions since the convolution with the flux would bring the same factor everywhere, that then cancels out in the ratio. 
The statistical uncertainty on the NC/CC ratio for observed events is expected to be lower than \SI{5}{\percent} while the systematic uncertainty on the unfolded ratio amounts to about \SI{10}{\percent}~\cite{snd_technical_proposal}.

%By extracting the Weinberg angle from the NC/CC ratio, we can check whether our detector and analyses are performing as intended.

\subsection*{Feebly Interacting Particles}
The SND@LHC experiment is also capable of performing model-independent direct searches for FIPs. The background from neutrino interactions can be rejected by making a time-of-flight (TOF) measurement. With a time resolution of $\sim\!200\,$ps, it will be possible to disentangle the scattering of massive FIPs and neutrinos, 
with a significance that depends on the particle mass~\cite{snd_technical_proposal}.
The hybrid nature of the apparatus, which combines emulsion trackers and electronic detectors, makes it possible to disentangle the scattering of massive FIPs and neutrinos, with a significance that depends on the particle mass. 

%The SND@LHC experiment can explore a large variety of Beyond Standard Model (BSM) scenarios describing Hidden Sectors. 
FIPs may be produced in the $pp$ scattering at the LHC interaction point, propagate to the detector and decay or scatter inside it. A recent work \cite{Boyarsky:2021moj} summarises SND@LHC sensitivity to physics beyond the Standard Model considering the  scatterings of light dark matter particles $\chi$ via leptophobic $U(1)_B$ mediator, as well as decays of Heavy Neutral Leptons, dark scalars and dark photons.
The elastic scattering was considered, showing an excess of neutrino-like elastic scatterings over the SM yield due to the $\chi+p$ process. 
%Two scattering signatures were considered: elastic, showing an excess of neutrino-like elastic scatterings over the SM yield due to the $\chi+p$ process, and inelastic where an excess of the ratio of neutral-to-charged current-like events over the SM prediction due to $\chi$+nucleus deep inelastic scattering is expected.
The excellent spatial resolution of nuclear emulsions and the muon identification system makes SND@LHC also suited to search for the decay of neutral mediators decaying in two charged particles.

\begin{figure}[b]
\centering
\includegraphics[width=0.9\columnwidth]{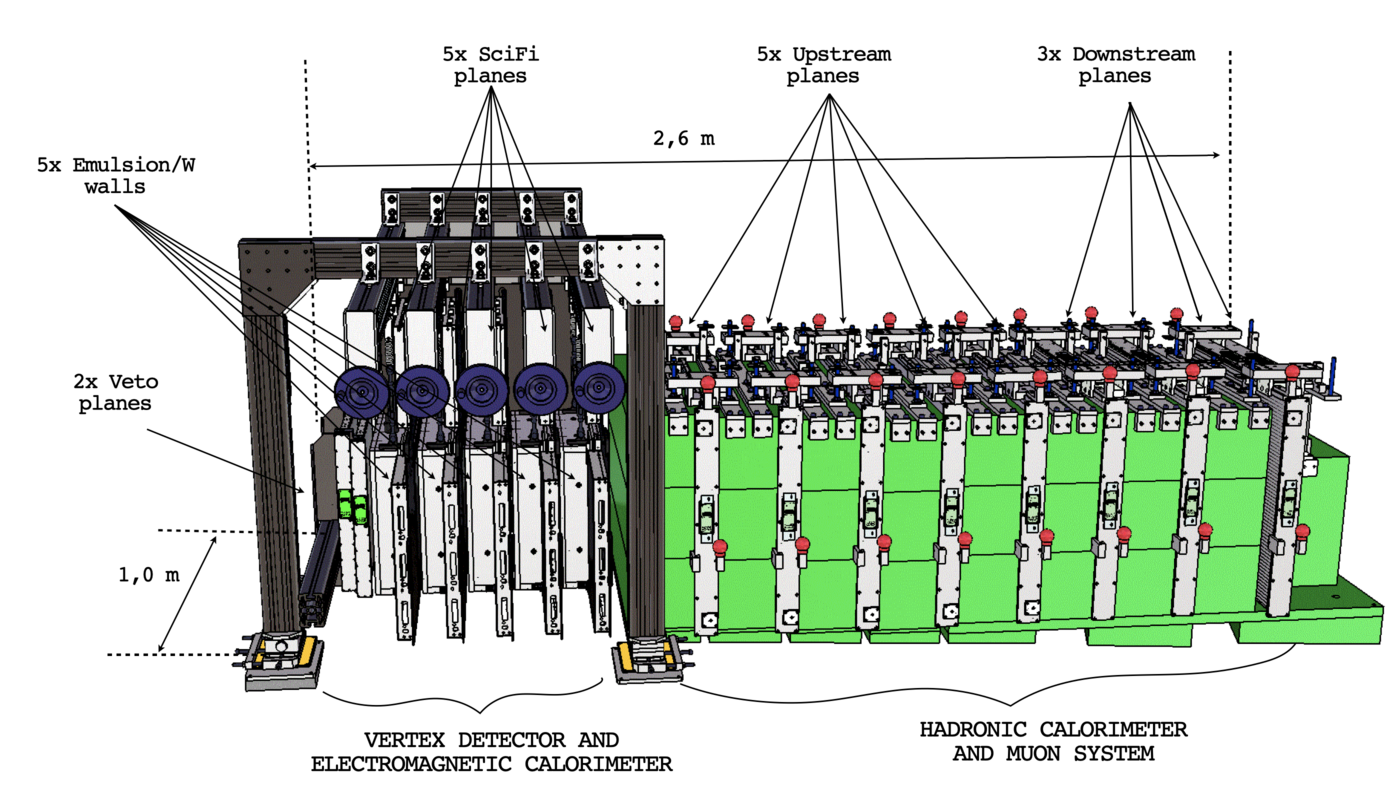}
\caption{Layout of the SND@LHC experiment.}
\label{fig:layout}
\end{figure}

\subsection{Detector layout}
\label{subsec:detector}
The detector layout was developed to allow for the identification of the three neutrino flavours and the direct search for FIPs.

%\begin{itemize}
%    \item a vertex detector, with a good enough resolution to disentangle the neutrino-interaction vertex from the one of the tau lepton decay;
%    \item a calorimeter, to measure both the electromagnetic and hadronic energy, with also  a good time resolution;
%    \item a muon system, to identify the muon produced in $\nu_\mu$ charged current interactions and in the muonic decay of the tau lepton.
%\end{itemize}

The layout of the detector, with the exclusion of the neutron shield, is shown in Figure~\ref{fig:layout}.
The apparatus is composed of a target region followed downstream by a hadronic calorimeter and a muon identification system.
Upstream of the target region, two planes of scintillator bars act as a veto for charged particles, mostly muons coming from IP1.
The target region, with a mass of about \SI{830}{kg}, is instrumented with five walls of Emulsion Cloud Chambers (ECC)~\cite{Acquafredda:2009zz}, each followed by a Scintillating Fibre (SciFi) plane
The ECC technology alternates emulsion films, acting as tracking devices with micrometric accuracy, with passive material acting as the neutrino target.
Tungsten is used as a passive material to maximize the mass within the available volume.
The SciFi planes provide the timestamp for the reconstructed events and have an appropriate time resolution for the time-of-flight measurements of particles from IP1.
The combination of the emulsion target and the target tracker also acts as an electromagnetic calorimeter, with a total of 85 $X_0$.

Veto, emulsion target and target tracker are contained in a 30\% borated polyethylene and acrylic box which has the dual function of acting as a neutron shield from low energy neutrons and maintaining controlled temperature and humidity levels in order to guarantee optimal conditions for emulsion films.

The hadronic calorimeter and muon identification system are located downstream of the target and consist of eight \SI{20}{cm}-thick thick iron slabs (green) making up 9.5 interaction lengths $\lambda_{\mathrm{int}}$ in total, each followed by one or two planes of \SI{1}{cm}-thick scintillating bars. 

The hadronic shower starts developing already in the target region, which adds on average 1.5 $\lambda_{\mathrm{int}}$, for an average total length of 11 $\lambda_{\mathrm{int}}$, thus providing a good coverage of the hadronic showers. 
The muon identification is mainly based on the last three planes of scintillator bars. These planes have double layers with narrower bars located both vertically and horizontally for higher granularity.

%of eight iron slabs (9.5 $\lambda_{\mathrm{int}}$ in total), each followed by one or two planes of scintillating bars.
%The resolution in the time-of-flight measurement will be limited by the \SI{200}{ps} temporal spread due to the length of the luminous region at IP1.

\begin{figure}
\begin{center}
\includegraphics[width=1.0\linewidth]{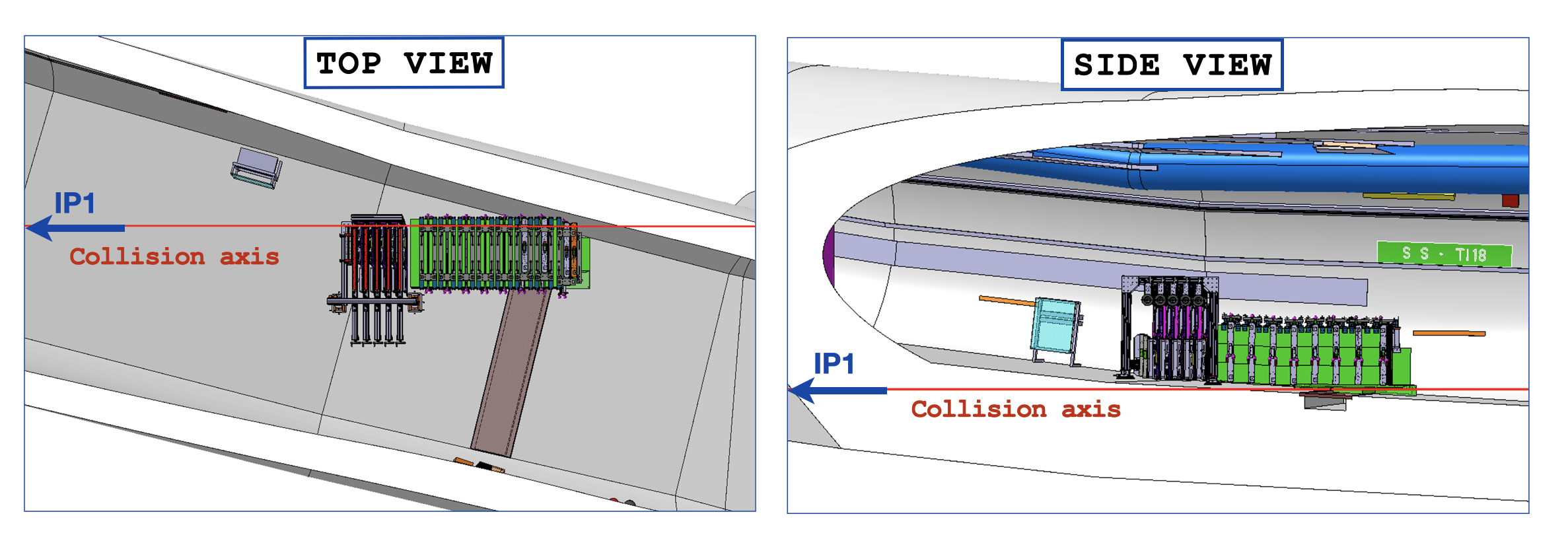}
\caption{Side and top views of the SND@LHC detector in the TI18 tunnel~\cite{snd_technical_proposal}.}
\label{fig:sideview}
\end{center}
\end{figure}

The detector, including the neutron shield described in Section~\ref{subsec:coldbox}, exploits all the available space in the TI18 tunnel to cover the desired range in pseudo-rapidity. 
Figure~\ref{fig:sideview} shows the side and top views of the detector positioned inside the tunnel. 
The size of the tunnel, the tilted slope of the floor, as well as the distance of tunnel walls and floor from the nominal collision axis, imposed several constraints to the detector design since no civil engineering work could have been done in time for the operation in Run 3. 
The detector layout was therefore optimised in order to find the best compromise between geometrical constraints and the following physics requirements: a good calorimetric measurement of the energy requiring about 10 $\lambda_{\mathrm{int}}$, a good muon identification efficiency requiring enough material to absorb hadrons, a transverse size of the target region having the desired azimuthal angular acceptance. 
The energy measurement and the muon identification set a constraint on the minimum length of the detector. 
With the constraints from the tunnel, this requirement competes with the azimuthal angular acceptance that determines the overall flux intercepted and therefore the total number of observed interactions. 
%The combination of position and size of the proposed detector is an optimal compromise between these competing requirements. 
The geometrical constraints also restrict the detector to the first quadrant only around the nominal collision axis.

%The result is a compact detector, 2.6 m in length. The energy measurement and the muon identification limit the target region to a length of about 80 cm. The transverse size downstream of about 80(H) $\times$ 60(V) cm$^2$ is limited by the constraint of the tunnel side wall. The transverse size of the target region is proportionally smaller in order to match the acceptance of the energy measurement and the muon identification for the vertices identified in the target volume. 

 The identification of the neutrino flavour is done in charged current interactions by identifying the charged lepton produced at the primary vertex (see Section \ref{sec:offline}).
 Electrons will be clearly separated from neutral pions thanks to the micrometric accuracy and fine sampling of the Emulsion Cloud Chambers, which will enable photon conversions downstream of the neutrino interaction vertex to be identified.
 % The left panel of Figure~\ref{fig:neutrinos} shows a $\nu_e$ interaction in the OPERA emulsion cloud chamber.
 % The electron produced at the primary vertex is clearly separated from the electromagnetic shower induced by the two photons produced by the $\pi^0$ decay.
 Muons will be identified by the electronic detectors as the most penetrating particles.
 Tau leptons will be identified topologically in the ECCs, through the  observation of the tau decay vertex, together with the absence of any electron or muon at the primary vertex, following the technique developed by OPERA~\cite{Agafonova:2010dc,Agafonova:2018auq}.
 %The right panel of Figure~\ref{fig:neutrinos} shows a $\nu_\tau$ candidate detected in OPERA~\cite{Agafonova:2014bcr}. SND@LHC will have the same granularity as in OPERA, alternating emulsion films with 1\,mm thick passive material.
 %This is motivated by the need to keep high tracking and vertexing performance in an environment with a high density of tracks, rather than by the $\tau$ flight length which is a few cm long at the LHC energies.
 %Figure~\ref{fig:first} shows the first OPERA tau neutrino candidate~\cite{Agafonova:2010dc} where the decay channel $\tau \rightarrow \rho \nu_\tau$ with subsequent $\rho \rightarrow \pi^0 \pi$ decay was identified. 

FIPs will be identified through their scattering off electrons and nuclei of the emulsion target material.
In the case of a FIP elastic scattering off atomic electrons, the experimental signature consists of an isolated recoil electron that can be identified through the development of an electromagnetic shower in the target region.
For FIPs interacting elastically with a proton, instead, an isolated proton will produce a hadronic shower in the detector.
In both cases the background can be reduced down to a negligible level by topological and kinematic selections. 
The timing information will be used to confirm any excess of events with the expected signature~\cite{snd_technical_proposal}.

% \begin{figure}[bh]
% \centering
% \includegraphics[width=1\columnwidth]{figs/Overview/neutrino.png}
% \caption{Display of reconstructed tracks in the OPERA emulsion detector for a $\nu_e$ (left) \cite{Agafonova:2013xsk} and a $\nu_\tau$ (right) \cite{Agafonova:2014bcr} candidate event.\\
% \\}
% \label{fig:neutrinos}
\section{Veto system}
\label{sec:veto}
\begin{figure}[b]
    \centering
    \includegraphics[width=0.6\textwidth]{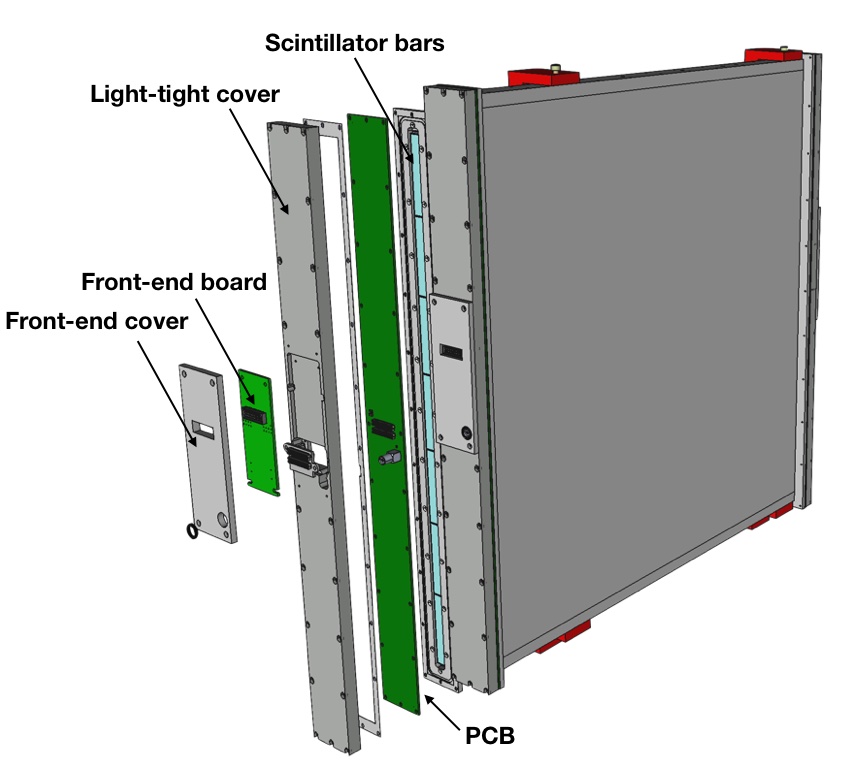}
    \caption{A rendering of the veto system illustrating the different components of the frames.}
    \label{fig:veto_parts}
\end{figure}

The veto system aims at rejecting charged particles entering the detector acceptance, mostly muons coming from IP1.
It is located upstream of the target region and comprises two parallel planes, located \SI{4.3}{cm} apart, of stacked scintillating bars read out on both ends by silicon photomultipliers (SiPMs) as shown in Figure~\ref{fig:veto_parts}.

One plane consists of seven \SI[product-units=power]{1 x 6 x 42}{cm} stacked bars of EJ-200 scintillator~\cite{ej200}.
EJ-200 is found to have the right combination of light output, attenuation length (\SI{3.8}{m}) and fast timing (rise time of \SI{0.9}{ns} and decay time of \SI{2.1}{ns}).
The emission spectrum peaks at \SI{425}{nm}, closely matching the SiPMs spectral response.
The number of photons generated by a minimum-ionising particle crossing \SI{1}{cm} scintillator is of the order of \num{e4}.
Bars are wrapped in aluminized Mylar foil~\cite{foil} to ensure opacity and isolate them from light in adjacent bars. 

\begin{figure}[t]
    \centering
    \includegraphics[width=\textwidth]{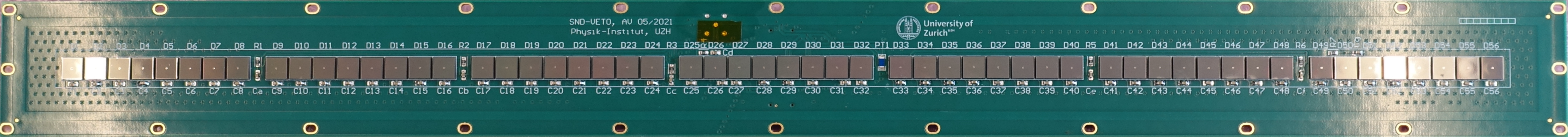}
     \includegraphics[width=\textwidth]{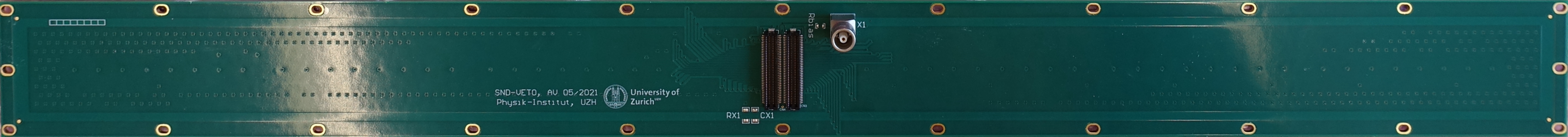}
    \caption{The two sides of the PCB for the veto system.}
    \label{fig:veto_pcb}
\end{figure}

Each bar end is read out by eight Hamamatsu S14160-6050HS~\cite{sipm_6050_url} (\SI{6 x 6}{mm^2} wide, \SI{50 x 50}{\micro m^2} pixel size) SiPMs.
The SiPMs are mounted on a custom built PCB, seen in Figure~\ref{fig:veto_pcb}, that covers all seven bars on each end of a plane.
A transparent silicone epoxy gel~\cite{silgel} fills the space of \SI{\sim 1}{mm} between the SiPMs and bars.
Each individual SiPM signal is read out by a single channel of the front-end (FE) board, containing two TOFPET2 ASICs (described in Section~\ref{subsec:tofpet}).
A DAQ board collects the digitized signals from four FE boards.
A CAEN mainframe, described in Section~\ref{subsec:readout}, which is shared with the muon system, houses low voltage (LV) and high voltage (HV) CAEN power supplies.
Details of the data acquisition (DAQ) system and of the boards are described in Section \ref{sec:daq}.
The total number of channels per PCB is 56, totaling 224 channels for the entire veto system.

The stacked bars for each plane are housed in an aluminum frame, with \SI{4}{mm} thick walls.
PCBs are mounted on \SI{4}{cm} wide rectangular flanges on both ends and act as end caps for the frame.
An aluminum cover on each end is used to ensure light tightness and also acts as a heat sink for the FE board, which generates \SI{\sim 3}{W} and is placed in a groove in the cover on the side opposite to the PCB.

\begin{figure}[t]
    \centering

    \includegraphics[height=0.36\textheight]{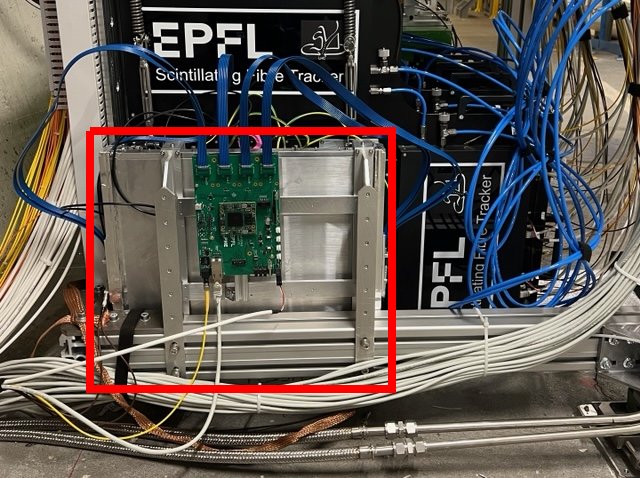}
    \includegraphics[width=0.36\textheight, angle=90]{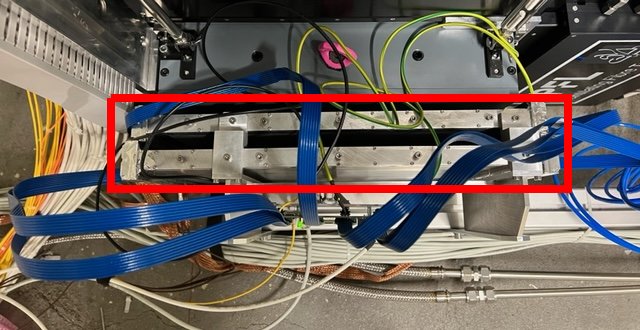}
    \caption{Front view of the veto system in the target region (left) and view from above (right) as seen in TI18. The veto planes are highlighted with a red rectangle.}
    \label{fig:veto_support}
\end{figure}

The two frames of the veto system are held together by a small support structure.
This in turn is attached to the support of the target region within \SI{1}{mm} accuracy, as shown in Figure \ref{fig:veto_support}.
A vertical shift of \SI{2}{cm} between the two frames allows for \SI{100}{\percent} coverage of the target region, compensating for inefficiency due to the dead area between bars introduced by wrapping material (\SI{\sim 60}{\micro m}) and variations in bar height (\SI{\sim 250}{\micro m}).
The DAQ board is mounted on the support frame directly in front of the veto planes.
Fine alignment is performed as part of the target region alignment as mentioned in Section~\ref{sec:target}.
    \section{Target Tracker and electromagnetic calorimeter}
\label{sec:scifi}

\subsection{Overview}

The Target Tracker system is made of five scintillating fibre (SciFi) planes interleaving the five target walls.
The SciFi technology is well suited to cover large surfaces in a low track density environment\footnote{
%The expected number of tracks is less than one per LHC bunch crossing.}
The expected rate of tracks from the ATLAS impact point is about \SI{0.8}{Hz/cm^2} at peak luminosity.}, where a \SI{\sim100}{\micro m} spatial resolution is required.  

The role of SciFi trackers is two-fold: assign a timestamp to neutrino interactions reconstructed in the ECC walls and provide an energy measurement of electromagnetic showers. 
Moreover, the combination of SciFi and scintillating bars of the muon detector will also act as a non-homogenous hadronic calorimeter for the measurement of the energy of the hadronic jet produced in the neutrino interaction and hence for the neutrino energy.

The matching with events reconstructed in the target walls is performed by connecting the centre of gravity of electromagnetic and hadronic showers, reconstructed in the SciFi immediately downstream of the ECC where the interaction occurred, with tracks reconstructed in emulsions. The large multiplicity of tracks produced in neutrino interactions and the high density of passing-through muons prevent a track-by-track matching between SciFi and ECC.

The measurement of electromagnetic shower energy is based on information provided both by ECC bricks and Target Tracker planes.
The five target walls (\SI{\sim 17}{X_0} each) interleaved with SciFi tracker modules, form a coarse sampling calorimeter.

The two main components employed in this SciFi tracker, the scintillating fibre mats and the multichannel SiPM photo-detectors, were developed by the EPFL group for the LHCb SciFi Tracker~\cite{LHCb_Tracker_TDR}.
The read-out electronics is different from the one used in LHCb and it has been optimised to have an improved time resolution and to detect electromagnetic showers.
%In contrast to the LHCb SciFi Tracker, where its readout electronics was optimized for single charged track information in a radiation environment, it serves different purposes in the SND@LHC detector.

%\paragraph{Connecting particle tracks between emulsion walls:}
%The lack of time information in the emulsion technology makes merging single tracks or EM showers between emulsion walls inefficient or almost impossible. The merging will be assisted by the position and time information recorded by the SciFi stations. Each $xy$-SciFi module acts as a seed in space and time for single particle or centre of gravity of EM showers.
%The expected single particle spatial resolution is of order of \SI{\sim 50}{\micro m} and the time resolution for a particle crossing a $x$ and $y$ plane \SI{\sim 210}{ps}.
%Multiple simultaneous tracks from EM showers will lead to an improved time resolution.

%\paragraph{EM and hadronic sampling calorimeter:} 
%The five emulsion walls, each $17 X_0$ thick, followed by the SciFi tracker modules form a coarse sampling calorimeter.
%The primary neutrino interaction will take place at an average distance of $42 X_0$, half of the target region thickness.
%The EM showers will therefore typically extended only over two SciFi tracker planes, resulting in a coarse energy information by the SciFi tracker.
%The energy calibration will be based on the minimum ionizing particle (MIP) signal amplitude and a GEANT4 simulation.     

\begin{figure}[h]
\centering
\includegraphics[width=0.49\linewidth]{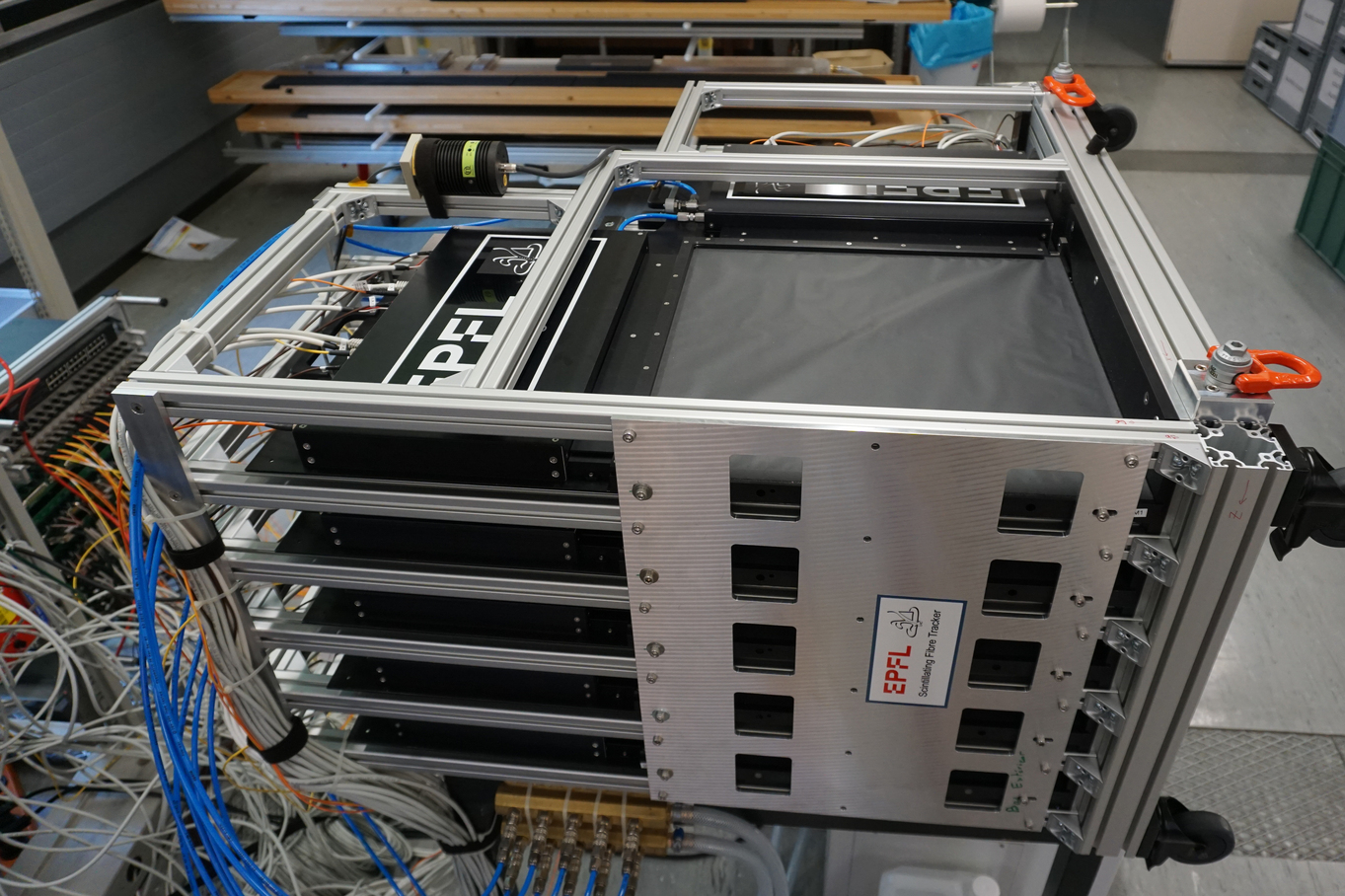}
\caption{The SciFi tracker detector setup in the lab for cosmic ray test. The active area of a tracker module is \SI{40 x 40}{cm} and consists of \num{5} $xy$ fibre planes.}
\label{fig:scifi_cosmics_lab} 
\end{figure}
  
\subsection{The SciFi modules}
\begin{figure}[h]
\centering
\includegraphics[width=0.49\linewidth]{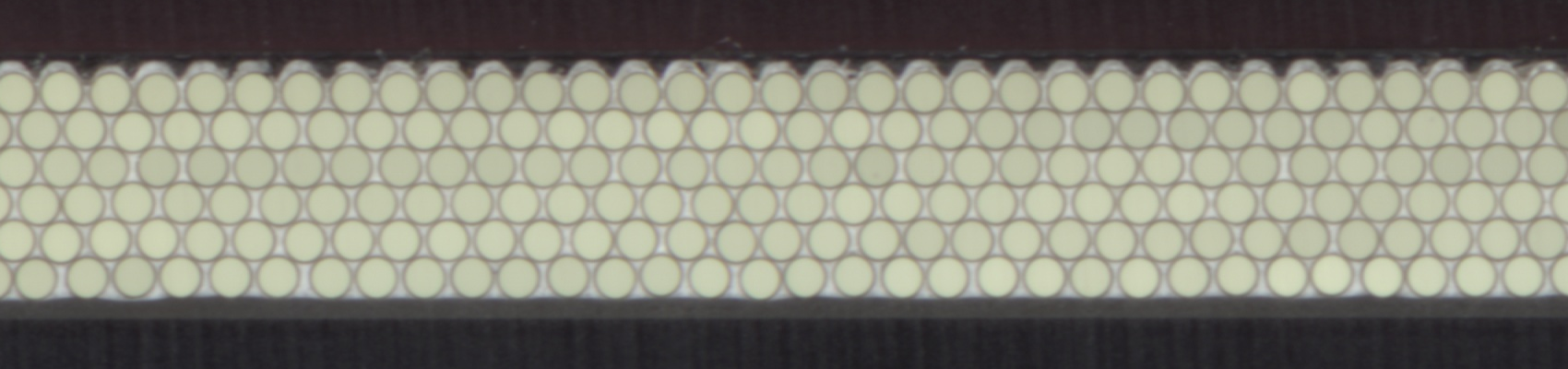}
\caption{The fibre mat is composed of six layers of fibres glued with a titan oxide loaded epoxy glue to suppress cross talk between fibres.}
\label{fig:six_layer_mat}
\end{figure}

The SciFi modules for SND@LHC, shown in Figure~\ref{fig:scifi_cosmics_lab}, are closely following the design of the \SI{2.5}{m} long modules built for LHCb.
The double-cladded polystyrene scintillating fibres from Kuraray (SCSF-78MJ), with a diameter of \SI{250}{\micro m}, are blue emitting fibres with a decay time of \SI{2.8}{ns}.
The fibres are arranged in six densely-packed staggered layers, forming fibre mats of \SI{1.35}{mm} thickness. A picture of the cross section of such a mat is shown in Figure~\ref{fig:six_layer_mat}.

\begin{figure}[h]
\centering
\includegraphics[width=0.7\linewidth]{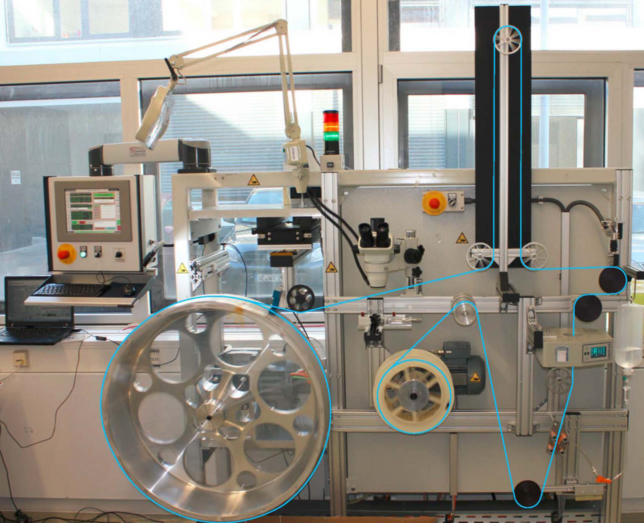}
\caption{\label{fig:fibre_machine} The winding wheel with a circumference of \SI{2.5}{m} allows to wind five \SI{40}{cm} mats. The winding process has been refined and adjusted in order to obtain precise and regular fibre mats. The path of the fibre in the machine is highlighted in cyan.}
\end{figure}

The fibre winding and gluing process has been developed within the LHCb SciFi collaboration.
A dedicated winding machine, shown in Figure~\ref{fig:fibre_machine}, with tension and position control as well as optical feedback has been engineered.
Fibre mats produced for the SND@LHC tracker are \SI{133}{mm} wide and \SI{399}{mm} long; they are integrated into a fibre plane with less than \SI{500}{\micro m} dead zones.% between mats. 

A polycarbonate end-piece is glued and an optical surface cut is applied to each end of the fibre mat.
One side of the mat is brought in direct contact with the epoxy entrance window of the photo-detector and the other end can optionally have a mirror or a light injection fibre coupling.
%For the SND@LHC modules, a light-leaking fibre is inserted for calibration purposes. 

\subsection{The SiPM photo-detectors and readout electronics}
\begin{figure}[h]
     \centering
     \begin{subfigure}[b]{0.49\textwidth}
         \centering
         \includegraphics[width=1.1\textwidth,angle=270]{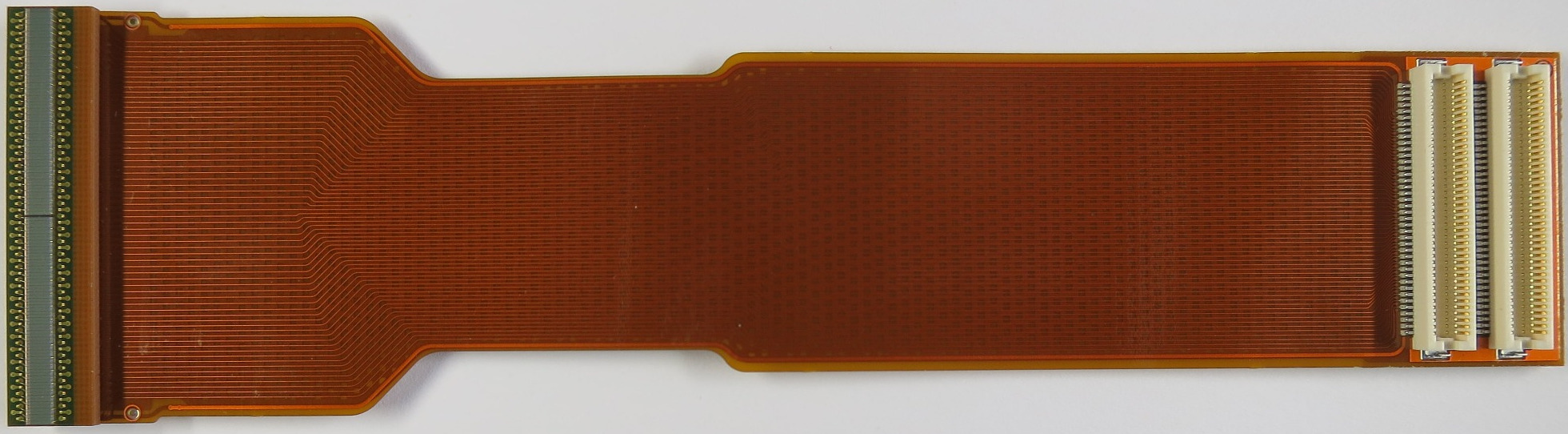}
         \caption{}
         \label{fig:scifi_sipm_flex}
     \end{subfigure}
     \hfill
     \begin{subfigure}[b]{0.49\textwidth}
         \centering
         \includegraphics[width=\textwidth]{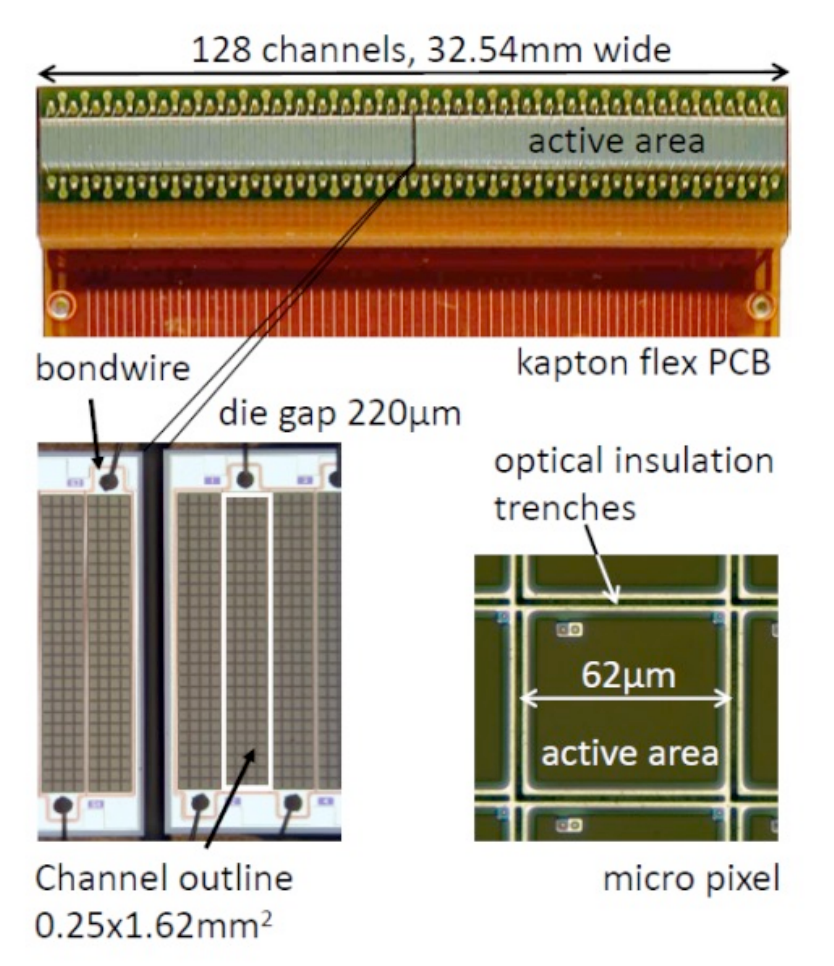}
         \caption{}
         \label{fig:scifi_sipm_details}
     \end{subfigure}
        \caption{SiPM arrays used for the light detection in the SciFi, mounted on the flex (a). The connector on the bottom mates to the front-end board, shown in Figure~\ref{fig:fe-board}. Details of the SiPM array (b).}
        \label{fig:scifi_sipm}
\end{figure}

The readout consists of the photo-detector (S13552 SiPM multichannel arrays by Hamamatsu) at the end of the fibre module, a short Kapton flex PCB holding the photo-detector and signal connectors and the front-end electronics board, shown in Figure~\ref{fig:scifi_sipm_flex}.
The light tightness of the module is ensured by a seal on the flat Kapton flex, the aluminium module frame and an opaque Tedlar sheet on both sides of the module.
This encloses the photo-detector and the entire fibre region. The light tightness is evaluated during the assembly phase and leaks closed with glue.  
The photo-detectors are not actively cooled, as their heat dissipation is low and the expected noise is acceptable at the operation temperature of \SI{15}{\degreeCelsius}. 

The SiPM multichannel array is optimised for low light-intensity detection.
For application in SND@LHC, the SciFi performance has to be tuned for maximising the hit detection efficiency at an acceptable noise rate.
The final threshold chosen for operation produces a noise rate of \SI{25}{Hz} per channel, which poses no problem to the event builder (described in Section~\ref{subsec:daq-events}) and can be efficiently suppressed by the online noise filter. 
The SciFi detector with this configuration features an efficiency of \SI{\sim 99}{\%}, as discussed in Section~\ref{subsubsec:commissioning_scifi_result}.

The array used in SND@LHC, shown in Figure~\ref{fig:scifi_sipm}, has an active channel area of \SI[product-units=power]{0.25 x 1.625}{mm}, a peak photo-detection efficiency (PDE) of \SI{47}{\%} and a gain of \num{3.6e6} at \SI{3.5}{V} over-voltage.
To obtain a high PDE, the choice of large pixels \SI{62.5 x 57.5}{\micro m}, leading to an array of \SI{26 x 4}{} pixels per channel, significantly limits the linearity and the dynamic range to about 50 photo-electrons per channel.
%A new sensor design with smaller pixels is in preparation but was not available at the construction time of this detector.
For this operation condition, the light yield (LY) for a minimum ionising particle (MIP) traversing the fibre plane in the center of the module, is \num{\sim 25} photo-electrons (PE).
%With an average number of channels collecting the light of a single particle (cluster size) of \num{2.5} and a desired hit detection efficiency of \SI{>99}{\%} the noise rejection threshold required is \num{5.5} photo-electrons.
%For this threshold setting, a noise rate of \SI{100}{Hz} per channel is measured.
%Due to dead regions at the photo-detector edges, the hit detection efficiency is reduced at the every boundary of every \num{64} channels.
%The inefficiency at the edges can be partially recovered by the neighbouring channels but at the boundaries of fibre mats a dead region of \SI{500}{\micro m} is present. 

The readout chip chosen for this tracker is the TOFPET2 ASIC~\cite{tofpet2_url,Schug:2018klm,Nadig:2019rjw}, described in Section~\ref{subsec:tofpet}.
%It has been designed for time of flight positron emission tomography (TOFPET).
%In comparison to the SciFi tracker application, the signal of TOFPET systems is as large as \SI{1500}{PE} and requires a large dynamic range and high noise thresholds.
Its power consumption is \SI{1.5}{W} per \num{64} readout channels, including the loss for linear voltage regulation.
A water cooling system has been chosen to counteract limited convection due to dense packing between modules.
To simplify the mechanical design of the water cooling, the FPGA of the DAQ boards is connected to the large aluminium support of the module and not to the water cooling.
The thermal design has been verified and the temperature lies within the required range during operation.
The heat dissipation of the electronics into the target enclosure is \SI{\sim 24}{W} per board or a total of \SI{\sim 720}{W} for the complete SciFi tracker.

\subsection{Low voltage and SiPM bias voltage}
The power for each DAQ board is provided with a CAEN A2519, an 8-channel, \SI{15}{V}, \SI{5}{A} power supply module, hosted in one of the CAEN mainframes (see Section~\ref{subsec:readout} for more detail).
To optimise the cost for the bias voltage of the SiPMs, a single channel per DAQ board is used.
A group selection  of SiPM arrays allows to minimise the break down voltage spread among SiPMs biased by the same bias voltage.

\subsection{Calibration}
The DAQ electronics provide an electrical injection signal, synchronous to all TOFPET2 FE chips on one board.
This allows for a first-order time calibration between channels on the same board.
Subsequently, a fine time calibration based on muon tracks among different boards and layers can be used to correct and verify the time calibration based on the collected data during the runs.
The studies from a DESY test beam in October 2019 show that, based on the initial time alignment, the time calibration for channels can be improved using data with single tracks producing multiple hits in all ten SciFi layers. %This procedure requires propagation time compensation in the SciFi.

%The SciFi technology employed for the target tracker has been developed and optimised for the LHCb tracking purposes where a small angle between two detection layers is used to reduce the number of ghost hits.
When detecting electromagnetic showers in a $x$-$y$ detector layout, a large number of tracks are produced in a small region of space and therefore only a projection of the shower profile can be obtained.
Additionally, the pixelised silicon photomultiplier (SiPM) suffers from non-linear amplitude response due to the limited number of pixels. 
With a light yield of \SI{25}{PE} and a total of \num{104} pixels per channel, a pixel occupancy of almost \SI{50}{\percent} is expected for a shower track density of \num{2} tracks per channel (\SI{250}{\micro m}).
Beyond this track density, a strong non-linear response of the detector signal is expected.
The saturation is of statistical nature and can be corrected to make the detector response linear. 
To obtain a correlation between the measured signal amplitude by the TOFPET2 electronics and the number of MIP tracks in the detector, a \textsc{Geant4} simulation will be used to model the EM shower development and the SiPM saturation.

\subsection{Alignment}
The mechanical alignment between the SciFi planes and the emulsion boxes is ensured with mechanical precision pins, constraining the relative position between the two objects.
%The two objects are fixed with screws. 
%The large number of high momentum muon tracks present in the target region allows spatial alignment using single tracks between SciFi planes.
Because of the large number of tracks from high-momentum muons traversing the target (a few thousands/\SI{}{cm^2}/\SI{}{fb^{-1}}), an accurate spatial alignment between SciFi planes can be obtained by using the tracks themselves. 
%An online alignment is not required for data acquisition and noise suppression.

Each SiPM array of \num{128} channels is expected to have a constant shift relative to the nominal position 
%(only shifts in $x$ for $x$-layers and $y$ for $y$ layers are important) 
and each fibre mat (three per detection plane) has to be corrected for its constant rotation angle.
These corrections have been studied during the commissioning in the SPS H6 beam line, presented in Section~\ref{subsec:commissioning_h6}.
%Typical values for the SiPMs shift are \SI{\pm 500}{\micro m} and rotation angle of the fibre mats \SI{\pm 0.5}{mrad}. The muon track data will also be used for relative alignment between emulsion and target tracker. 

%\clearpage \newpage

\section{Target and vertex detector}
\label{sec:target}

\subsection{Overview}

The emulsion target is based on the Emulsion Cloud Chamber (ECC) technique, that makes use of nuclear emulsion films interleaved with passive layers to build up a tracking device with sub-micrometric spatial and milliradian angular resolution, as demonstrated by the OPERA experiment~\cite{Acquafredda:2009zz}.
It is capable of detecting $\tau$ leptons~\cite{Agafonova:2018auq} and charmed hadrons~\cite{Agafonova:2014khd} by disentangling their production and decay vertices.
It is also suited for FIP detection through the direct observation of their scattering off electrons or nucleons in the passive plates.
%The high spatial resolution of  nuclear emulsion films allows for identifying electrons by observing electromagnetic showers in the ECC~\cite{Agafonova:2018dkb}.

The ECC technology alternates 1-mm thick tungsten plates acting as the neutrino target with $\sim$300-micron thick films, each made of two sensitive emulsion layers poured on a plastic base, acting as tracking devices with micrometric accuracy.
%In addition to the micrometric position resolution, each emulsion layer provides track segments of charged particles tracks with a few milliradian angular accuracy. 
The reconstruction of track segments in consecutive films provide the vertex reconstruction with an accuracy at the micron level. The fine segmentation of active films interleaving tungsten plates is motivated by the longitudinal resolution required to observe the tau lepton track and by the need to keep the combinatorial background in the association of track segments sufficiently low over an integrated luminosity of about 25 fb$^{-1}$ (corresponding to  $\sim 45$ days of data taking in nominal conditions), after which the emulsion films are replaced. It also makes the emulsion-tungsten ECC a high-sampling electromagnetic calorimeter with more than three active layers every radiation length, $X_0$, essential for electron identification and discrimination against neutral pion decays~\cite{Agafonova:2018dkb}.

\begin{figure}[htbp]
\centering
\includegraphics[width=0.9\columnwidth]{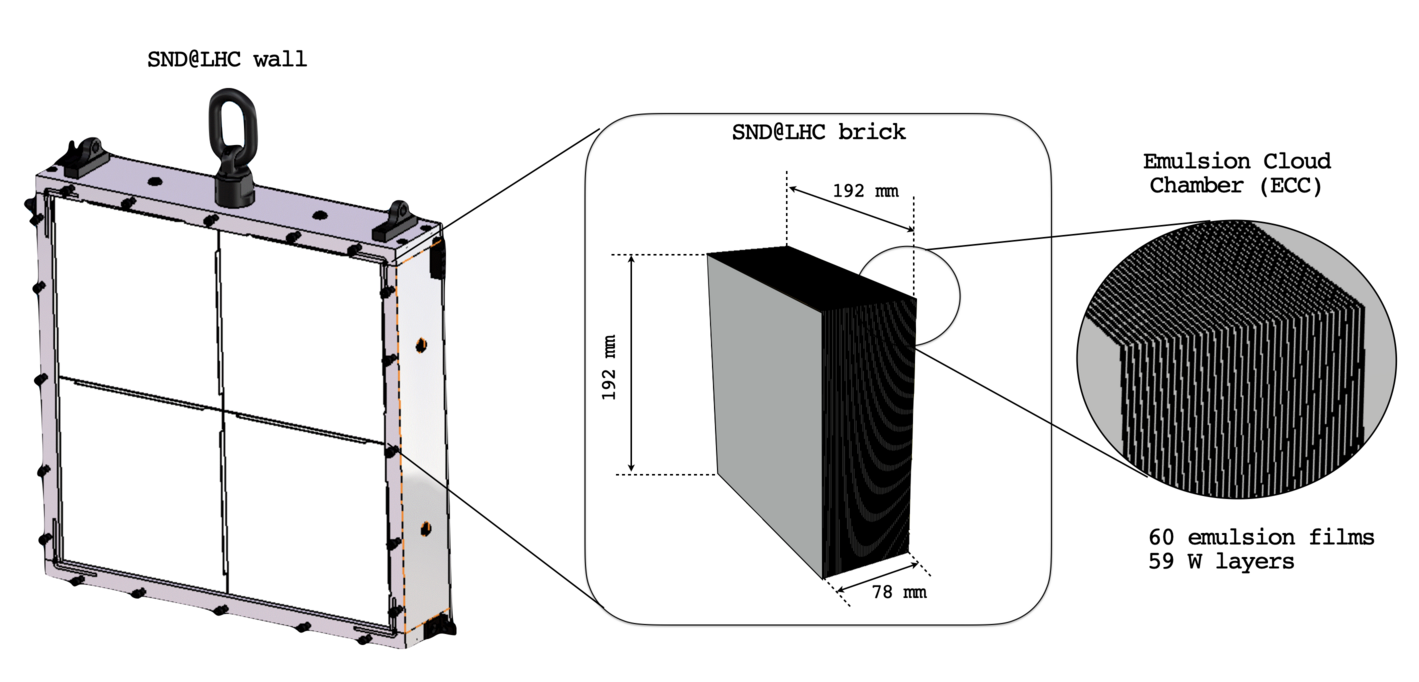}
\caption{An emulsion wall is composed of four bricks, each consisting of 60 emulsion films interleaved with 59 tungsten sheets.}
%\caption{Layout of the emulsion target, consisting of five walls. Each wall is made by four bricks.}
\label{fig:emulsion_target}
\end{figure}

The emulsion target is made of five walls with a sensitive transverse size of \SI[product-units=power]{384 x 384}{mm}.
Each wall consists of four cells, called \emph{bricks} as illustrated in Figure~\ref{fig:emulsion_target}.
Each brick is made of 60 emulsion films with a transverse size of \SI[product-units=power]{192 x 192}{mm}, interleaved with 59 \SI{1}{mm}-thick tungsten plates.
The resulting brick has a total thickness of \SI{\sim 78}{mm}, making \SI{\sim 17}{X_0}, and a mass of \SI{41.5}{kg}.
The overall target mass with five walls of \num{2 x 2} bricks amounts to \SI{830}{kg}.

The layout of the target was optimised to fulfill conflicting requirements: overall dimensions that cover the desired pseudo-rapidity region and maximse the azimuthal angular acceptance, large emulsion surface to maximise the event containment in the brick and reduced number of bricks per wall to minimise the dead area between adjacent cells.

%Nuclear emulsion films are produced by Nagoya University  and by the Slavich Company in Russia. For the long-term stability of the emulsion films, the temperature of the target will be  kept at 15$^\circ$C. 

\subsection{Target walls}
Nuclear emulsion films are the most compact, thinnest and lightest three-dimensional tracking detectors with sub-micrometric position and milliradian angular resolution.
A nuclear emulsion film has two sensitive layers (\SI{70}{\micro m}-thick) on both sides of a transparent plastic base (\SI{170}{\micro m}-thick).
By connecting the two hits generated by a charged particle on both sides of the base, the slope of the track can be measured with milliradian accuracy. 
%Nuclear emulsion films used in SND@LHC will have a transverse size of $192\times192\,$mm$^2$.
The whole detector will contain 1200 emulsion films, for a total of \SI{44}{m^2}. 
Emulsion films will be produced by the Nagoya University in Japan and by the Slavich Company in Russia.\footnote{Slavich Company, Yaroslavskaya Region, Peresavl-Zalessky, Russia.}
%For the long-term stability of the emulsion films, the temperature of the target is  kept at \SI{15}{\celsius}.

Emulsion films are analysed by fully automated optical microscopes~\cite{Arrabito:2006rv,Armenise:2005yh}.
The scanning speed, measured in terms of film surface per unit time, was significantly increased in recent years~\cite{Alexandrov:2015kzs,Alexandrov:2016tyi,Alexandrov:2017qpw}, reaching \SI{\sim 180}{cm^2/h}.
R\&D is still ongoing~\cite{Alexandrov:2019dvd} to further increase the scanning speed.

Tungsten was selected as target material in order to maximise the interaction rate per unit volume.
Its small radiation length (\SI{\sim 3.5}{mm}) allows for good performance in the electromagnetic shower reconstruction in the ECC.
The low intrinsic radioactivity makes tungsten a suitable material for an emulsion detector.\footnote{Tungsten supplied by Luoyang Sifon Electronic Material Co Ltd, China, through the INTENT Company, Torino, Italy.}
%Tungsten plates have a transverse size of \SI[product-units=power]{192 x 192}{mm}, the whole detector will contain 1180 passive layers.

\begin{figure}[htbp]
\centering
\includegraphics[width=1.0\columnwidth]{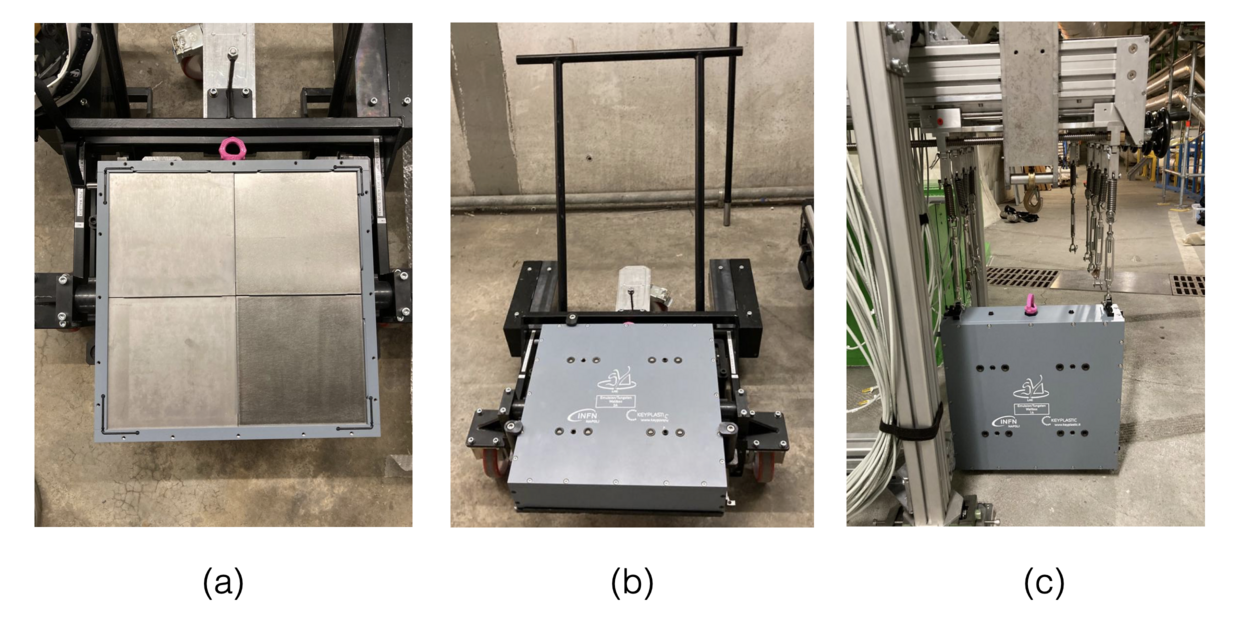}
\caption{Target wall during the assembly with tungsten plates (a), on the transportation trolley (b) and suspended from the mechanical structure (c).}
\label{fig:wall}
\end{figure}

%The different assembly phases are sketched in Figure~\ref{fig:assembly}\footnote{Design of wall structure and assembly tools performed in collaboration with the GWM Company, Suisio (BG), Italy.}. 
An ECC wall is contained in an aluminum box that hosts the four bricks, which are assembled one after the other by piling up 60 emulsion films and 59 tungsten sheets.
The box is then closed using a semi-automatic tool that keeps the necessary pressure to avoid relative displacements between emulsion films.
Once closed, the box is light tight.
Each wall is transported from the dark room where it is assembled to the TI18 tunnel by means of a custom trolley and, once there, inserted into the mechanical structure of SND@LHC.
The different phases of the wall assembly, transportation and installation are described in Figure~\ref{fig:wall}.\footnote{Design of wall structure, transportation trolley and assembly tools performed in collaboration with the KeyPlastic Company, Montale (MO), Italy.}

%\begin{figure}[htbp]
%\centering
%\includegraphics[width=0.9\columnwidth]{images/Target/trolley.png}
%\caption{Wall transportation trolley in the travelling %(top) and loading (bottom) positions. }
%\label{fig:trolley}
%\end{figure}

The emulsion target will be replaced every $\sim$\SI{25}{fb^{-1}}.
The exchange of target walls will be performed during LHC Technical Stops.
Since it is not assured that the integration of the target luminosity will be in coincidence with Technical Stops, the Collaboration has developed a  procedure for a fast brick replacement (about \SI{5}{hour} shift), that could fit within shorter accesses to the LHC tunnel.

\subsection{Target mechanics}
\begin{figure}[htbp]
\centering
\includegraphics[width=1.0\columnwidth]{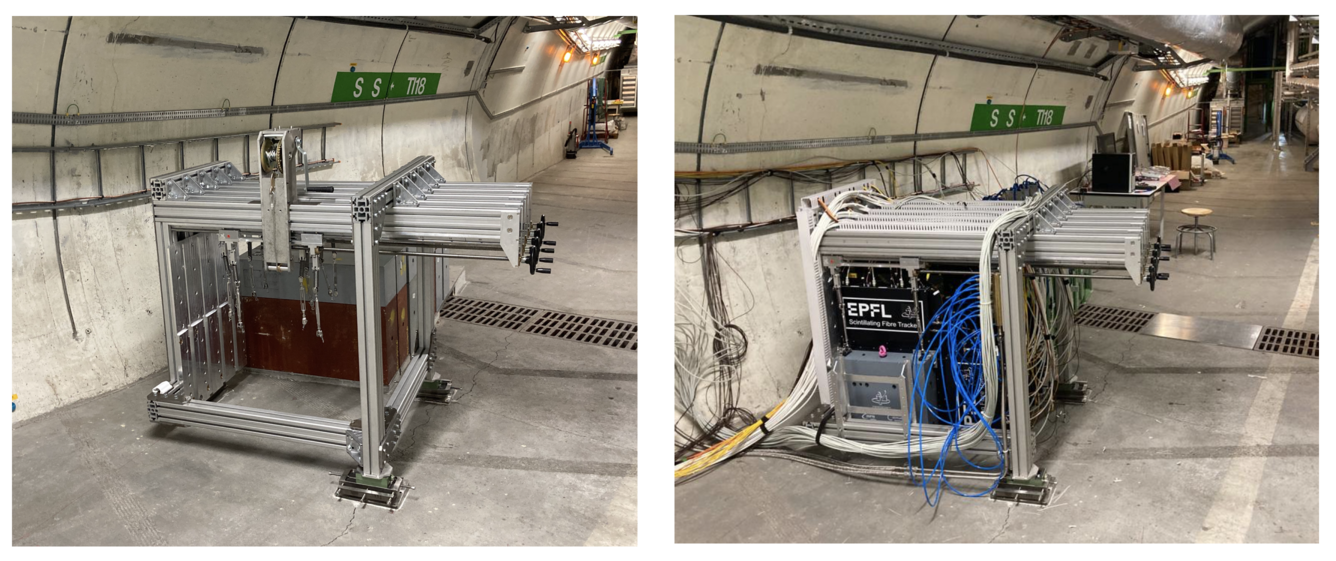}
\caption{Mechanical support of the target system after the installation (left) and fully loaded with wall boxes and SciFi planes (right).}
\label{fig:structure}
\end{figure}

The mechanical structure of the SND@LHC target was designed to have a single support structure for both the five emulsion/tungsten walls and the five SciFi planes.
It is made of a vertical rectified aluminum plate, that guarantees a fine mechanical alignment of target walls, and of five aluminum horizontal profiles, each sustaining a target wall, as shown in Figure~\ref{fig:structure}.
Each SciFi plane is fixed to the upstream wall box via three pins.
Wall boxes are suspended to two horizontal profiles by two rope tensioners, two springs and a pendulum link.
Each wall box is placed into the \emph{loading position} with the transportation trolley, it is then suspended to the structure and translated to the \emph{final position} via recirculating ball guides.
Finally, the wall box is secured to the vertical plate with two screws.   

The whole structure is supported isostatically on three points.
Alignment feet are used to adjust the height of the structure, to compensate for the inclined floor.
Horizontal plates located below each foot are used for fine adjustment of the target position on the tunnel floor. The alignment of the target system is performed via three alignment spheres located at the rear of the vertical plate. The required mechanical tolerances are of the order of the millimeter, being the fine alignment performed with penetrating tracks coming from the ALTAS impact point.

\subsection{Neutron shield and cold box}
\label{subsec:coldbox}
\begin{figure}[h]
     \centering
     \begin{subfigure}[b]{0.46\textwidth}
         \centering
         \includegraphics[width=\textwidth]{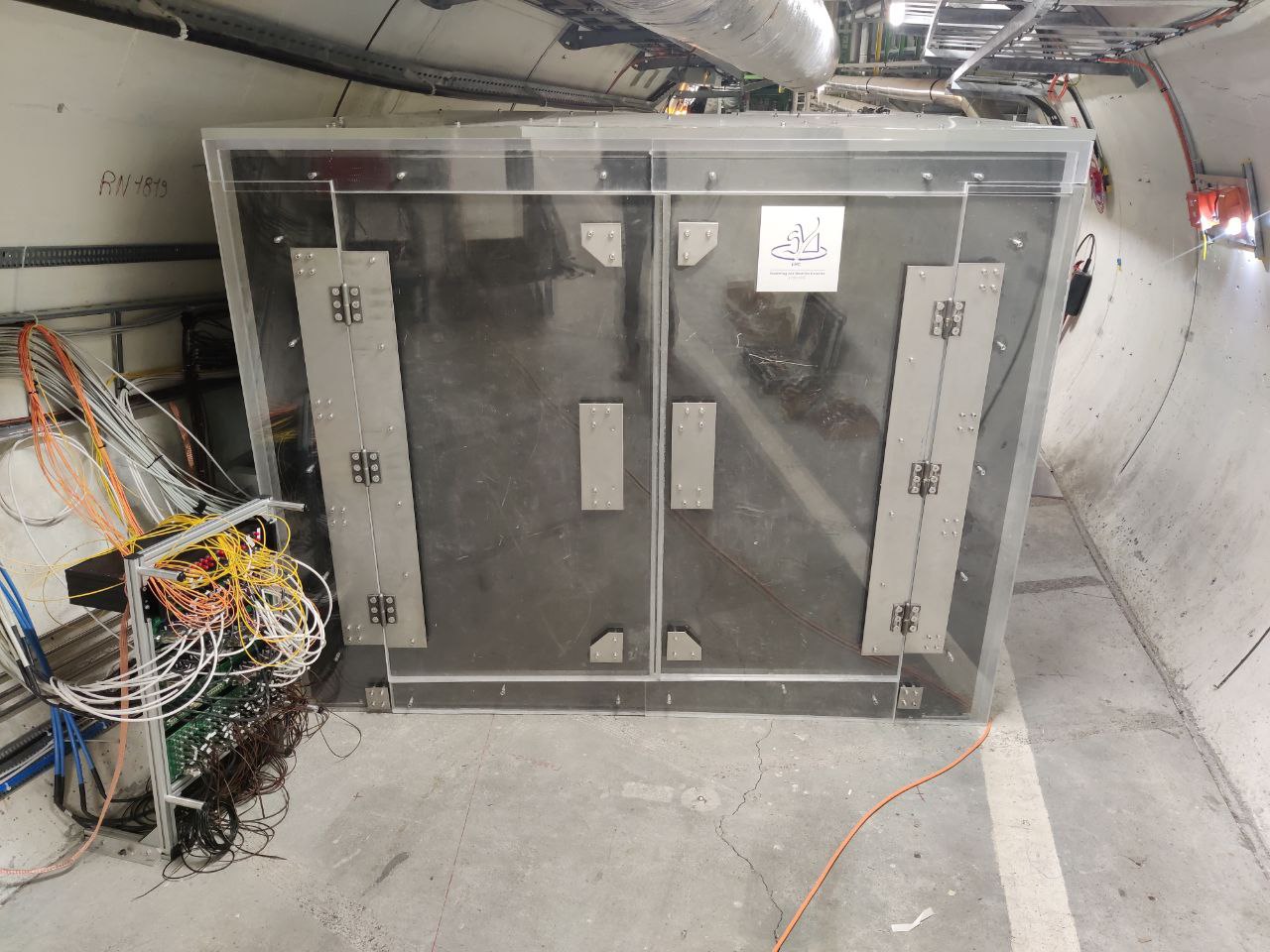}
     \end{subfigure}
     \begin{subfigure}[b]{0.53\textwidth}
         \centering
         \includegraphics[width=\textwidth]{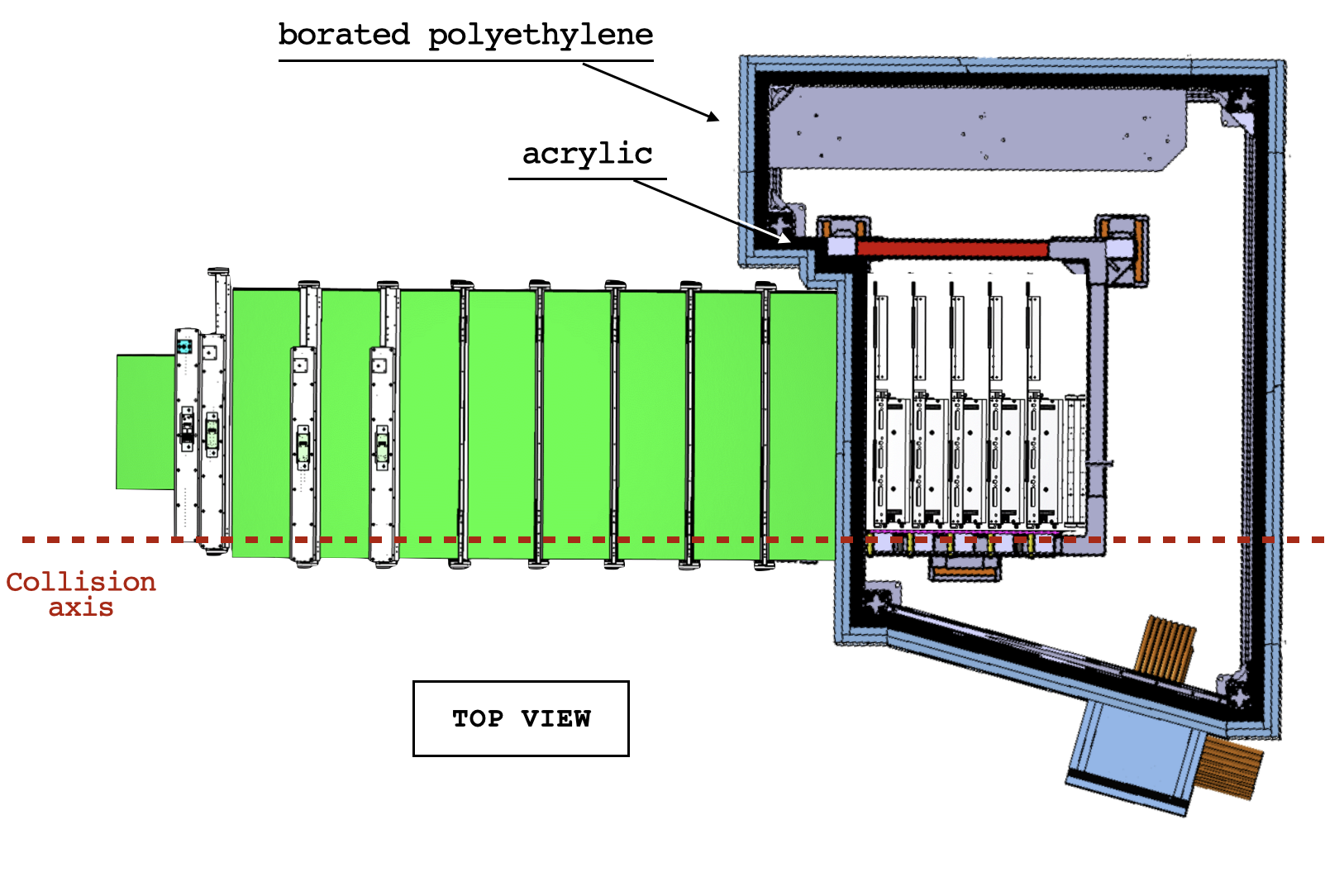}
     \end{subfigure}
        \caption{Neutron-shielded box surrounding the target region. On the left a picture, taken from upstream, of the assembled shield and on the right the top view of a schematic drawing.}
\label{fig:coldbox}
\end{figure}

The interaction of proton beams with the residual gas inside the LHC beam pipe produces low energy neutrons, with a spectrum ranging from a few meV to a few hundreds of MeV, about half of them being thermal neutrons. 
The neutron flux expected in the TI18 tunnel is predominantly produced by the counterclockwise beam (beam 2) that passes by TI18 while moving towards IP1.
Thermal neutrons would cause an increase of {\it fog}, $i.e.$ number of developed grains per unit volume, in the emulsion films. A shielding box was therefore built around  the target region. It is made of \SI{4}{cm}-thick 30\% borated polyethylene and \SI{5}{cm}-thick acrylic layers, as shown in Figure~\ref{fig:coldbox}.
It provides a background reduction of a factor of \num{e-7}~\cite{snd_technical_proposal}.

%For this purpose, an insulated box was built around the target region and a cooling system was installed, as shown in Figure~\ref{fig:coldbox}. 
%The walls of the insulated box are made of \SI{4}{cm}-thick 30\% borated polyethylene and \SI{5}{cm}-thick acrylic layers, to protect the apparatus from the neutron flux.
%and absorb low-energy neutrons originated from beam-gas interactions:

The box acts also as an insulation chamber. For the long-term stability of emulsion films, a cooling system was installed to keep the temperature of the target at ($15\pm1$)\SI{}{\celsius} and the relative humidity in the range \num{40} to \SI{50}{\percent}. 

% \begin{figure}[htbp]
% \centering
% \includegraphics[width=1.0\columnwidth]{images/Target/cold_box.png}
% \caption{Neutron-shielded box surrounding the target region. On the left a picture, taken from upstream, of the assembled shield and on the right the top view of a schematic drawing.}
% \label{fig:coldbox}
% \end{figure}

%%%%%%

%\input{sections/hcal}
\section{Hadronic calorimeter and muon system}
\label{sec:muon}
\subsection{Overview}
\begin{figure}[t]
    \centering
    \includegraphics[width=0.7\textwidth]{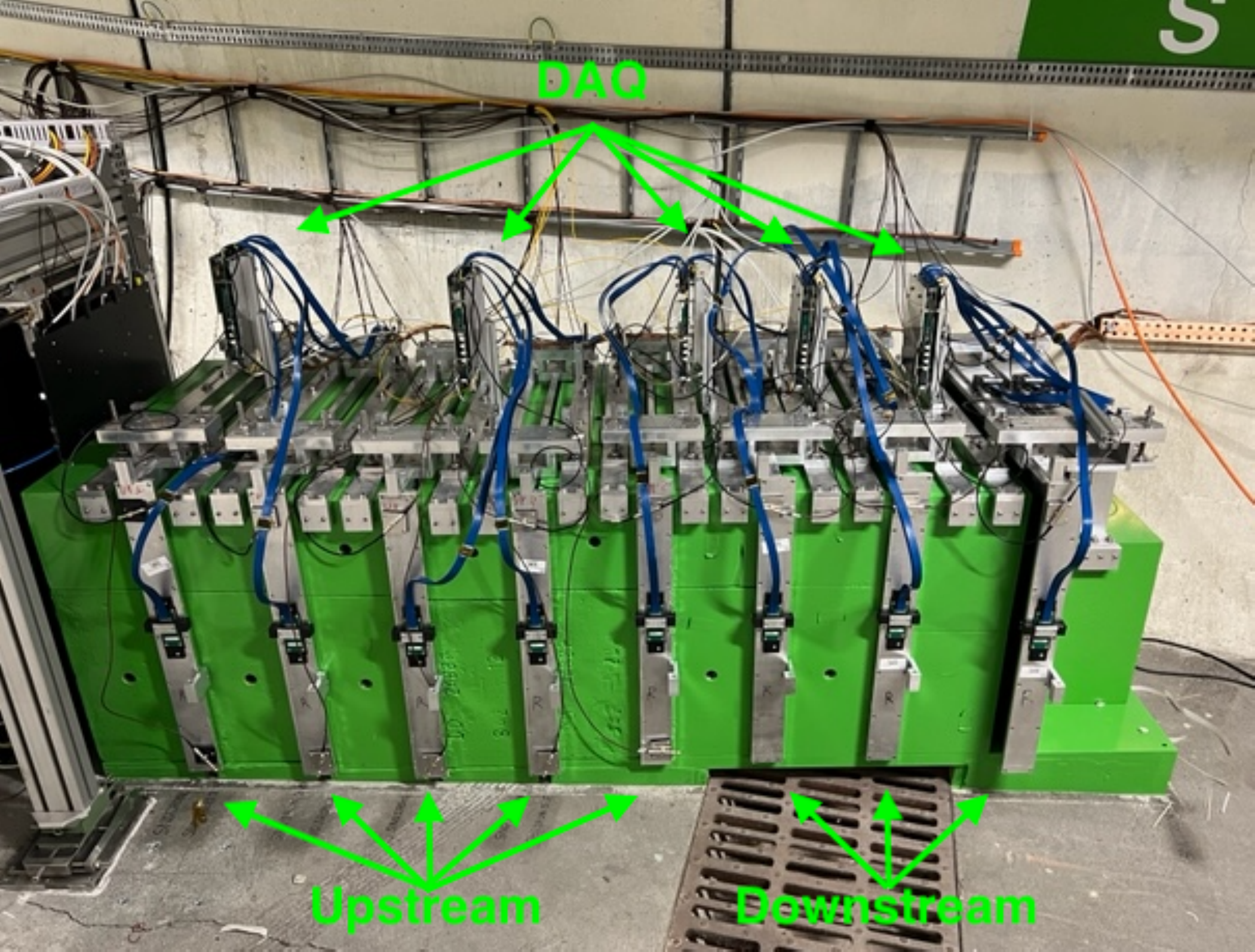}
    \caption{Picture of the hadronic calorimeter and muon system installed in TI18.}
    \label{fig:muon_tot}
\end{figure}

\begin{figure}[t]
    \centering
    \includegraphics[width=0.8\textwidth]{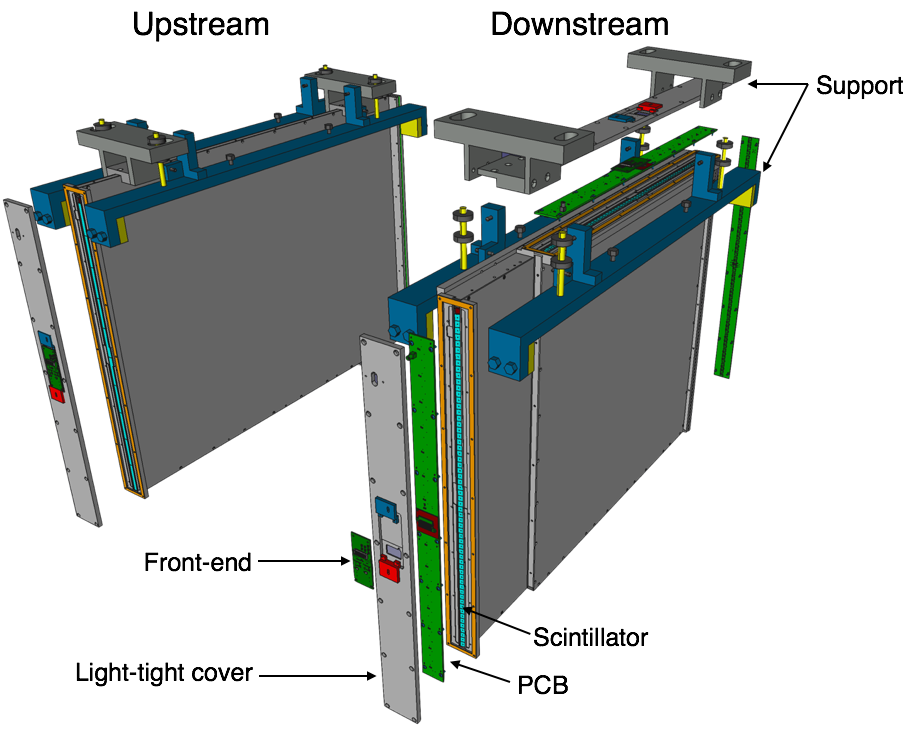}
    \caption{Illustration of an upstream and a downstream planes.}
    \label{fig:muon_us_ds}
\end{figure}

Downstream of the target region lies the hadronic calorimeter and muon system, shown in Figure \ref{fig:muon_tot}.
Its primary purpose is to identify passing-through muons and, together with the SciFi, it serves as a sampling hadronic calorimeter, enabling measurement of the energy of hadronic jets.
It comprises eight layers of scintillating planes interleaved with \SI{20}{cm}-thick iron slabs, which acts as passive material with a thickness of $9.5\lambda_{\rm int}$. This adds up to an average total of $11\lambda_{\rm int}$ for a shower originating in the target region.
Muons are identified as being the most penetrating particles through all eight planes.
The system is further divided in two sections: the first five upstream layers (US), made of \SI{6}{cm}-thick horizontal scintillating bars, and the last three downstream layers (DS), made of fine-grained horizontal and vertical scintillating bars, illustrated in Figure~\ref{fig:muon_us_ds}.

\subsection{Upstream system}

The first five US layers are similar to the veto planes, albeit with different dimensions. Each layer consists of ten stacked bars of EJ-200, each bar having dimensions \num{1 x 6 x 82.5} \SI{}{cm^3}.
The length was chosen to be longer than the iron blocks to allow the FE to be placed outside the gap between them, reducing the space needed for the gap and overall length of the muon system along the collision axis, a critical parameter in the apparatus design as described in Section~\ref{sec:introduction}. The bars are wrapped in aluminized Mylar foil in the same fashion as the veto system.

\begin{figure}[t]
    \centering
    \includegraphics[width=\textwidth]{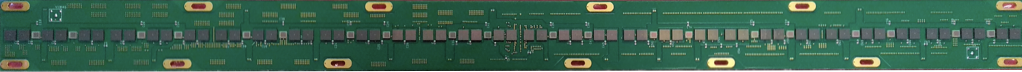}
    
    \includegraphics[width=\textwidth]{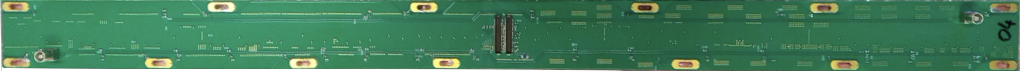}
    \caption{The two sides of the PCB for the US muon system.}
    \label{fig:PCB_US}
\end{figure}

\begin{figure}[t]
    \centering
    \includegraphics[width=0.49\textwidth]{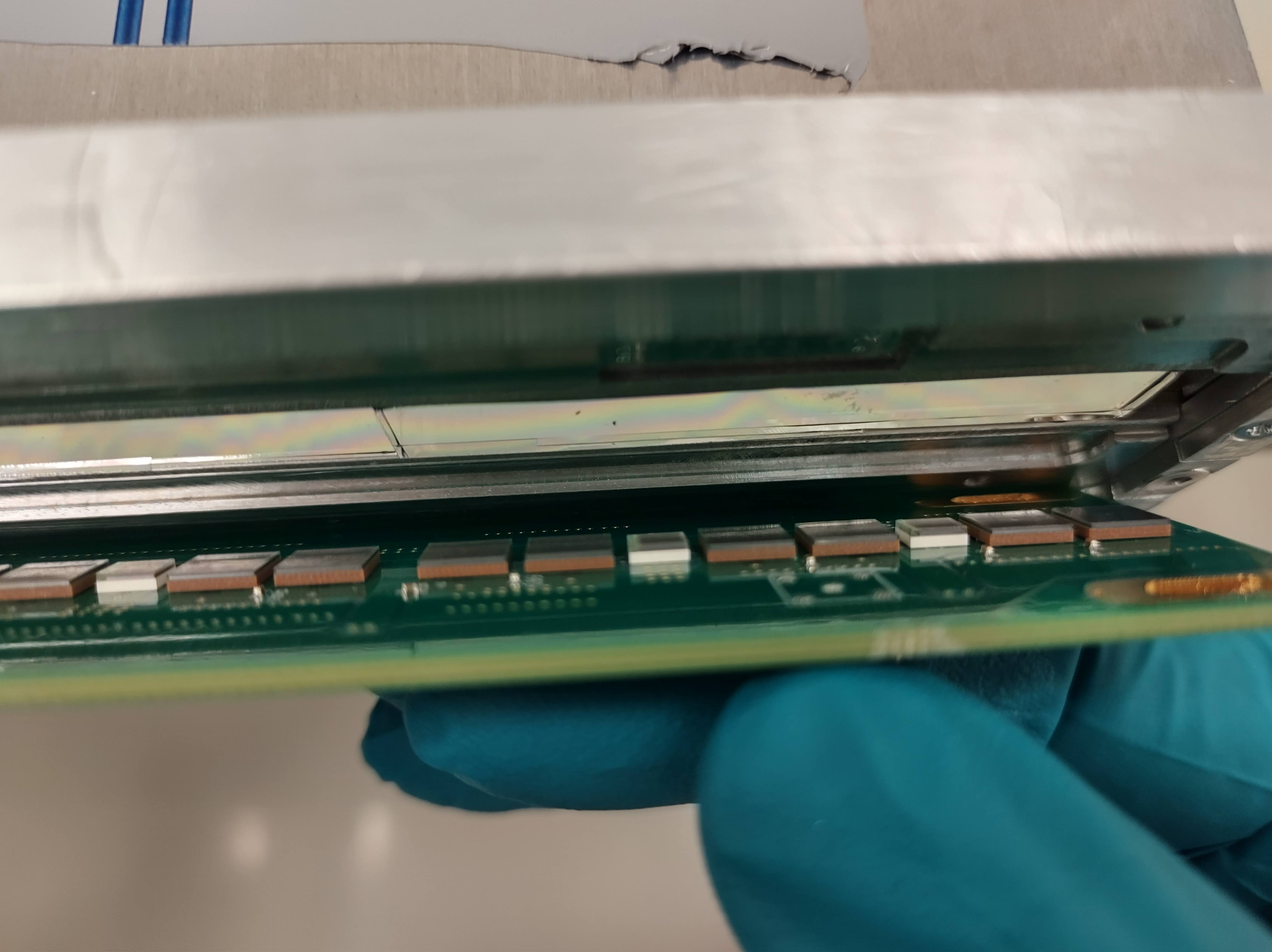}
    \includegraphics[width=0.49\textwidth]{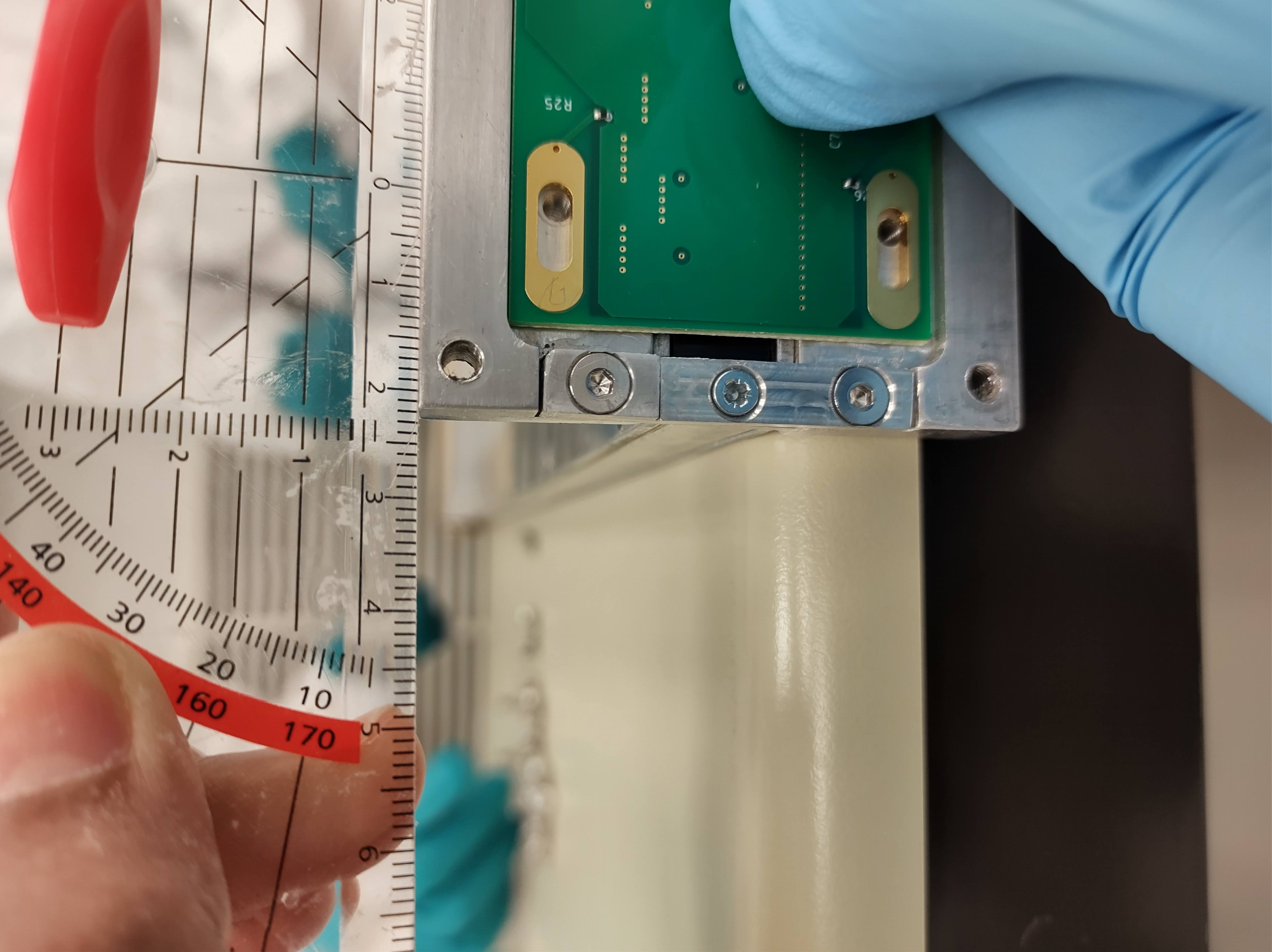}
    \caption{Placement (left) and alignment (right) of an US PCB.}
    \label{fig:US_pcb_align}
\end{figure}

Every bar end is viewed by eight SiPMs; six Hamamatsu S14160-6050HS (\num{6 x 6} \SI{}{mm^2}, \SI{50}{\micro m} pitch) and two Hamamatsu S14160-3010PS~\cite{sipm_3010_url} (\num{3 x 3}\SI{}{mm^2}, \SI{10}{\micro m} pitch) SiPMs.
The SiPMs are arranged on a custom PCB as shown in Figure \ref{fig:PCB_US}, which is read out by a front-end TOFPET2 board (see Section~\ref{sec:daq}).
The placement of SiPMs along a bar can be seen on the left of Figure~\ref{fig:US_pcb_align}.
The two smaller-size SiPMs are used to increase the dynamic range for each bar, which has to cover the low light yield generated by minimum ionizing particles and the large light yield in case of hadronic showers created in the target region or iron blocks.
The latter can lead to large charged-particle fluxes through the bars, and hence to large signals, which can saturate the larger SiPMs but not the smaller ones.
Each SiPM is read out as a single channel, giving 80 channels per PCB totaling 800 channels for all US layers. 

The PCBs are aligned to the bars within \SI{1}{mm}, as shown on the right in Figure~\ref{fig:US_pcb_align}.
The space between the SiPMs and bars on one side is filled with the same silicon epoxy gel (\SI{\sim 1}{mm} thick) as in the veto, while the PCB on the opposite end is pressed against the bars to minimize the air gap. This leads to a small asymmetry between left and right side signals for a given plane, which however does not affect the detection efficiency.

\subsection{Downstream system}

\begin{figure}[hbtp]
    \centering
    \begin{subfigure}{0.49\textwidth}
    \includegraphics[width=\textwidth]{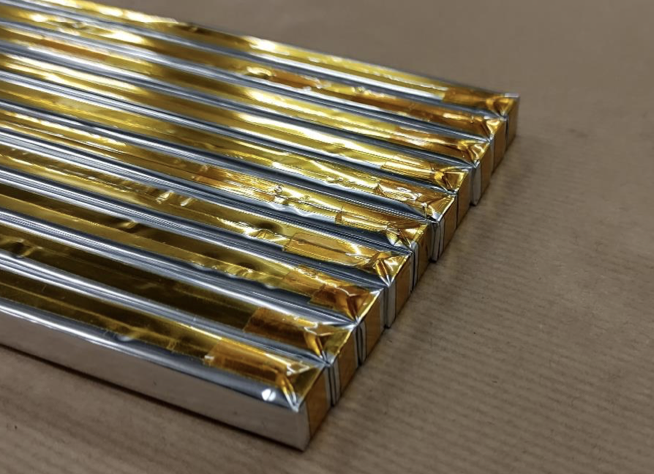}
    \caption{}
    \end{subfigure}
    \begin{subfigure}{0.49\textwidth}
    \includegraphics[width=\textwidth]{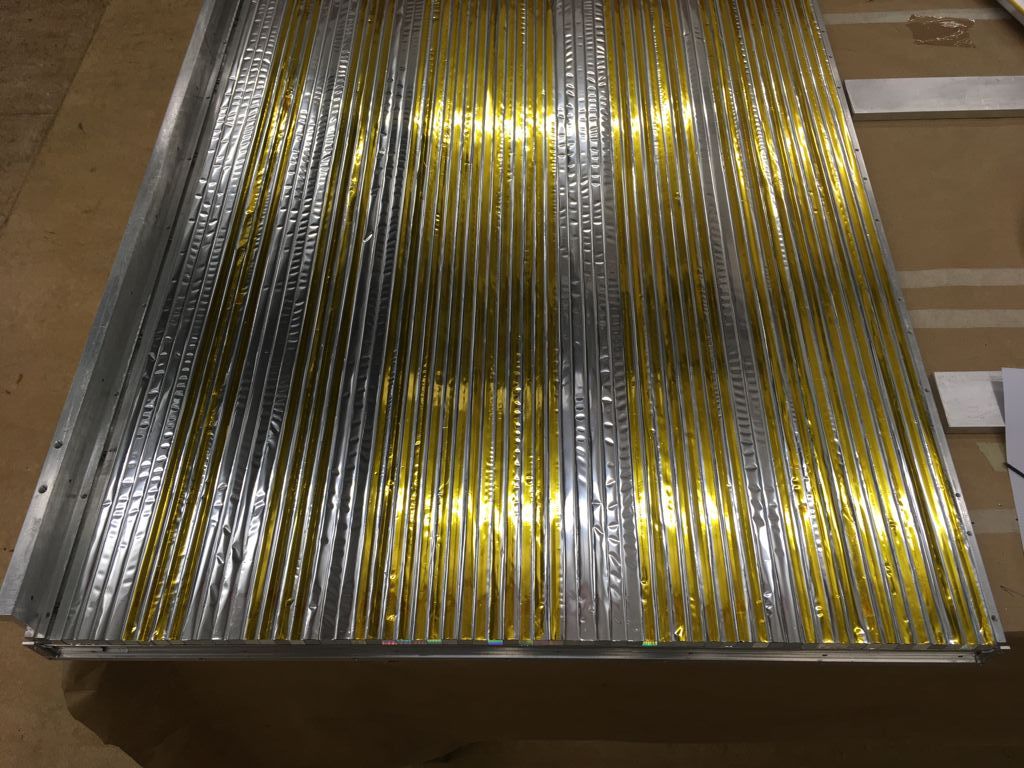}
   % \includegraphics[width=0.45\textwidth]{images/DSPCBdetail.jpg}
%    \vspace{0.1in}
   \caption{}
    \end{subfigure}
    \begin{subfigure}{\textwidth}
    \includegraphics[width=\textwidth]{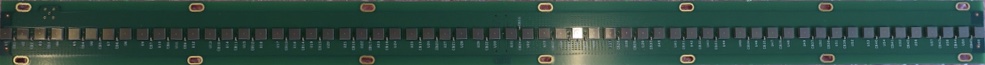}
    \end{subfigure}
    \begin{subfigure}{\textwidth}
    \includegraphics[width=\textwidth]{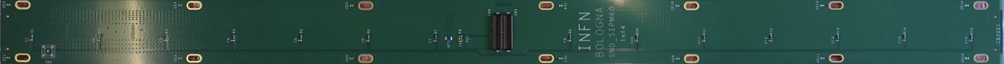}
%    \vspace{0.1in}
   \caption{}
    \end{subfigure}
    \begin{subfigure}{0.5\textwidth}
    \includegraphics[width=\textwidth]{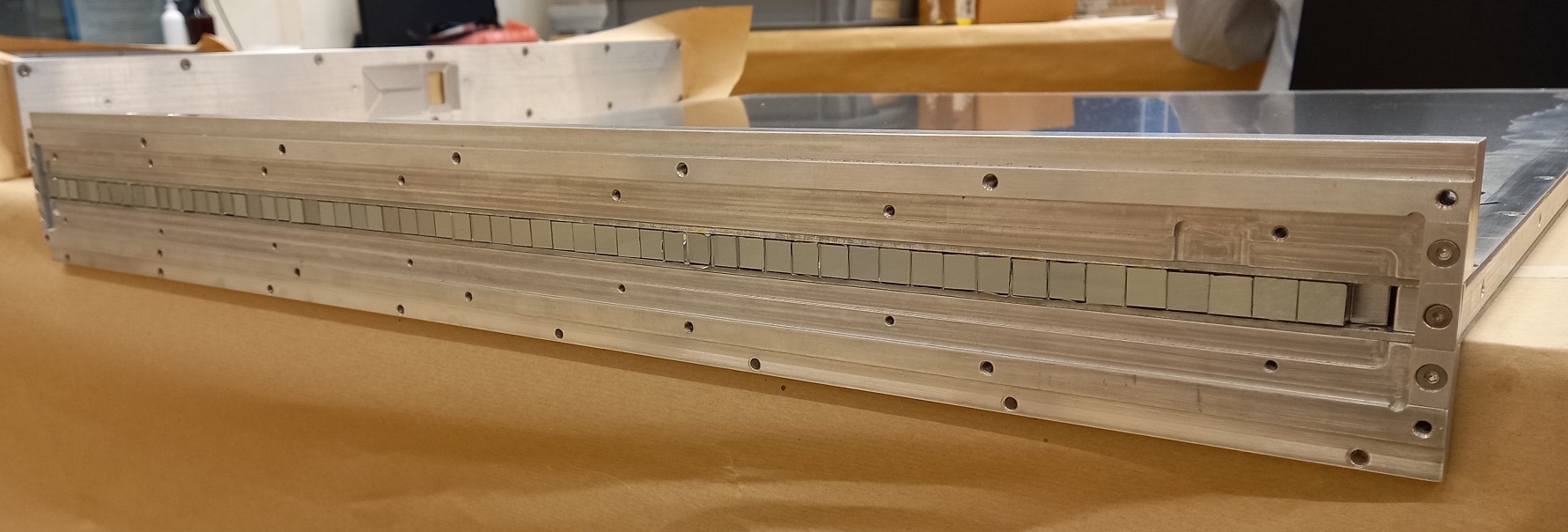}
   \caption{}
    \end{subfigure}
    \begin{subfigure}{0.4\textwidth}
    \includegraphics[width=\textwidth]{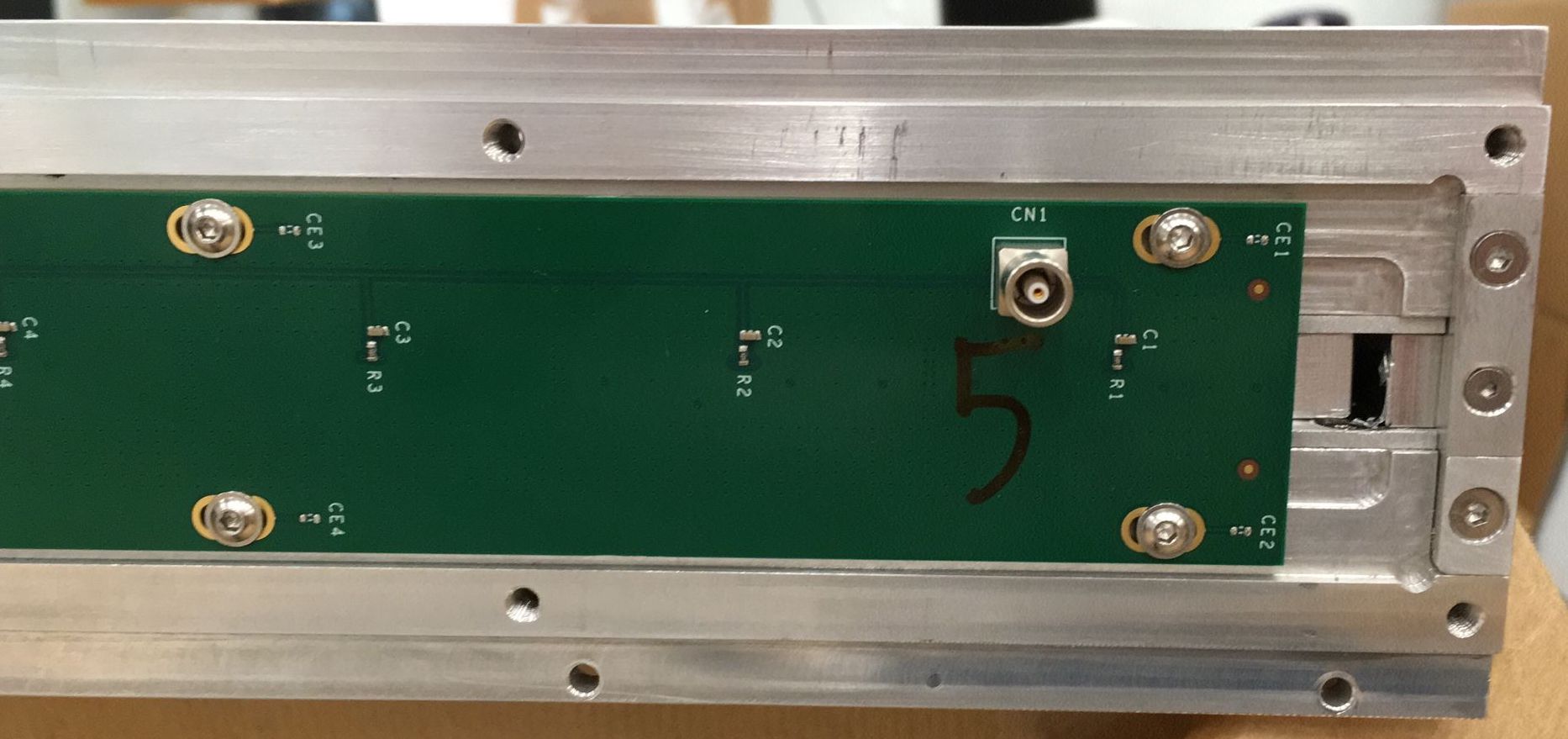}
   \caption{}
    \end{subfigure}
    \begin{subfigure}{0.55\textwidth}
    \includegraphics[width=\textwidth]{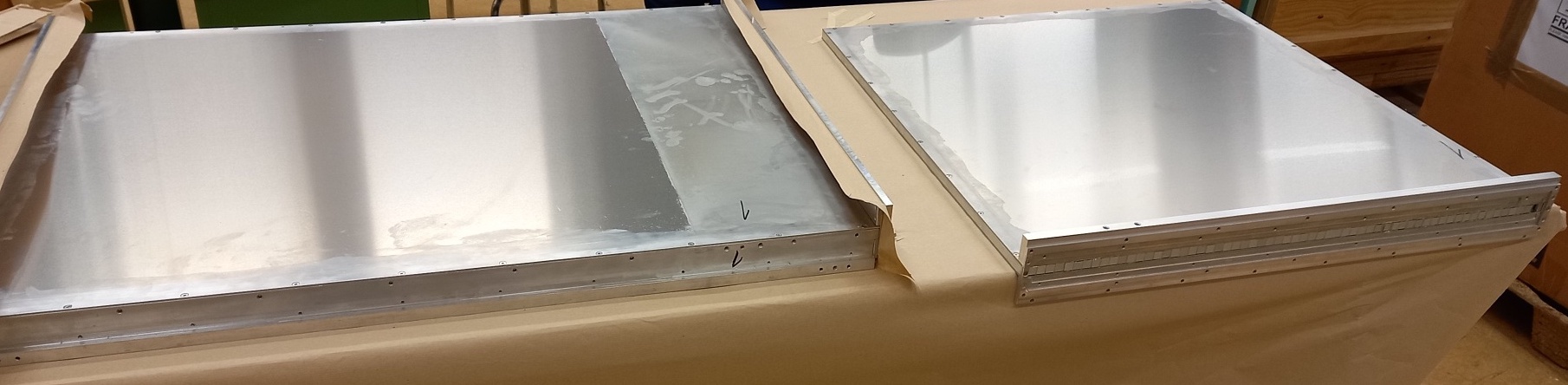}    \caption{}
    \end{subfigure}
\begin{subfigure}{0.35\textwidth}
    \includegraphics[width=\textwidth]{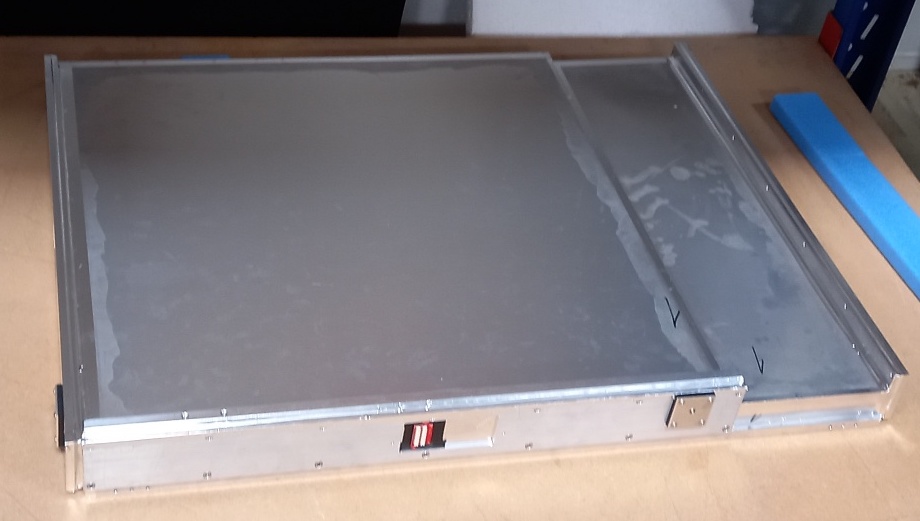}  
   \caption{}
    \end{subfigure}
    \caption{DS fabrication overview. From top left to bottom right: a)  stack of a few and b) a plane of \num{60} stacked scintillating bars, individually wrapped in aluminized mylar foils; c) the two sides of the PCB hosting the SiPMs, which are positioned at regular \SI{1.1}{cm} intervals; d) view of bar ends of a stack in its aluminum frame; e) SiPMs PCB covering the stack bar ends; f) and g) horizontal and vertical bar planes before and after assembly to form a DS muon station.}
    \label{fig:DS}
\end{figure}

Muon identification is completed with three high-granularity DS stations, placed further downstream, for providing the muon position with a resolution of better than \SI{1}{cm}
(Figures~\ref{fig:muon_tot},~\ref{fig:muon_us_ds}). 
Each station consists of two planes of thin scintillating (EJ-200) bars
(Figure~\ref{fig:DS}):
one of \num{60} horizontal bars (\num{1 x 1 x 82.5} \SI{}{cm^3} each), and one of \num{60} vertical bars (\num{1 x 1 x 63.5} \SI{}{cm^3} each). The third station has an additional plane of vertical bars. 

Every horizontal bar is read out by one SiPM (Hamamatsu S14160-6050HS) on either end; verticals bar have one SiPM only on the top edge, since the bottom is end is located as close as possible to the floor in order to maximise the detector acceptance.
Sixty SiPMs are mounted at regular intervals of \SI{1.1}{cm} on a single PCB, placed at the edge of the bar stack. The total number of channels for DS part of the muon system is \num{600}. 

Bars are individually wrapped in aluminized mylar foil. 
Because of the bar shape, most of the light collected by the edge SiPM is indirect. Therefore a tool was developed in order to ensure that the aluminized mylar foil is very tightly wrapped around the scintillating bar, minimising the light loss because of multiple reflections. For vertical bars, the wrapping of the foil at the bar end without SiPM was terminated with an additional flat layer that optimises reflection.

Scintillating bars in the same plane can differ in dimensions and be out of square from one edge to the other within fabrication tolerances. Since the SiPMs are locked in position on the PCB, assembling tools have been developed to ensure that the one-to-one alignment of SiPMs and bar edges is preserved along the entire stack: \SI{6}{mm} wide SiPMs are centred on the \SI{10}{mm} wide bar edge with an uncertainty of \SI{1}{mm}.

The quality of the contact between SiPMs and bar edges can differ because of differences in bar lengths.
%Insertion of a gel for optical contact would have complicated the replacement of individual bars, in case they turned out to be defective after being coupled with the SiPMs.
Thus it was decided to sort bars in groups of similar lengths, maximising uniformity in the same plane, and adjust the distance of the PCB, so that the SiPM-to-bar-edge gap was measured to be less than \SI{100}{\micro m} for all bars.

%\begin{figure}[t]
%    \centering
%    \includegraphics[width=\textwidth]{images/pcb_ds_front.jpg}
%    \includegraphics[width=\textwidth]{images/pcb_ds_back.jpg}
%    \caption{The two sides of the PCB for the DS muon system.}
%    \label{fig:PCB_DS}
%\end{figure}

\subsection{Low voltage and SiPM bias voltage}

The low voltage for powering the DAQ boards and the bias voltage for the SiPMs are provided by CAEN power supplies, described in Section~\ref{subsec:readout}.
For the US system, two separate HV bias lines are used for the two types of SiPMs, connected with two LEMO connectors on each PCB, as seen in Figure~\ref{fig:PCB_US}, while for the DS a single LEMO connector is used to power all SiPMs. %Each SiPM is read out as a single channel, giving 80 channels per PCB totaling 800 channels for all US layers. 

\begin{figure}[t]
    \centering
    \includegraphics[height=0.35\textheight]{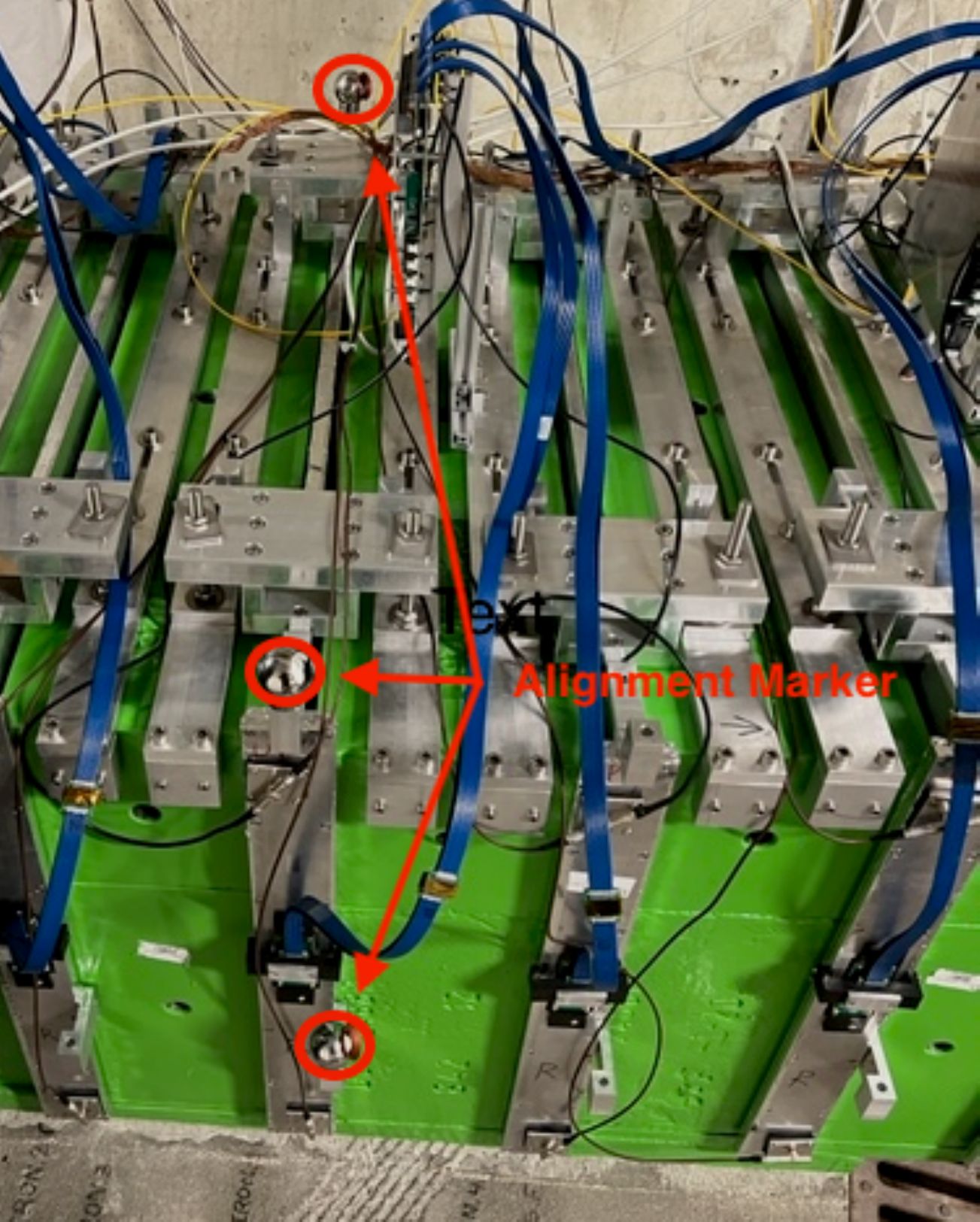}
    \includegraphics[height=0.35\textheight]{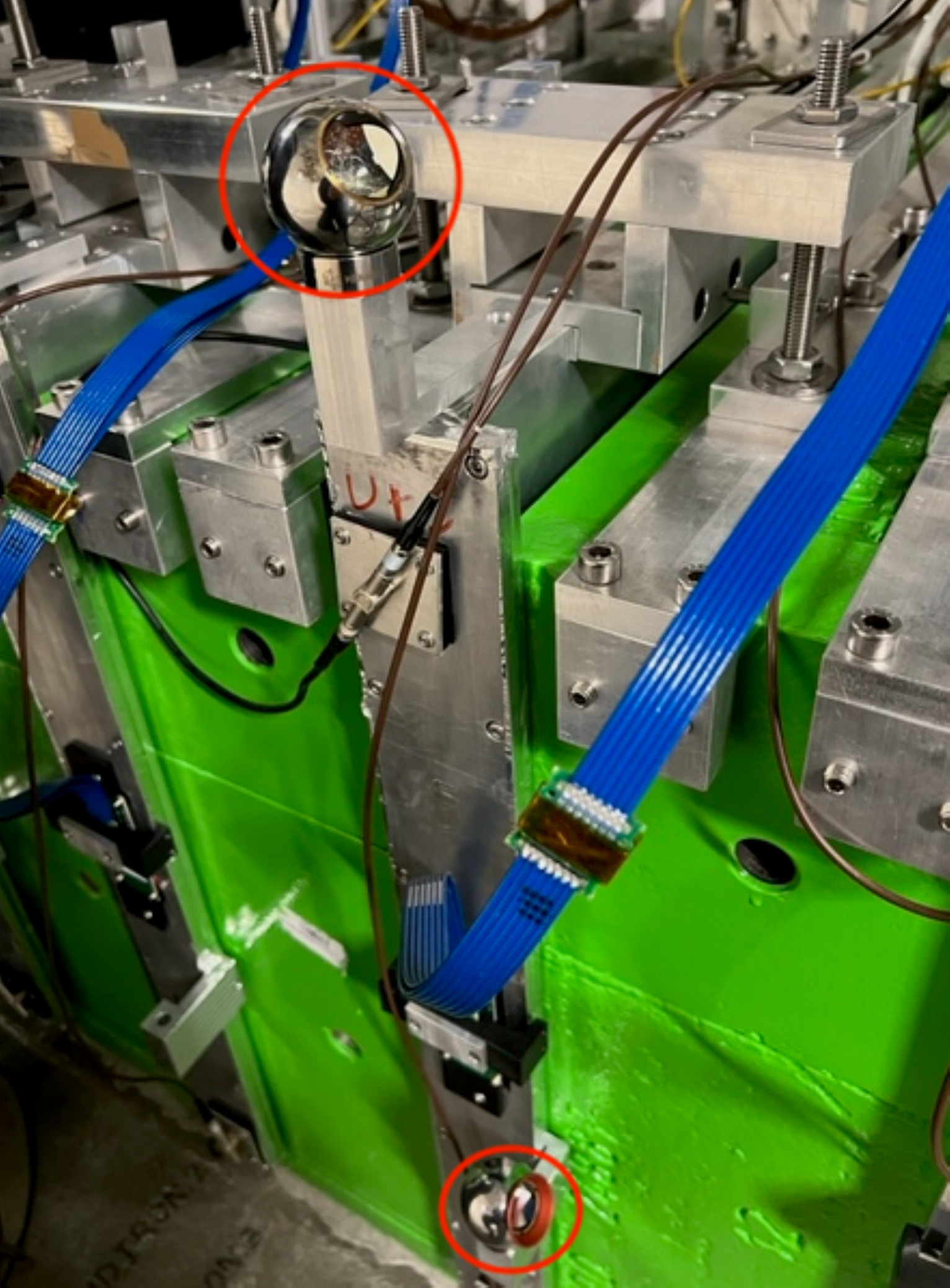}
    \caption{Pictures indicating the location of the alignment spheres on a muon plane, circled in red.}
    \label{fig:muon_align}
\end{figure}

\subsection{Mechanical support}

Bars and PCBs are housed in aluminum frames that provide light tightness.
The thickness of the frames is \SI{2}{mm} and the rectangular flanges are \SI{4}{cm} wide.
An aluminum cover shields the PCBs from outside light and protects it from heat generated in the FE board, located on the opposite side of the cover.
A Kapton gasket between the PCB and flange prevents electrical shorts between the electronics and frame.
The inside of the aluminum cover is also lined with Kapton to electrically isolate the PCB.

\subsection{Alignment}
Frame support mechanics are mounted on the iron blocks; adjustment screws allow for correcting the placement in position, rotation and tilt. Individual frames are installed in gaps between the iron blocks, and then fixed to the support mechanics. Three spherical alignment markers, shown in Figure~\ref{fig:muon_align}, are mounted on each frame for global survey measurements. Alignment was performed by the Geodetic Metrology Group within the Beams Department (BE-GM) at CERN, with each marker aligned with respect to the nominal positions within 1--\SI{2}{mm}.
\section{Online system}
\label{sec:daq}

%\subsection{The online architecture}
The SND@LHC online system includes all components involved in operating the experiment, i.e. the timing and the data acquisition hardware and software that realise the data flow from the detector to the storage, the detector control system (DCS) that controls and monitors the detector services, such as power supplies, cooling system, etc, and the data quality monitoring (DQM) and real-time analysis (RTA) system, necessary to ensure a good quality of the collected data.
Globally, the top-level software, the experiment control system (ECS), encompasses all the sub-components above, together with the system of logs and databases in order to store information about the state, configuration and conditions of the data taking.
The ECS is constructed to allow full automation of the data taking.

\begin{figure}[h]
    \centering
    \includegraphics[width=1.0\textwidth]{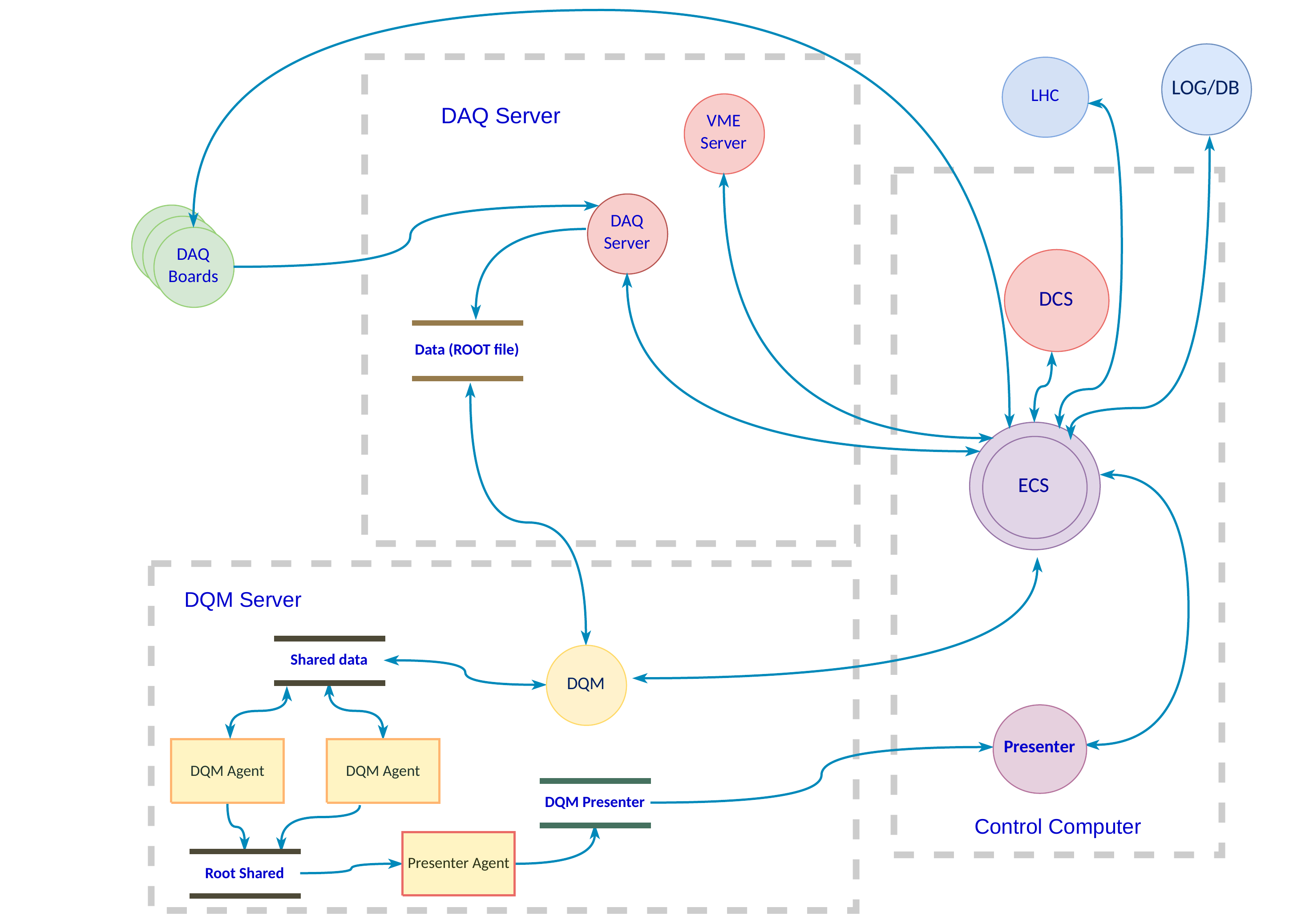}
    \caption{Simplified scheme of the SND@LHC online system.}
    \label{fig:daq-scheme}
\end{figure}

The different components, shown in the scheme in Figure~\ref{fig:daq-scheme}, and the readout system are described in more details in the following sections.

\subsection{Readout system}
\label{subsec:readout}

As discussed in Sections~\ref{sec:veto}, \ref{sec:scifi} and~\ref{sec:muon}, the SND@LHC experiment features two types of electronic detector systems: scintillator bars read out by SiPMs in the Veto, the hadronic calorimeter and  muon system, and scintillating fibres read out by SiPMs in the Target Tracker.
These sub-systems are read out with the same data acquisition (DAQ) electronics, consisting of front-end (FE) boards, described in Section~\ref{subsec:tofpet}, and DAQ readout boards, described in Section~\ref{subsec:daq-electronics}.
They read out the signals from the SiPMs, digitize them and send the recorded data (including timestamp and integrated signal charge) to a DAQ server.

The detector uses in total 37 DAQ boards, each of which is connected to four FE boards.
The system runs synchronously with the LHC bunch crossing clock, and operates in a trigger-less fashion, i.e.\ all hits recorded by each board are transmitted to the DAQ server.
Noise reduction is performed at the front-end level by setting an appropriate threshold for each channel, and on the DAQ server after event building.

A Trigger Timing Control (TTC) crate\cite{ttc_url}, described in Section~\ref{subsec:lhc-signals}, is responsible for receiving the LHC clock and the orbit signals from the LHC Beam Synchronous Timing (BST) system and distribute them to the DAQ boards.

The detector is powered using CAEN A2519 modules for the DAQ readout boards power, requiring \SI{12}{V} and \SI{2}{A} each, and A1539B modules for the bias voltage of the SiPMs, requiring up to \SI{60}{V} and up to \SI{300}{\micro A} per channel.
These modules are hosted in two SY5527 mainframes.

The control of power supplies is performed by the detector control system (DCS, in the control computer in Figure~\ref{fig:daq-scheme}) discussed in Section~\ref{subsec:daq-dcs}, which also monitors the voltages and currents drawn on both the LV and HV channels and monitors the presence of alarms.

The online system (DAQ Server, DQM server and Control computer shown in Figure~\ref{fig:daq-scheme}) includes two servers located on the surface.
One of them receives data from the DAQ readout boards, combines the data into events, and performs the online processing of the detector data, as described in Section~\ref{subsec:daq-events}, before saving the data to disk.
The other one runs the ECS and the other elements of the online system.

%Details about each component are given in the following sections. A schematic representation of the DAQ system is shown in Figure~\ref{fig:daq-scheme}.

\subsubsection{Timing system}
\label{subsec:lhc-signals}

The LHC clock (\SI{40.079}{MHz} bunch crossing frequency) and orbit clock (\SI{11.245}{kHz} revolution frequency of the LHC) signals are obtained from the LHC BST system via optical fibres based on the TTC system.%, along with other information such as the machine mode, the beam type and energy, the GPS absolute time, etc.

\begin{figure}[h]
    \centering
    \includegraphics[width=\textwidth]{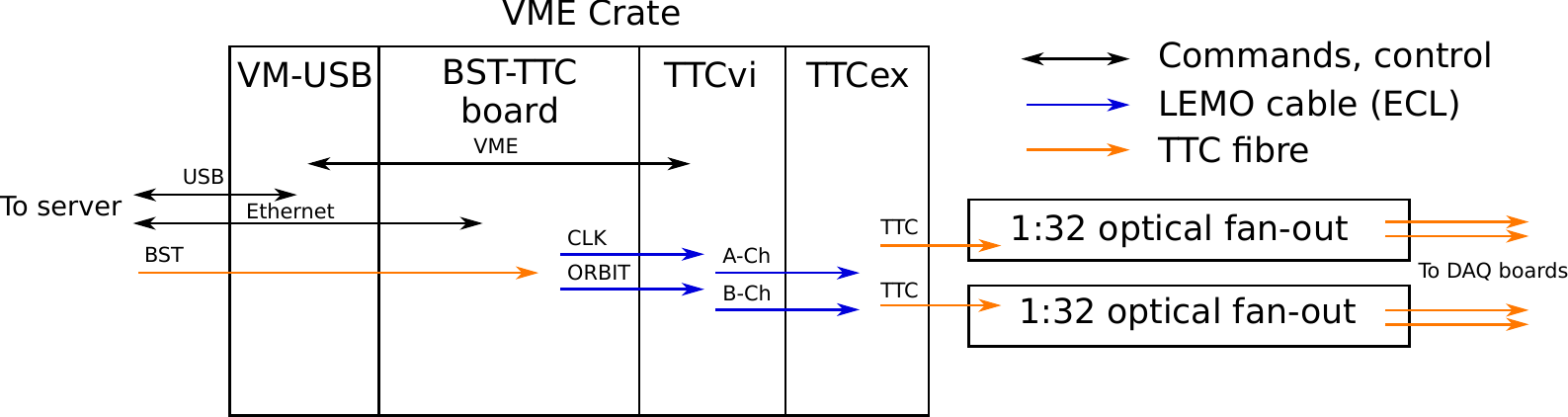}
    \caption{Simplified scheme of the SND@LHC TTC system.}
    \label{fig:ttc-vme-scheme}
\end{figure}

A scheme of the system used in SND@LHC is shown in Figure~\ref{fig:ttc-vme-scheme}.
The BST signal is received by a dedicated board, BST-TTC, that extracts the clock and orbit signals, cleans the clock using a Phase Lock Loop, and distributes them to the detector using the TTC system.
The board is the same that is used for the read-out of the detector, described in Section~\ref{subsec:daq-electronics}, with the addition of a mezzanine card to generate the correct signal levels for the TTC modules.
%In case of absence of the BST signal, the board provides a substitute clock to the system.

The clock and synchronous commands are distributed to the DAQ boards using the TTCvi and TTCex modules~\cite{ttc_web}, housed in a VME crate.
The TTCvi receives the clock and orbit signals, and generates the A-channel (trigger) and B-channel (synchronous and asynchronous commands) signals, which are encoded and transmitted by the TTCex.

The TTCvi module can be programmed and controlled using the VME bus.
A USB-to-VME converter allows it to be programmed from the computer server.

Variations of several nanoseconds in the phase of the clock are to be expected due to temperature changes. For this reason, the absolute timing offset will be calibrated with the timestamps of the muons generated by $pp$ collisions at the ATLAS interaction point and detected in SND@LHC.

\subsubsection{Front-end electronics}
\label{subsec:tofpet}

\begin{figure}[h]
    \centering
    \includegraphics[width=0.6\textwidth]{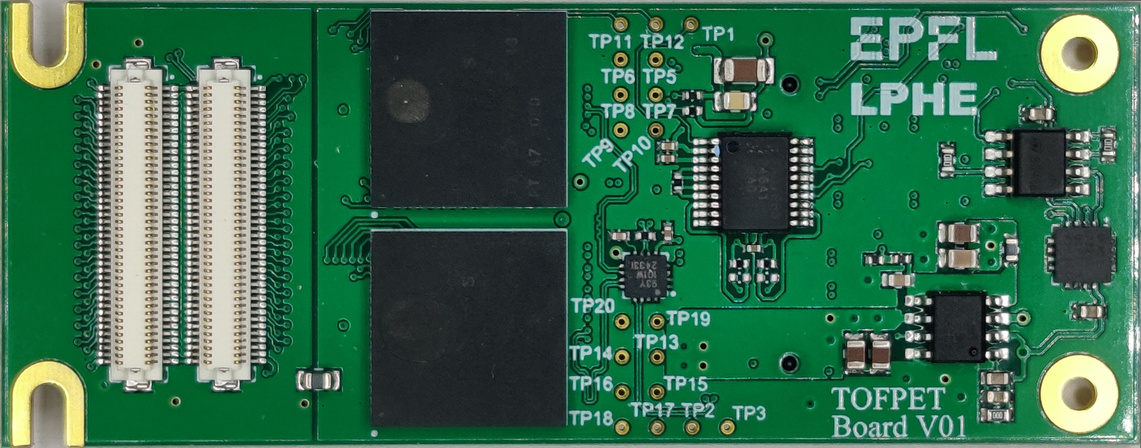}
    \caption{The FE board. The two TOFPET2 ASICS (centre) and the SiPM connectors (left) are visible.}
    \label{fig:fe-board}
\end{figure}

The front-end (FE) boards, shown in Figure~\ref{fig:fe-board}, are based on the TOFPET2 ASIC by PETsys~\cite{tofpet2_url}.
The TOFPET2 is a 64-channels readout and digitization ASIC designed for time of flight positron emission tomography systems~\cite{tofpet2_datasheet}.
It incorporates signal amplification circuitry, discriminators, charge integrators, analog-to-digital converters (ADC, in this case QDCs as in charge-to-digital converters) and time-to-digital converters (TDC).
The ASIC has been found to be perfectly suitable to measure signals produced by SiPMs, and to record their timestamp and charge. The FE boards contain two TOFPET2 ASICs each for a total of 128 channels, and has temperature monitoring capabilities of both the SiPMs and the boards themselves.

Each channel of the TOFPET2 features a preamplifier and two amplifiers, one optimized for the timing measurement and the second for the charge measurement.
A combination of up to three discriminators with configurable thresholds can be used. The first one is mainly used for timing measurements, and  normally has the lowest threshold, while the other two are used to reject low amplitude pulses and to start charge integration.
The TDCs feature a time binning of \SI{\sim 40}{ps} and the QDCs have a linear response up to \SI{1500}{pC} input charge.
In addition, the gain of the QDC branch can be selected to have a value between \num{1.00} and \num{3.65} to achieve the best resolution and dynamic range, depending on the signal generated by the SiPMs~\cite{tofpet2_datasheet}.

The TOFPET2 ASIC requires calibration in a three-step procedure. The first two steps include adjusting the input level of each amplifier just above the electronic noise produced by the detector, and then calibrating the TDCs and QDCs with the help of pulses with known duration and phase relative to the clock injected from the FPGAs.  This is performed with the bias voltage of the SiPMs below the breakdown. Last, the dark count rates of the SiPMs are measured as a function of the thresholds of the three discriminators in each TOFPET2 to determine the best settings to achieve an optimal efficiency and data rate. The last step is performed at the nominal operating voltage of the SIPMs. The calibration procedure has been implemented in an automated way, and the parameters obtained from it are stored in configuration files.% for each DAQ board.
%When data acquisition is started, these parameters are saved alongside the data, as they are needed to process the raw values of timestamp and charge associated to each hit.

\subsubsection{DAQ electronics}
\label{subsec:daq-electronics}

\begin{figure}[h]
    \centering
    \includegraphics[width=0.7\textwidth]{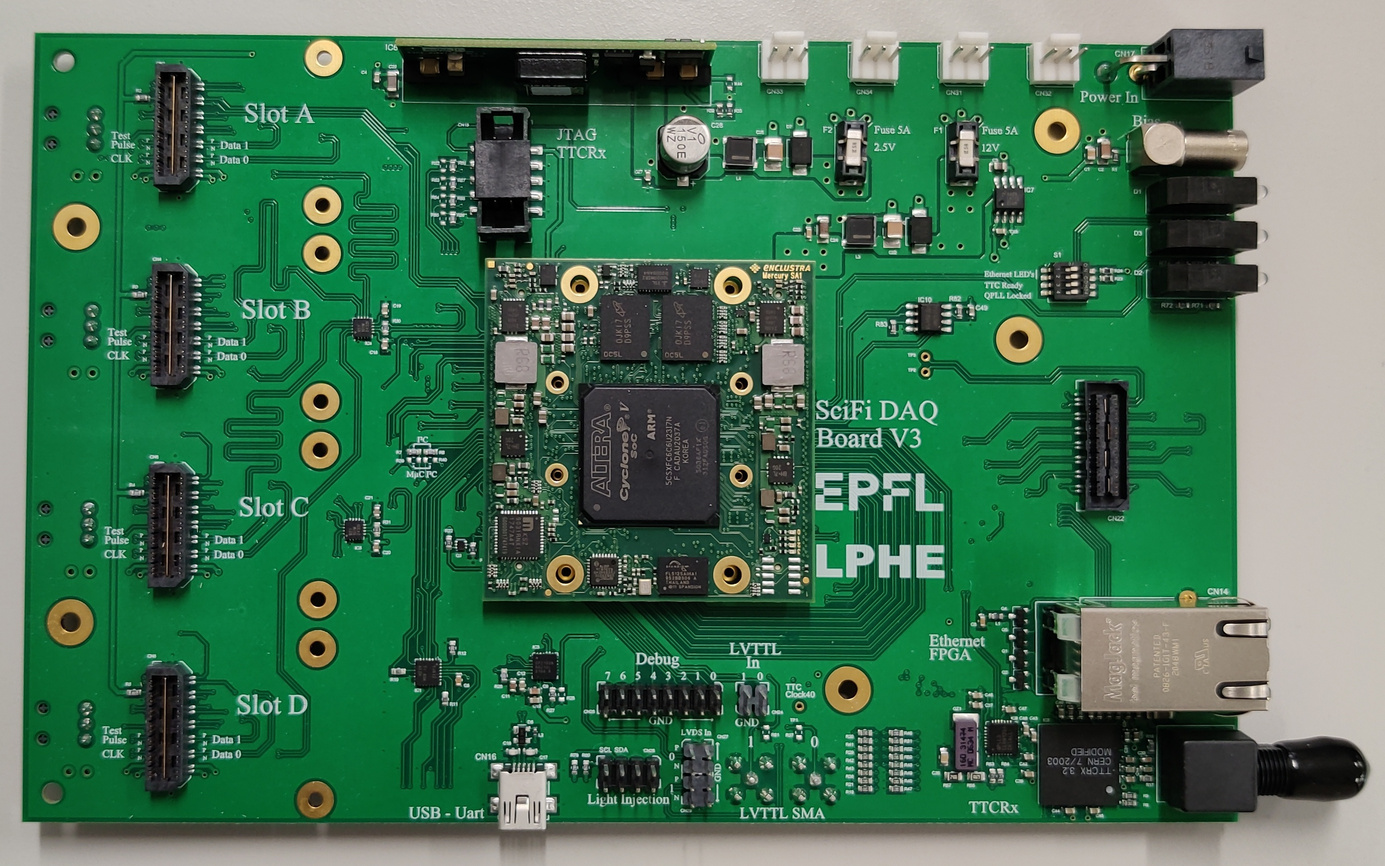}
    \caption{Photo of the DAQ board. The four FE board connectors are visible on the left, the TTCrx and optical receiver on the bottom-right, and the Enclustra Mercury SA 1 module in the centre.}
    \label{fig:daq-board}
\end{figure}

The DAQ readout boards, shown in Figure~\ref{fig:daq-board}, are based on the Mercury SA1 module from Enclustra~\cite{enclustra_mercury_sa1}, featuring an Altera Cyclone~V FPGA. 
This board is equipped with four high-speed connectors for the FE boards, a TTCrx ASIC~\cite{ttc_web} with an optical fibre receiver to receive the clock and synchronous signals from the TTC system, a \SI{1}{Gb} Ethernet port used for data and command transmission, and a coaxial LEMO connector to deliver the bias voltage to SiPMs.

Each DAQ board collects the data digitized by four FE boards, i.e. 512 channels, and transmits it to the DAQ computer server located on the surface.

\subsubsection{Readout software}
\label{subsec:daq-events}

Each DAQ board transmits all the recorded hits to the DAQ computer server, where event building is performed.
The hits are grouped into events based on their timestamp, and saved to disk as a ROOT file.

The DAQ boards also transmit periodic triggers received from the TTC system. These heartbeat triggers are used by the event building software in the DAQ server to verify that all the boards are  running synchronously, and operating properly even when there is no data.

The readout process from starting servers to starting the data taking, sending periodic triggers, monitoring the status of each element, etc, is fully controlled by the ECS, described in Section~\ref{subsec:ecs}. 

\begin{figure}[h]
    \centering
    \includegraphics[width=0.7\textwidth]{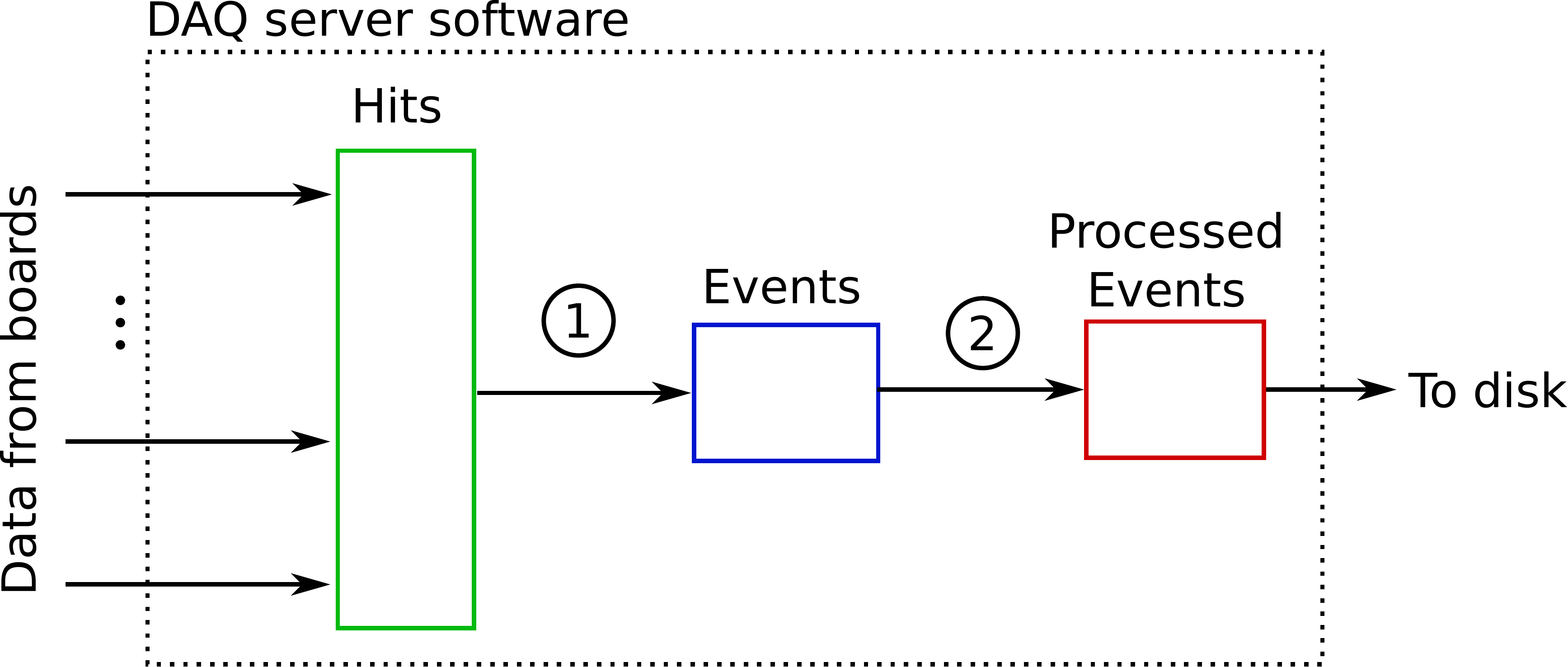}
    \caption{Schematic view of the event builder process. The colored rectangles represent queues of data being processed and the numbers identify the steps of event building, discussed in the text.}
    \label{fig:evt-builder}
\end{figure}

The event building process is structured in two main steps, shown in Figure~\ref{fig:evt-builder}.
In the first step, hits collected by all boards and belonging to the same event, i.e.\ with time stamps within \SI{25}{ns},  are grouped into ``events''.
The event timestamp corresponds to the timestamp of the earliest hit within the event. The events are then filtered and processed online, before being written to disk. The details of the processing depend on the chosen settings, but it always contains an online noise filter, described below.

The noise filtering is performed in two steps. In the first one, events are required to have a minimum number of DAQ boards that have detected a certain number of hits. This is fast and eliminates all the events generated by single noise hits. In the second step, the hits are grouped by the plane that generated them. This allows more advanced requirements on the topology of the events to be imposed. 

The system includes a number of additional configurable data processors, such as the FE calibration that may be applied during the data acquisition.

The DAQ server writes the recorded data to a local disk. At the end of each run the data is transferred to a permanent storage and converted to the format used by the reconstruction software.

The DAQ server can cope with a maximum rate of \SI{1}{M hits/s}.
The dark rate in the whole detector produces \SI{\sim 400}{k hits/s}, the muon flux at the highest instantaneous luminosity (\SI{0.8}{Hz/cm^2}) produces \SI{\sim 450}{k hits/s}, leaving \SI{\sim 150}{k hits/s} of spare bandwidth.

\subsection{Experiment control}

\subsubsection{Detector control and safety monitoring}
\label{subsec:daq-dcs}

The detector control system (DCS) controls and monitors the status of all the detector services, i.e.\ the detector and electronics power supplies, the cooling system and the environmental sensors within the neutron-shielded box surrounding the target system, as well as the safety system.  The voltage, currents and channel status of the power supplies are continuously monitored and transmitted to the ECS. The ECS then acts accordingly, logging the events or raising an alarm in case of problems. 

\begin{figure}[h]
    \centering
    \includegraphics[width=0.8\textwidth]{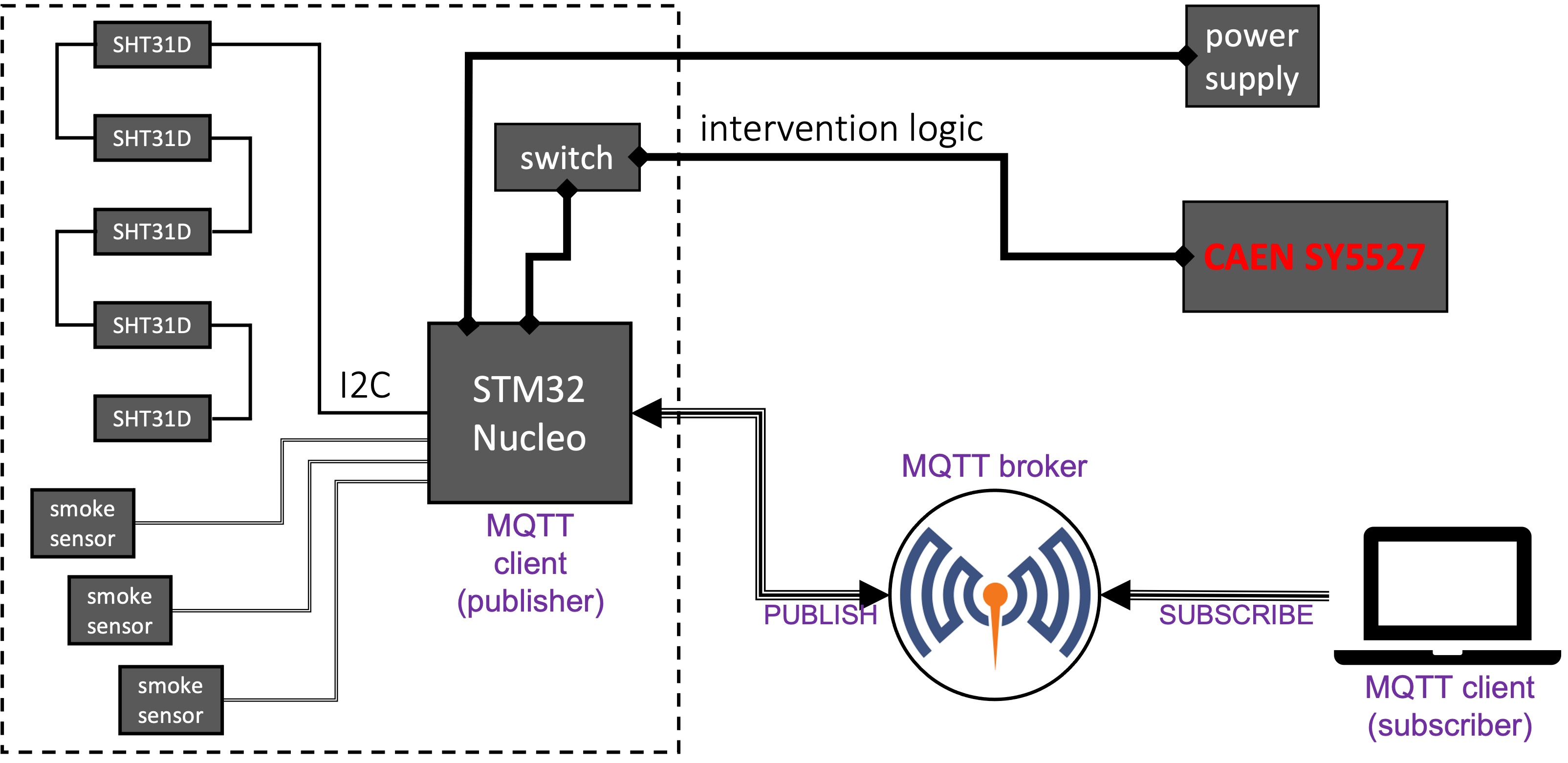}
    \caption{Schematic view of the safety and environment monitoring system. The items included in the dashed line are those physically installed inside the neutron-shielded box around the targe system.}
    \label{fig:sms}
\end{figure}

The neutron-shielded box surrounding the target system is equipped with sensors for temperature, humidity and smoke. The safety and environment monitoring system (SMS) monitors these environmental parameters inside the box, detects the presence of smoke and monitors the status of the cooling system. The monitoring and safety decison logic is implemented on a NUCLEO-H743ZI development board~\cite{nucleo_url}, featuring an ARM Cortex-M7 microcontroller.
A schematic illustration of the SMS is depicted in Figure~\ref{fig:sms}.

Temperature and humidity are measured in five different locations, using digital sensors (Sensirion SHT31~\cite{sensirion_trh_url}) which guarantee an accuracy of \SI{\pm 0.2}{\celsius} and \SI{\pm 2}{\percent} respectively, and a reliable I2C communication with the host microcontroller.
The positions of the sensors have been chosen to minimize the interference with the emulsion replacement procedure. Three of them are positioned on the back metal plane, monitoring the temperature and humidity as close as possible to the emulsion boxes, while the remaining two sensors are placed on the opposite side, facing the cold box.
This configuration assures a comprehensive temperature and humidity mapping of the target system.

The SMS is also equipped with three smoke sensors with relay output.
They are used for fire and smoke detection and are connected to the microcontroller digital inputs with interrupt capability.
In addition, a signal from the cooling system is used to monitor its status.

The microcontroller firmware has been developed using the Mbed OS~\cite{mbed_os_url}, an ARM Real Time Operating System.
It continuously monitors its inputs, evaluates possible alarm conditions, takes the safety actions according to the alarm levels and manages the data communication with the DCS. The MQTT protocol~\cite{mqtt_url} is used to transmit data. % the SMS acts as a publisher for sensors data and alarm messages, which are received by the DCS, while it can receive configuration messages from the DCS.
%The system is designed to be in operation 24/7 and the frequency with which data are posted can be configured by the operator.A watchdog has been implemented to strengthen the robustness and reliability of the system. If the microcontroller freezes for any reason, it is automatically reset by dedicated hardware.
The DCS monitors and logs the temperature inside the neutron shield. If it exceeds \SI{18}{\celsius}, an alarm is raised and the power to the DAQ boards is cut, to protect the emulsion films.
Alarms are also raised if a failure is detected in the cooling plant, i.e.\ if it turns off and stays off for more than \num{10} minutes.

The SMS also acts as an interlock for the CAEN power supplies. If an alarm condition is detected, it can turn them off without relying on the DCS being functional.
% If the environmental conditions in the cold box do not meet some predefined requirements, an alarm is raised.
% When an alarm occurs, the SMS sends an alarm message to the MQTT broker and undertakes the needed action to % maintain the safety of the infrastructure according to the level of the alarm.
The control algorithm is designed to classify alarms according to three levels referred to as low, medium and high.
A low level alarm occurs when the temperature or humidity readings of one sensor exceeds the set thresholds, or if only one of the three smoke sensors is triggered. %This is considered a low-level alarm because if only one smoke sensor is triggered, it is most likely a fault in the sensor rather than the actual presence of smoke. This assumption is based on the fact that the three smoke sensors are all inside the cold box and smoke would trigger all of them) or if the cooling system turns off.
An alarm is posted but no further action is taken.

A medium level alarm occurs when the temperature or humidity readings of two or more sensors exceed the set thresholds or if the cooling system has been turned off for more than 10 minutes.
An alarm is posted and the power supplies are immediately turned off, to avoid the temperature inside the cold box to raise further and potentially affect the emulsion films.

A high level alarm occurs when two or more smoke detectors are triggered.
An alarm is posted and the power supplies are immediately turned off. In addition, it could be setup to trigger a response of the fire brigade.

%\subsubsection{Control software}

\subsubsection{Data processing and quality monitoring}

The data quality monitoring (DQM) is fundamental to ensure that useful data is recorded and to verify that all the sub-systems of the experiment operate correctly.

The DQM process runs on the second computer server located at the surface and reads in real-time the data file that is being written by the DAQ server.
The process performs the conversion to the offline data format and makes this data available to the DQM agents, which process it and displays the results in the ECS.

Several levels of complexity of the processing have been implemented from simple hit maps to ensure that all detector channels work as expected, to full real-time reconstruction that allows high-lever quantities to be checked, such as efficiencies and resolutions.

\subsubsection{Experiment control system}
\label{subsec:ecs}

The ECS is the top-level control of the experiment online system, providing a unified framework to control the hardware and software components, and to sequence all data taking operations.

A hierarchical architecture has been implemented in which the ECS is a layer above the other online systems, preserving their autonomy to operate independently.
With this architecture, the various online components do not strictly require the ECS to operate, e.g. detector calibration and data taking are stand-alone processes. The ECS also performs the logging of the relevant detector information, either in ELOG or in databases depending on the type of information.

The ECS consists of two main software components: the ECS Process Manager (EPM) and the ECS Graphical User Interface (GUI).
The software is written in C++ and the inter-communication with the DAQ and the DCS system is done with Python scripts.

The EPM is a process which runs on the main server and acts as the communication link between the different online system components. It takes care of starting them and continuously monitoring their status.
The EPM process is kept alive by a cron-based watchdog. The status of the process is monitored at fixed time intervals, and restarted if it is not running. 
The EPM and its slave processes are associated with state machines that are driven by the status of the process activities.
The information of these state machines is stored in a shared memory supervised by the EPM.

\begin{figure}[h]
    \centering
    \includegraphics[width=0.7\textwidth]{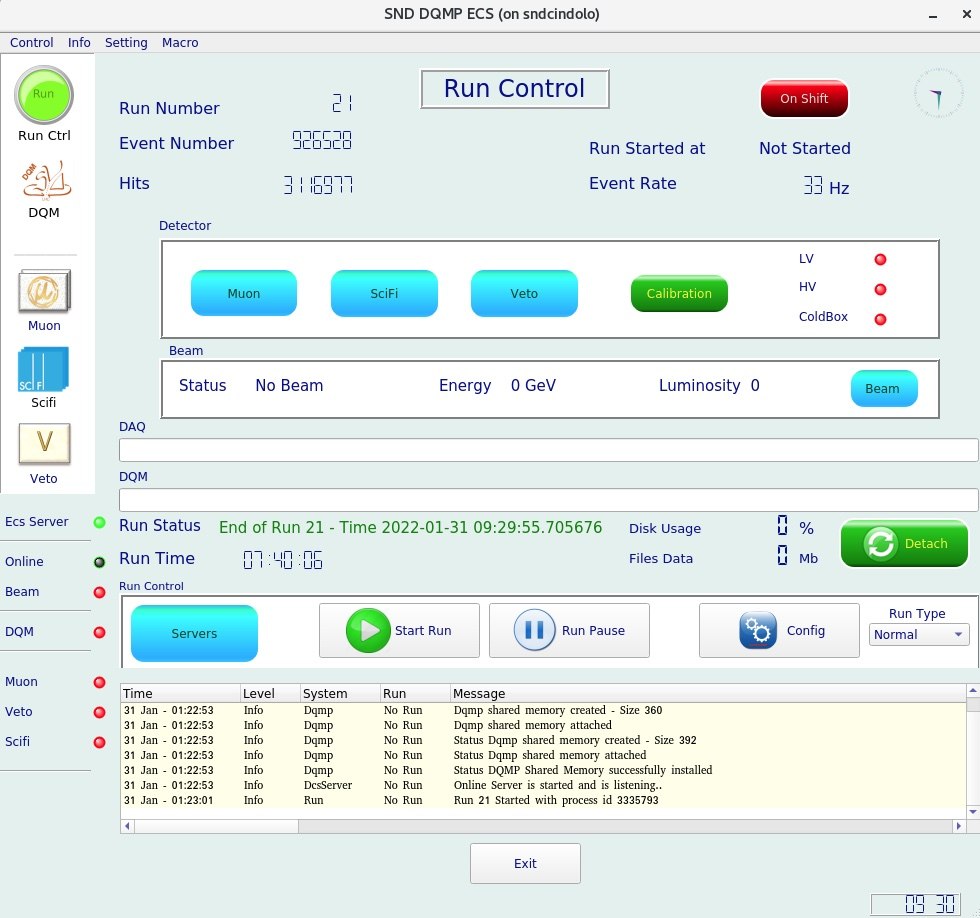}
    \caption{The main window of the ECS graphical interface.}
    \label{fig:ecs-gui}
\end{figure}

The ECS GUI has been designed to ensure a simple and compact view of the run control, status of the sub-detector and of the peripheral systems. The main windows of the ECS GUI is shown in Figure~\ref{fig:ecs-gui}.

The ECS is designed to operate the online system automatically, controlled by a global finite-state machine that receives the status of the LHC and of the detector to perform predefined actions in order to run the data taking and recover from errors. The accelerator states are received from the LHC Data Interchange Protocol system, the power supply and environment conditions from the DCS, and the data acquisition status from the DAQ boards and the DAQ server. As an example, the ECS automatically starts the data acquisition when the LHC declares stable beams, stops it when the beams are dumped, and logs any event that can be useful to analyse the data.
Furthermore, it can reboot a board that has become unresponsive, stop the DAQ and cut the power to the boards in the neutron-shielded box if the temperatures rises above the thresholds, try to restore a tripped SiPM bias channel, etc.

\section{Offline software and simulation}
\label{sec:offline}

The offline software framework, \texttt{sndsw}, is based on the FairRoot framework~\cite{FAIRROOT},
and makes use of the experience gained with the FairShip software suite,
developed within the SHiP collaboration~\cite{Anelli:2015pba}. The reconstruction and analysis tools developed by the SHiP collaboration had been successfully applied to the SND@LHC use cases and further improved.

The offline software is developed, maintained, and distributed on Github. \texttt{sndsw} and its dependencies are built from source and are configured using the AliBuild tool, developed within the ALICE Collaboration for their upgrade software. The recipes for the dependencies are shared with ALICE and SHiP, where possible, to reduce the maintenance of the framework. Specific patches and recipes for software uniquely used by SND@LHC are added, where required. Container images with the dependencies as well as an installation on the CVMFS are provided for various use cases.
Raw and reconstructed data from testbeam and TI18 commissioning are available on EOS for analysis.

\subsection{Detector geometry}
\begin{figure}
    \centering
    \includegraphics[width=0.8\textwidth]{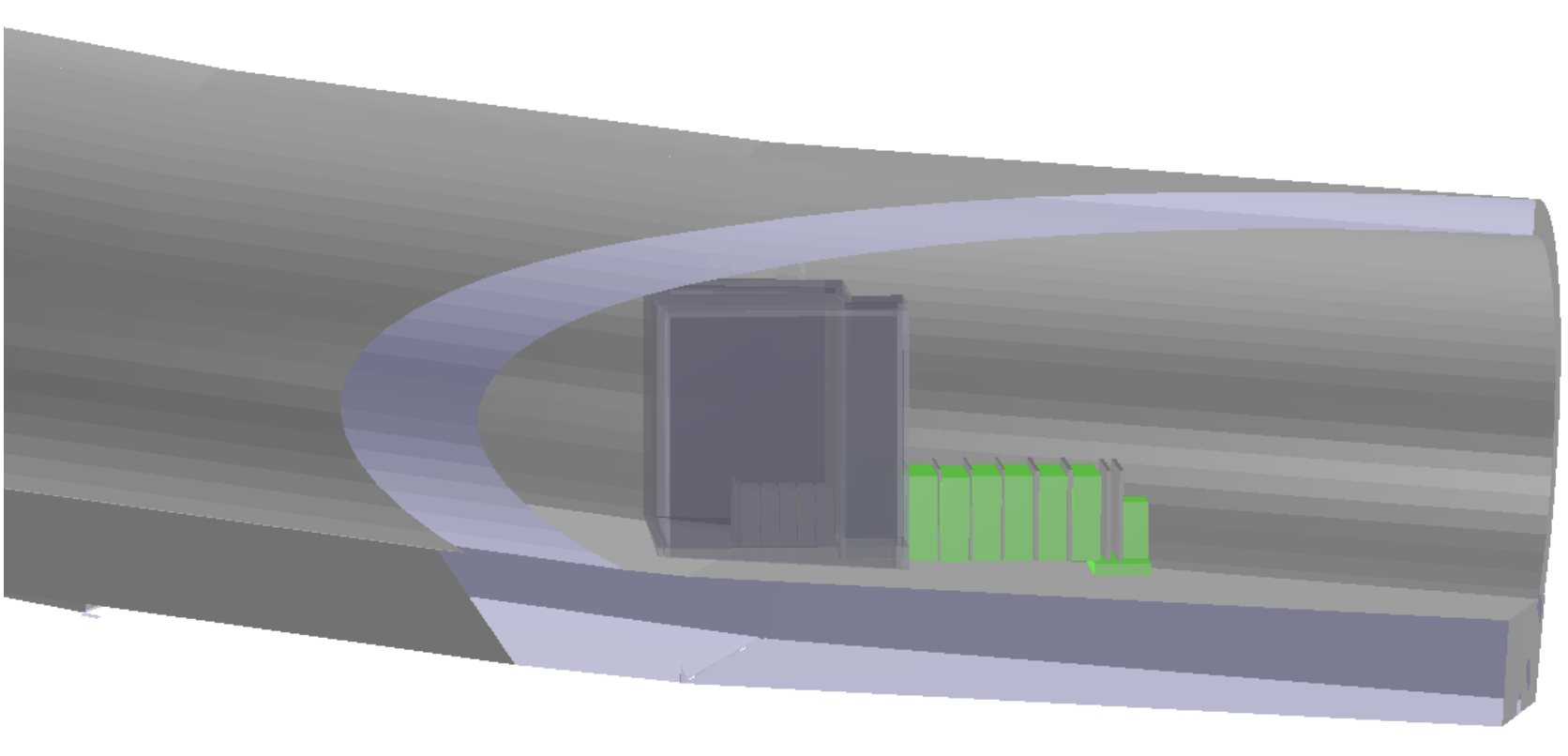}
    \caption{The SND@LHC detector layout and the TI18 tunnel geometry as implemented in the \textsc{Geant4} simulation.}
    \label{fig:sw_geant4}
\end{figure}

The detector geometry is implemented using the TGeo package of ROOT and used in the simulation by \textsc{Geant4} as well as in the reconstruction. A  model of the detector, the neutron shield and the surrounding tunnel can be seen in Figure~\ref{fig:sw_geant4}. Electronic detectors and emulsion films are implemented as sensitive volumes. For the electronic detectors, the full granularity is implemented, from scintillator bars to individual scintillating fibres. The \textsc{Geant4} simulation stops with the deposition of energy in the sensitive detectors. The digitisation step takes this energy and simulates an electronic signal, taking into account the transformation to photons, the light propagation and absorption along the scintillating fibre or bar, the photodetection efficiency of the SiPMs and the response of the front-end.

\subsection{Simulation}
Several simulation engines are available. Muons from IP1 simulated by FLUKA~\cite{Fluka2,Fluka3} and transported through the detector by \textsc{Geant4}~\cite{Geant4}, muon deep inelastic scattering using Pythia6~\cite{Pythia6}, DPMJET3 (Dual Parton Model, including charm)~\cite{Roesler_2001} or Pythia8~\cite{Sjostrand:2014zea} for neutrino production at IP1 and GENIE~\cite{cite:GENIE} for the neutrino interactions in the detector target. In addition, \textsc{Geant4} had been used to investigate the neutron shielding performance of various materials.

\subsection{Data reconstruction}
The event reconstruction is performed in two phases: the first one is performed during the data taking using the response of the electronic detectors.  The second phase incorporates the emulsion data, that will be available about six months after the exposure.

First data became available from the testbeam campaign in H8 for the hadronic calorimeter and muon system, using a pion beam with different energies for the energy calibration studies. Data from a parasitic run in H6 with in addition the SciFi and Veto detector installed is also available.
Theis data was used to make a first internal space alignment of the SciFi detector, with a subsequent alignment of the other detectors with respect to the SciFi. This will be repeated with the first data in TI18. The data was also used to determine the light propagation speed in the scintillator bars as well as the absorption length, as reported in Section~\ref{sec:commissioning}.

\begin{figure} [hbtp]
    \centering
    \includegraphics[width=1.0\textwidth]{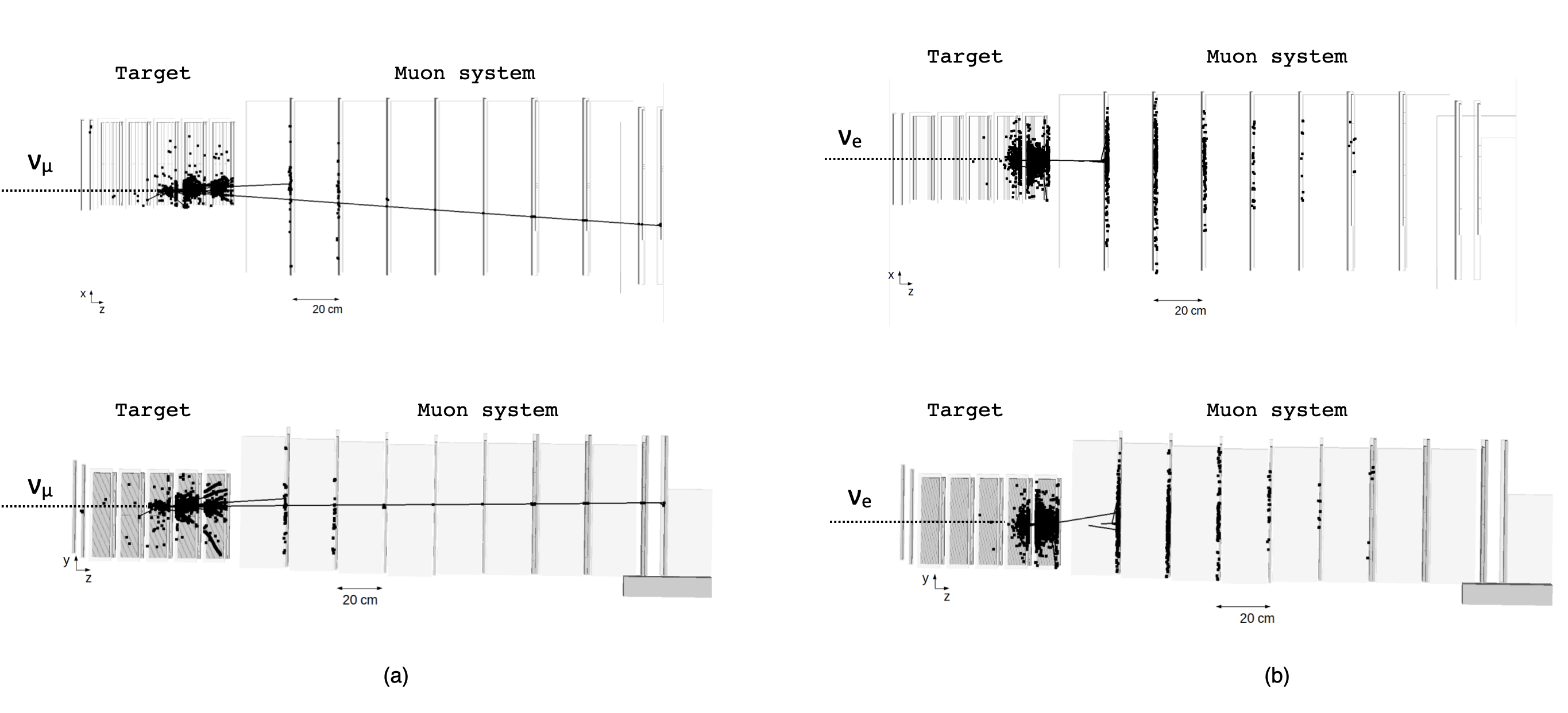}
    \caption{Event display of a $\nu_\mu$ (a) and a $\nu_e$ (b) simulated charged-current interaction in the SND@LHC detector. Detectors are shown in top view on the top and in side view on the bottom.}
    \label{fig:event_reconstruction}
\end{figure}

During Run~3 operation, the upstream veto planes will tag incoming muons that will be used for fine alignment between detector planes. The occurrence of a neutrino interaction or a FIP scattering 
will be first detected by the target tracker and the muon system. Electromagnetic showers are expected to be absorbed within the target region and will therefore be identified by the target tracker, while muons in the final state will be reconstructed by the muon system. In addition, the detector as a whole acts as a sampling calorimeter. The combination of data taken from both systems will be used to measure the hadronic and the electromagnetic energy of the event. A schematic representation of a $\nu_e$ and a $\nu_\mu$ charged-current interaction in the SND@LHC detector is shown in Figure~\ref{fig:event_reconstruction}.

The reconstruction of the emulsion data begins during the scanning procedure. Optical microscopes \cite{Alexandrov:2016tyi,Alexandrov:2015kzs,Alexandrov:2017qpw} analyse the whole thickness of the emulsion, acquiring tomographic images at equally spaced depths.
After digitizing the acquired images an image processor recognizes the grains as \textit{clusters}, i.e.~groups of pixels of a given size and shape. Thus, the track in the emulsion layer (usually referred to as \textit{micro-track}) is obtained by connecting clusters belonging to different levels. Since an emulsion film is formed by two emulsion layers, the connection of the two micro-tracks through the plastic base provides a reconstruction of the particle's trajectory in the emulsion film, called \textit{base-track}.  
The reconstruction of particle tracks in the full volume requires connecting base-tracks in consecutive films. In order to define a global reference system, a set of affine transformations has to be computed to account for the different reference frames used for data taken in different films. Muons coming from the IP will be used for fine film-to-film alignment.  Once all emulsion films are aligned, \textit{volume-tracks} (i.e., charged tracks which crossed several emulsion films) can be reconstructed. The offline reconstruction tools currently used for track finding and vertex identification are based on the Kalman Filter algorithm and are developed in \mbox{FEDRA} (Frame-work for Emulsion Data Reconstruction and Analysis)~\cite{Tyukov:2006ny}, an object-oriented tool based on C++ and developed in the ROOT \cite{Brun:2000es} framework.

\begin{figure}[htbp]
\centering
\includegraphics[width=1.01\columnwidth]{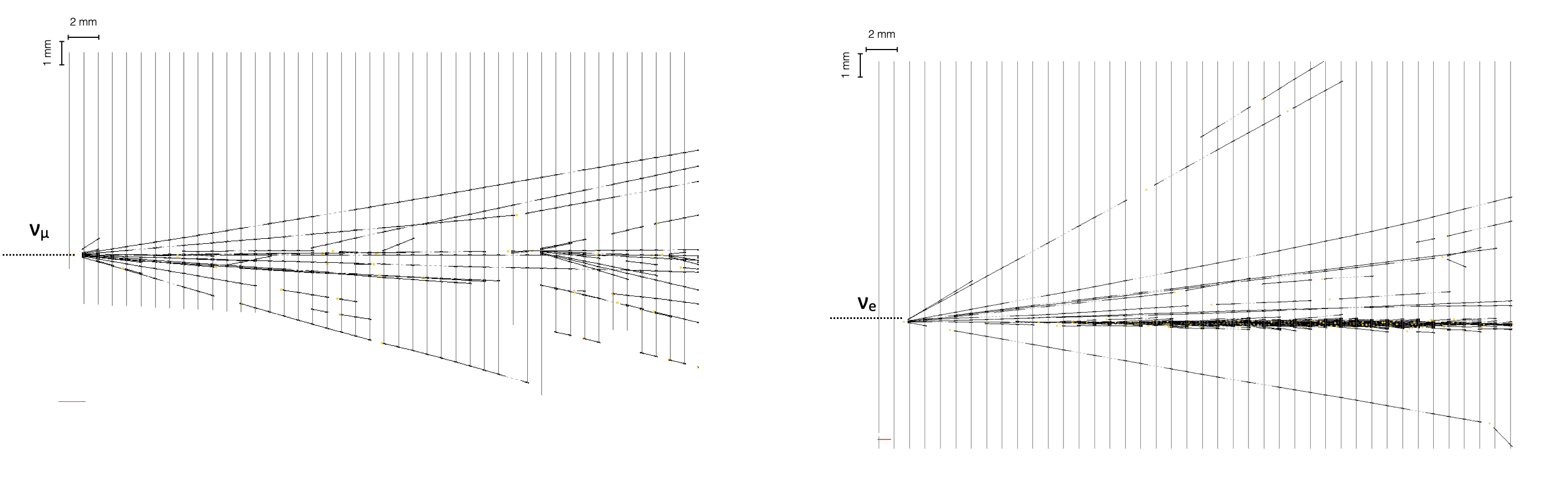}
\caption{Event display of a $\nu_\mu$ (a) and a $\nu_e$ (b) simulated charged-current interaction in the emulsion target}.
%\caption{Illustration of some of the signal topologies that can be reconstructed in the SND@LHC brick.}
\label{fig:neutrino_int}
\end{figure}

The topologies of some signal events that can be reconstructed in the SND@LHC brick are illustrated in Figure~\ref{fig:neutrino_int}.

About twenty neutrino interactions are expected in each brick, given the replacement at every $\sim$25fb$^{-1}$. The matching with the adjacent target tracker plane will be performed by aligning the centre-of-gravity of events reconstructed in the two detectors, thus assigning timing information to interactions reconstructed in the brick. The emulsion data will be also used to complement the target tracker system for the energy measurement of electromagnetic showers.

\subsection{Emulsion scanning system}

\begin{figure}[htbp]
\centering
\includegraphics[width=0.8\columnwidth]{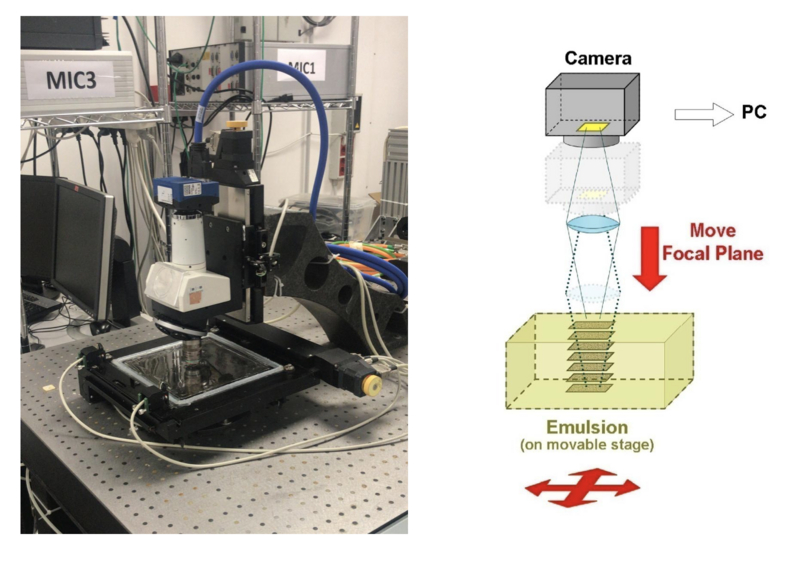}
\caption{(Left) Optical microscope used for the SND@LHC emulsion film scanning. (Right) Schematic drawing of the scanning procedure.}
\label{fig:microscope}
\end{figure}

The emulsion readout is performed in dedicated laboratories equipped with automated optical microscopes, as the one shown in the left panel of Figure~\ref{fig:microscope}. 
The system analyses the whole thickness of the emulsion, acquiring tomographic images at equally spaced depths by moving the focal plane along the vertical axis.

A recently developed upgrade of the European Scanning System (ESS) \cite{Armenise:2005yh, Arrabito:2006rv, DeSerio:2005yd} combines the use of a faster camera with smaller sensor pixels and a higher number of pixels, a lower-magnification objective lens and a new software LASSO \cite{Alexandrov:2016tyi,Alexandrov:2015kzs}, allowing to increase the scanning speed to 180\,cm$^2$/h \cite{Alexandrov:2017qpw}, more than a factor ten faster than before. 
The lens of the microscope guarantees a sub-micron resolution and, having a working distance longer than 300\,$\upmu$m in the direction perpendicular  to the film, allows for a scan of both sides of the emulsion film. 
In order to make the optical path homogeneous in the film, an immersion lens in an oil with the same refraction index of the emulsion is used. A single field of view is 800$\times$600\,$\upmu$m$^2$. 
Larger areas are scanned by repeating the data acquisition on a grid of adjacent fields of view. 
The images grabbed by the digital camera are sent to a vision processing board in the control workstation to suppress noise.
%An implementation of the scanning system with different objective lens and camera and with polarised light, profiting of the plasmon resonance effect~\cite{GarciadeAbajo:2009zz}, has achieved a resolution on the nanometric scale~\cite{Alexandrov:2020gra}.

The total emulsion-film surface to be scanned in SND@LHC is expected to be about $44\,$m$^2$ every four months, thus requiring at least ten scanning systems fully devoted to this activity in order for the readout time to be approximately equal to the exposure time.

\section{Commissioning}
\label{sec:commissioning}
Commissioning of electronic detectors and target mechanics largely took place in the autumn of 2021 in the North Area of the SPS.
These tests included a test beam campaign of the hadronic calorimeter and muon system with hadrons using the H8 beamline as well as commissioning of all the electronics detectors with parasitic muons using the H6 beamline.

%Two test beam campaigns were carried out in September and October 2021 at the H8 beamline of the CERN SPS.
The first is necessary to tune Monte Carlo simulations for accurate shower reconstruction. 

The commissioning in H6 was used to evaluate the performance of all electronic sub-detectors when read out together.
In addition, part of the floor in the H6 beamline was inclined to reproduce the angle of the floor in TI18, to allow the commissioning of the mechanical support.

\subsection{Pion test beam in H8}
\label{subsubsec:h8}

During test beam campaign in H8, all five US station and two DS stations, including passive iron blocks, were tested.
%In the second test beam, an additional DS station was added.
A wall of iron \SI{80}{cm} wide, \SI{60}{cm} tall and \SI{29.5}{cm} thick was placed \SI{20}{cm} upstream of the first US iron block, reproducing the target region in terms of hadronic interaction lengths.

%The test beam campaigns in H8 were carried out to evaluate the detector performance and to provide an energy measurement as an input to simulations. In order to accurately reconstruct the deposited energy in the hadronic calorimeter and muon system, the response of the detector to hadrons of different energies must be evaluated.
%This is necessary to tune Monte Carlo simulations for accurate shower reconstruction. 

%Two test beam campaigns were carried out in September and October 2021 at the H8 beamline of the CERN SPS.
%During the first campaign, all five US station and one DS station, including  passive iron blocks, were tested.
%In the second test beam an additional DS station was added.
%In both configurations, a wall of iron \SI{80}{cm} wide, \SI{60}{cm} tall and \SI{29.5}{cm} thick was placed \SI{20}{cm} upstream of the first US iron block, mimicking the target region.

\begin{figure}[t]
    \centering
    \includegraphics[width=0.6\textwidth]{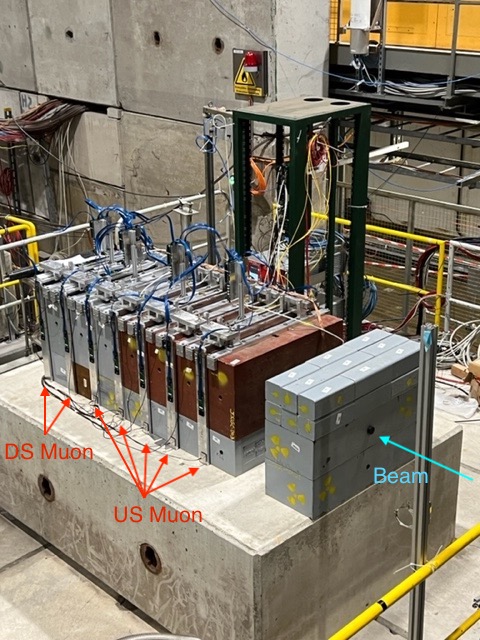}
    \caption{Picture of the second test beam configuration seen in H8 at the SPS. }
    \label{fig:muon_h8}
\end{figure}

\begin{figure}[t]
    \centering
    \includegraphics[width=0.5\textheight]{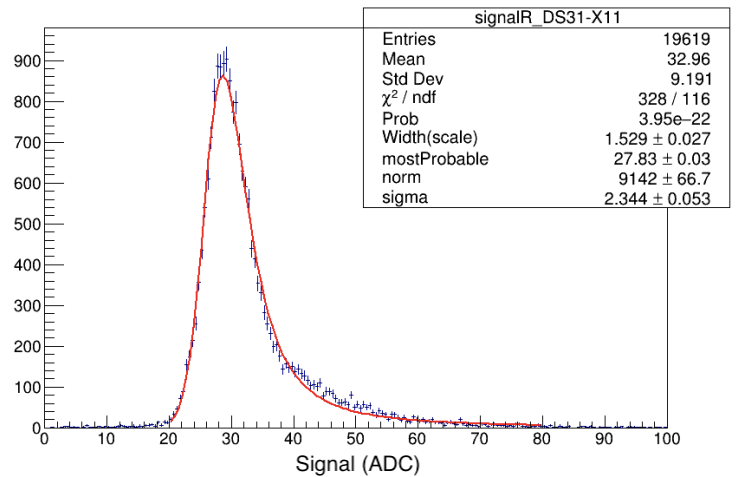}
  
    \caption{An example of the charge in ADC counts recorded by a single SiPM in a DS station at \SI{300}{GeV} fitted with a convolution of a Landau and a Gaussian function.}
    \label{fig:sipm_signal}
\end{figure}

\begin{figure}[t]
    \centering

    \includegraphics[width=0.7\textwidth]{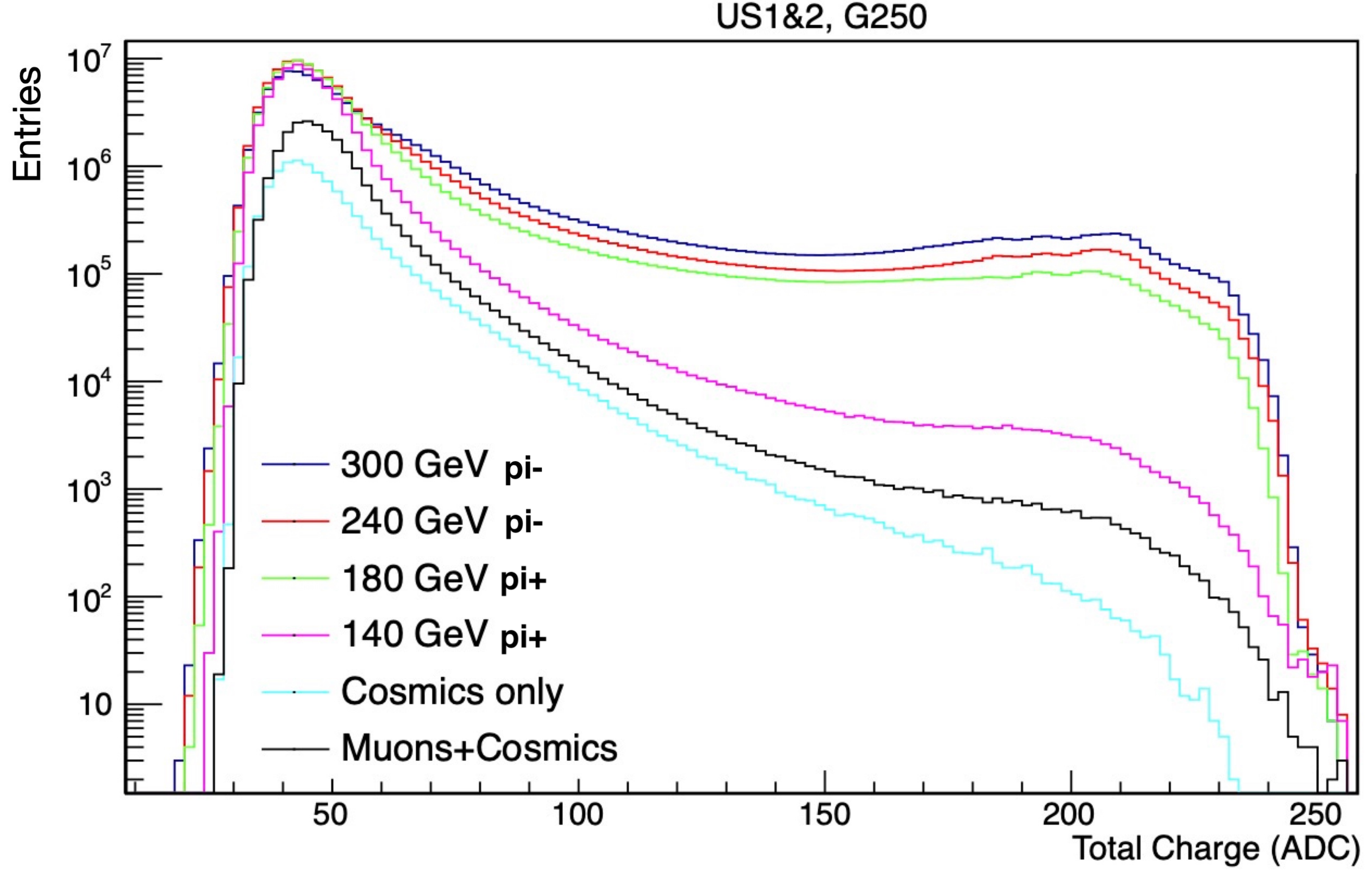}
    \caption{The distribution of total charge per SiPM at a QDC gain of \num{2.50} in the first two US stations recorded for different beam energies including background events from halo and cosmic muons.}
    \label{fig:charge}
\end{figure}

\begin{figure}[t]
    \centering
 
    \includegraphics[width=0.459\textwidth]{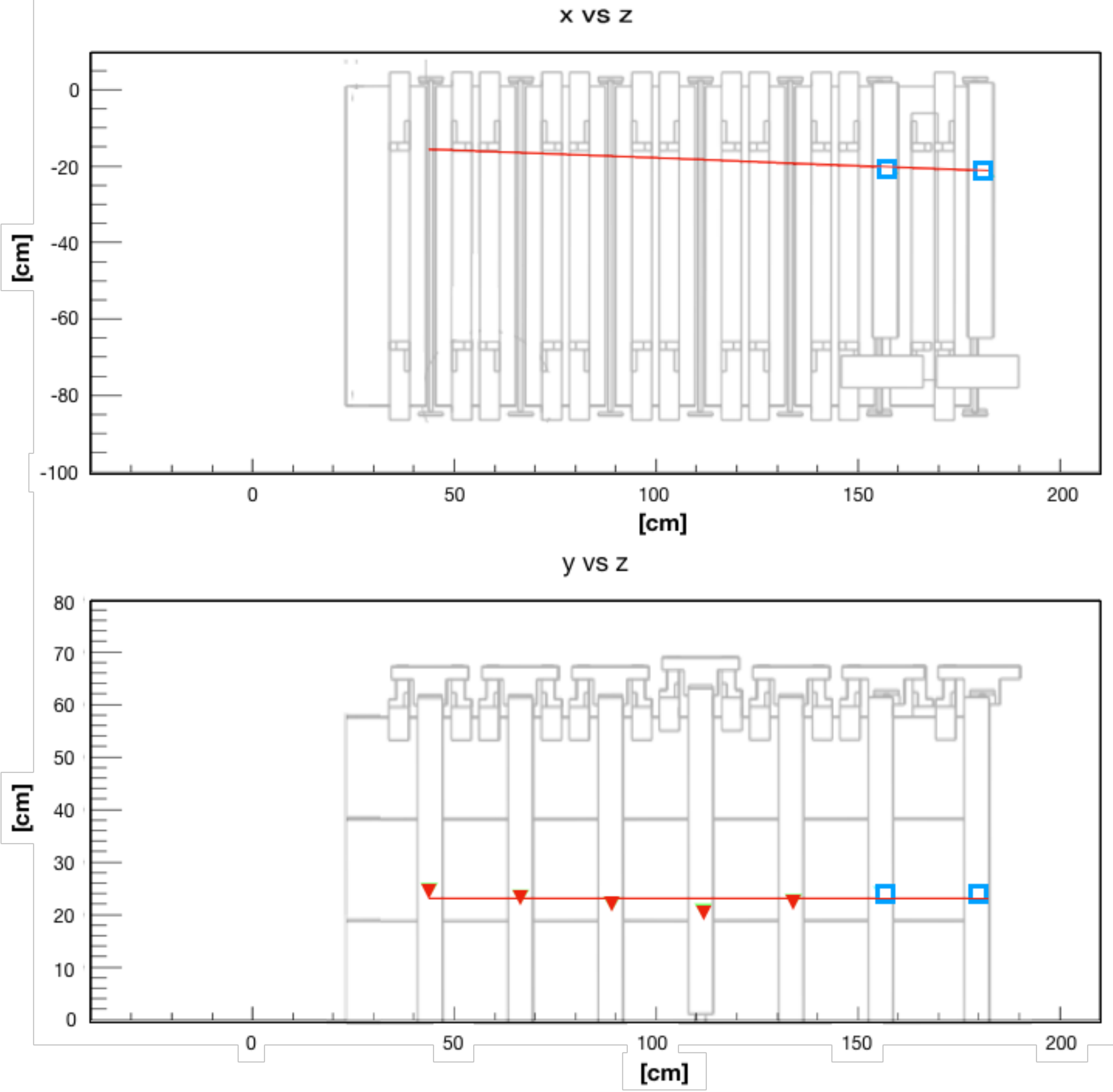}
    \includegraphics[width=0.531\textwidth]{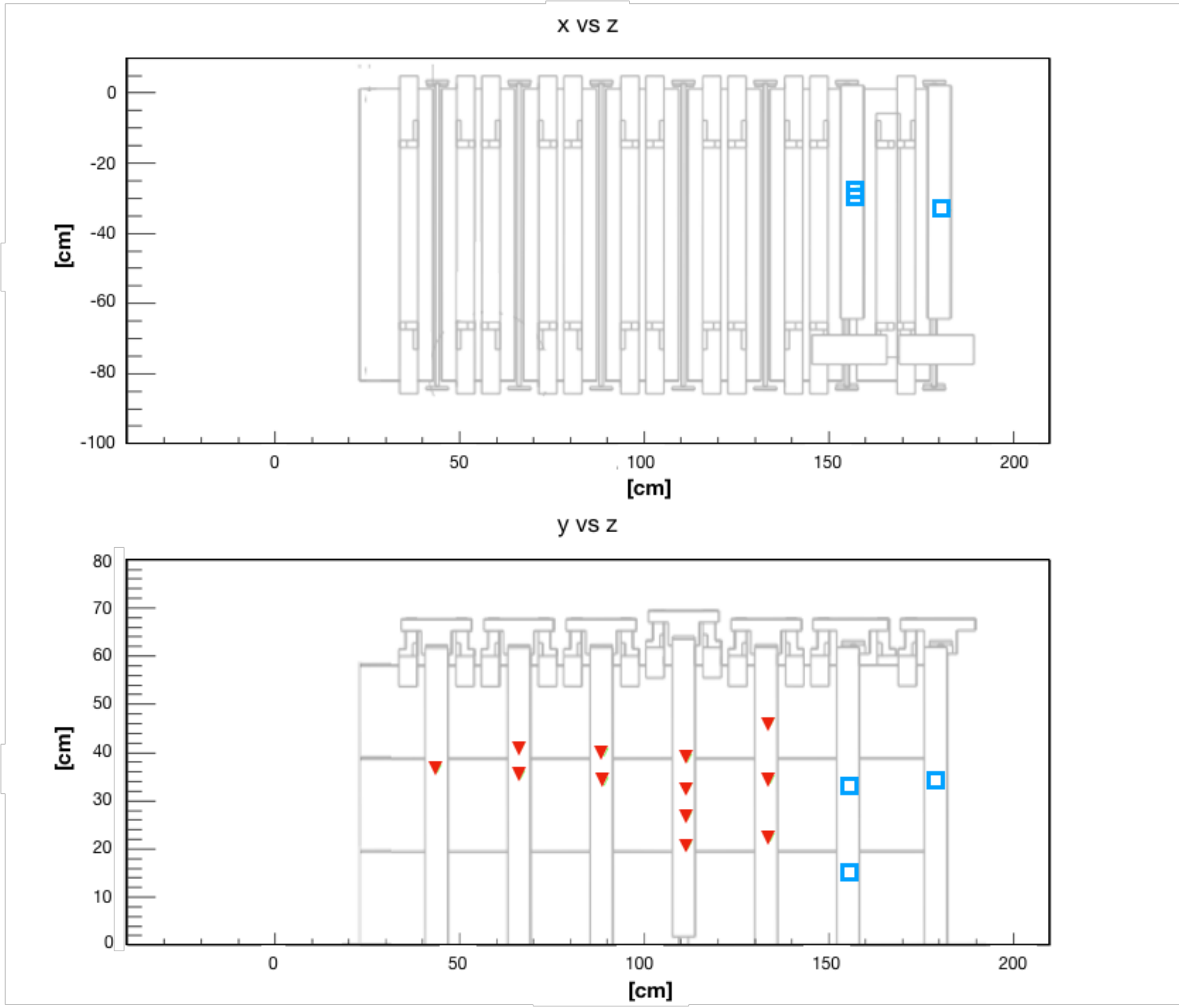}
    \caption{Event display from the second test beam at \SI{300}{GeV} at a gain setting of \num{3.65} with the location of the detector superimposed as seen from above (top) and the side (bottom). The left shows a single particle event with a fitted track while the right shows a multiple particle event.}
    \label{fig:event_h8}
\end{figure}

\begin{figure}[t]
    \centering
 
    \includegraphics[width=0.8\textwidth]{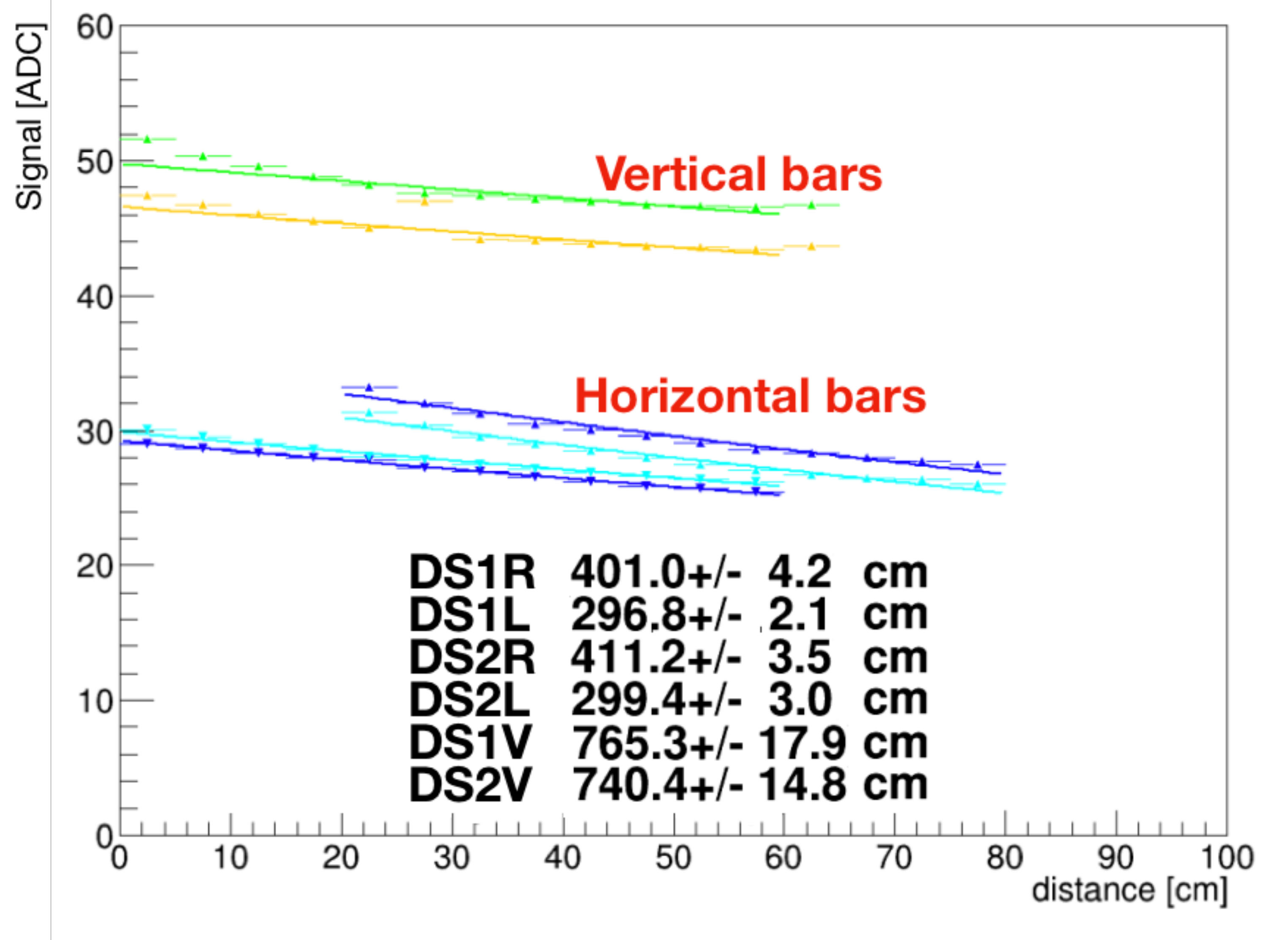}
    \caption{The average signal amplitude as a function of position along the DS bars used to determine the attenuation length. The labels identify the first and second downstream stations (DS1 and DS2 respectively), with the last letter referring to the SiPMs reading out the horizontal bars, left (L) and right (R) and the vertical bars (V). . Note the attenuation length for the vertical bars is nearly double the value for the horizontal bars due to the larger signal from collecting the additional light reflected from the bottom of the bars. The quoted attenuation length values do not include the systematic uncertainties, which are currently being evaluated.}
    \label{fig:att}
\end{figure}

\begin{figure}[t]
    \centering
 
    \includegraphics[width=0.8\textwidth]{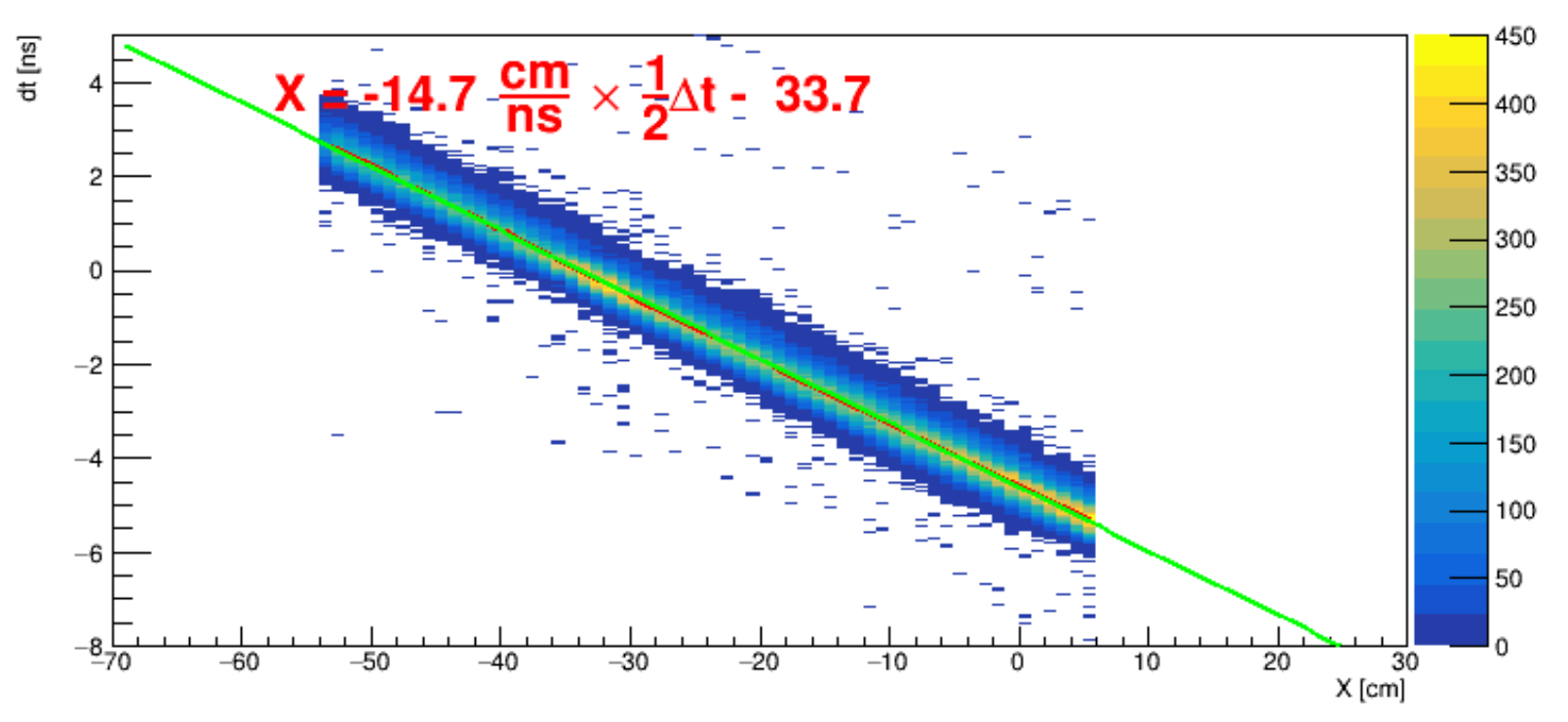}
    \caption{The time difference in the signal between the left and right side of a DS horizontal bar, as a function of the position. The $x$ position is determined from the position of the DS vertical plane right behind the horizontal plane. The slope is used to extract the light propagation speed.}
    \label{fig:prop_speed}
\end{figure}

\begin{figure}[t]
    \centering
 
    \includegraphics[height=0.6\textheight]{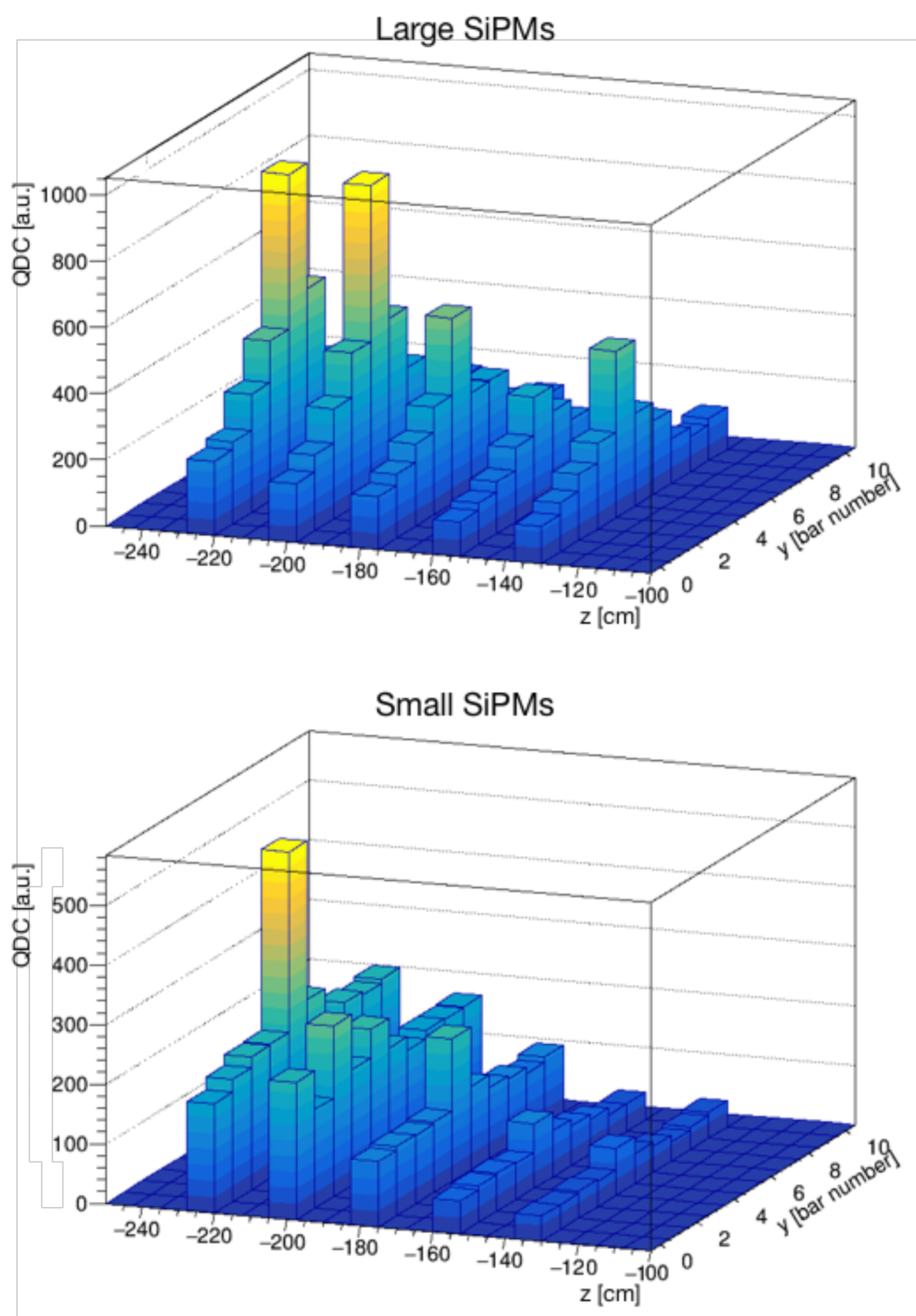}
    \caption{The sum of signals for each bar in the US planes showing the different response between the large (top) and small (bottom) SiPMs at \SI{300}{GeV} and a gain setting of \num{3.65}.}
    \label{fig:shower}
\end{figure}

Besides an energy calibration measurement, the test beam served to investigate the appropriate DAQ settings for data taking. Three different gain settings of the QDC were investigated, \num{1.00}, \num{2.50} and \num{3.65}. Calibration was performed for each gain setting before the beginnings of data taking. Subsequent tests on a spare PCB with a tuneable laser found that a gain setting of \num{2.50} provided the most linear behavior of the recorded signal as a function of  injected charge.

The system, shown in Figure~\ref{fig:muon_h8}, was exposed to 140 and \SI{180}{GeV} positive pions and 240 and \SI{300}{GeV} negative pions.
Additional runs were taken with cosmic muons when the SPS beam was off and with halo muons when a beam dump was placed upstream to obstruct the beamline.
During both test beam campaigns, the beam spot was about \SI{1}{cm} in diameter and the particle rate ranged from \SI{100}{Hz} to \SI{2}{kHz}.

Analysis of the test beam results is ongoing, with initial studies focusing on signal distributions, light attenuation lengths of the bars, detection efficiency, spatial and time resolution, timing calibration, signal propagation speed in the bar, event displays, saturation effects of the SiPMs, MC/data comparison, background estimation and hadronic shower evolution.

Some preliminary results from the test beams are presented here.
The average signal size (in ADC counts) in each SiPM follows the distribution of a Landau distribution convoluted with a Gaussian one, as seen in Figure~\ref{fig:sipm_signal}.
A comparison of the total charge (expressed in QDC units) in the first two US stations for different energies is shown in Figure~\ref{fig:charge}: a significant increase is noticeable in the step from \SI{140}{GeV} to \SI{180}{GeV}.
This is not fully understood and will be investigated with a follow-up testbeam.
Example event displays at \SI{300}{GeV} and at a QDC gain setting of \num{3.65} are displayed in Figure~\ref{fig:event_h8}.

The signal as a function of position along the DS bars is shown in Figure~\ref{fig:att} and the measured attenuation length of $3.6 \pm 0.1$ \SI{}{m}, obtained from the average of the listed values (with the values of the vertical bars taken as half the listed value, due to the presence light reflected off the bottom), is consistent with the value given by the manufacturer (\SI{3.8}{m}) \cite{ej200}.
The time difference between signals collected on opposite ends of a bar, as seen for a DS horizontal plane in Figure~\ref{fig:prop_speed}, can be used to calculate signal propagation speed along the bar which, at about \SI{15}{cm/ns}, closely matches the literature value \cite{betancourt:2017}.
The response of the different SiPM types at \SI{300}{GeV} and the highest gain setting for the US can be seen in Figure~\ref{fig:shower}, with the small SiPM response indicating that hadronic showers are mainly contained in the first three layers.
The drop seen in Figure~\ref{fig:shower} for the large SiPMs of the fourth US station is presumed to be due to dead channels on the PCB, although this is still under investigation.%Preliminary studies of the timing and spatial resolution of each plane is \SI{\sim 100}{ps} and \SI{\sim 1}{cm}, respectively. 

\subsection{Muon test beam in H6}
\label{subsec:commissioning_h6}
All electronic detectors have been accurately tested before being installed in the TI18 tunnel.
An important part of these tests has been performed in the H6 beamline of the CERN SPS, where all electronic subdetectors have been operated together for the first time.

\begin{figure}[h]
    \centering
    \includegraphics[width=0.8\textwidth]{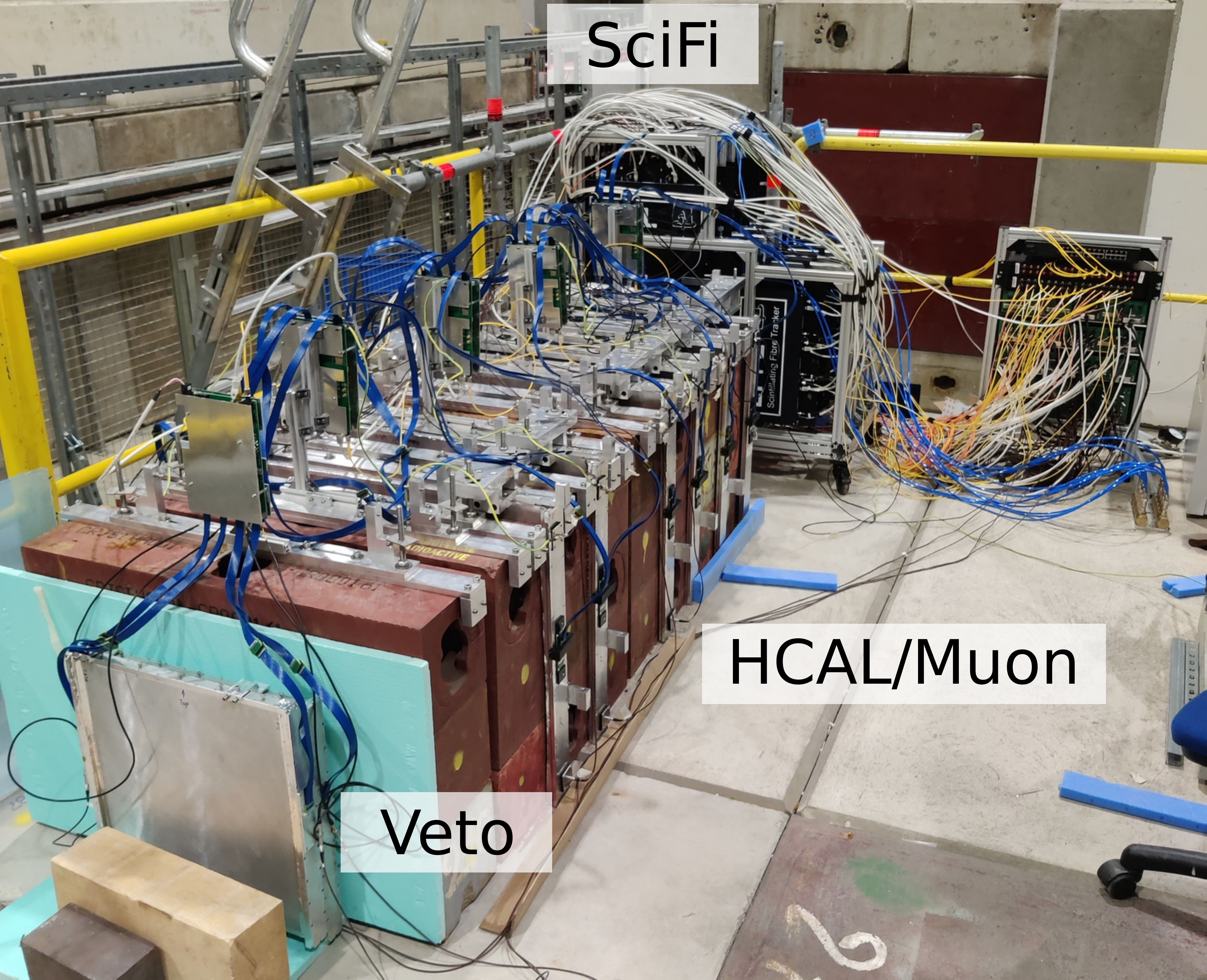}
    \caption{The setup used for the commissioning in H6. The veto planes are visible in the foreground, followed by the hadronic calorimeter and muon system, and finally the SciFi tracker.}
    \label{fig:setup_h6}
\end{figure}

A picture of the setup installed in H6 is shown in Figure~\ref{fig:setup_h6}.
Due to space constraints, the order of the detectors was not the same as in its TI18  configuration: the veto is placed right in front of the hadronic calorimeter, while the SciFi is located behind the hadronic calorimeter and muon system.

The measurements performed in H6 focused on the study of the performance and alignment of all subsystems.
%In addition, they have been a test for the DAQ software as well, helping to find bugs and making it more robust.
Several runs at different settings have been collected.
The details for each sub-system are given in the following sections.

\subsubsection{Veto, hadronic calorimeter and muon system results}

Commissioning of the veto and muon systems were carried out in two phases, before and after the test beam in H8. The commissioning tests in H6 represented the first test of the veto and its electronics.

During the first commissioning phase, five US and two DS stations of the hadronic calorimeter and muon system were tested along with the SciFi. Ground loops were also discovered, leading to a significant noise increase and difficulties with DAQ calibration. 
This led to the introduction of a grounding cable between the ground of the HV and ground of LV, which were ready in the second phase of commissioning. During this phase, dimensions and spacing of the veto system within the target structure mechanics were also checked. 

\begin{figure}[t]
    \centering
    \includegraphics[width=0.7\textwidth]{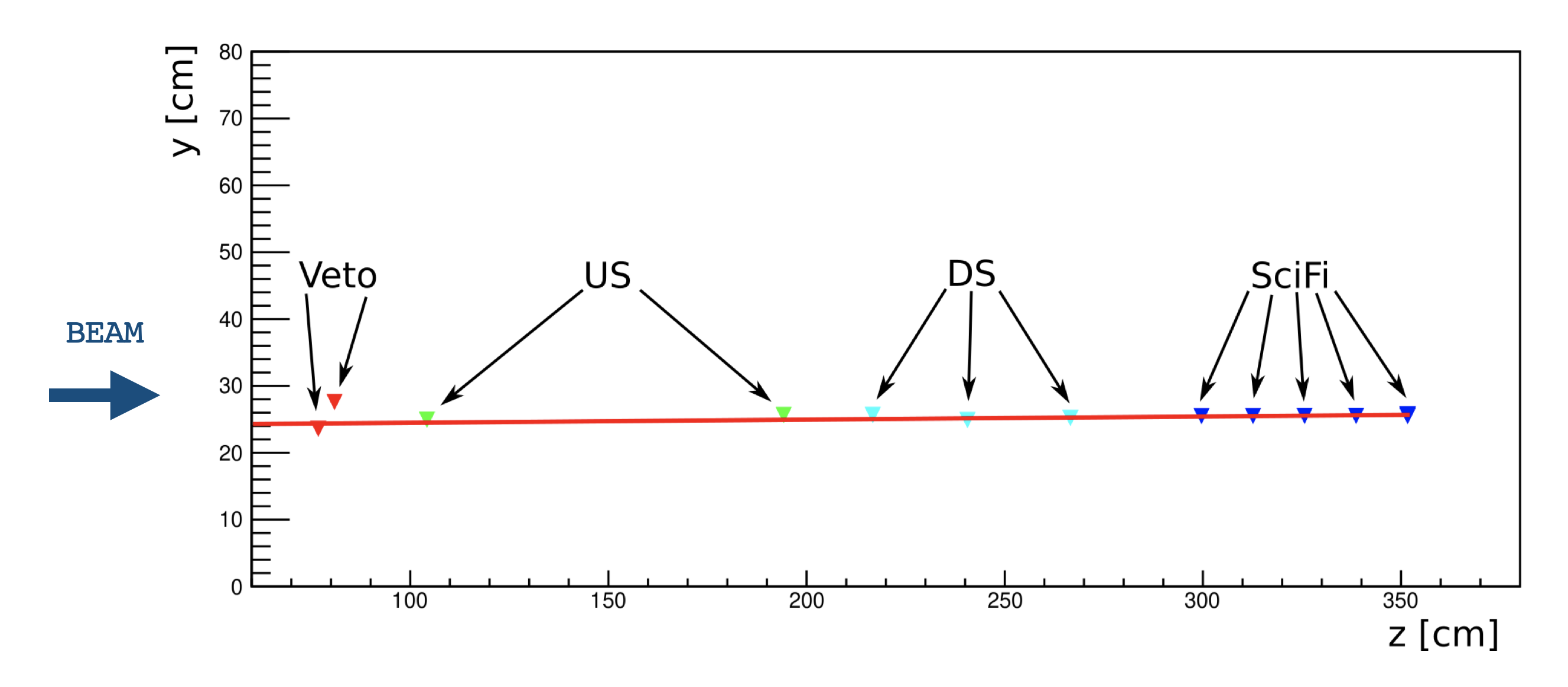}
    \caption{Event display showing the veto, muon system and SciFi as taken in H6.}
    \label{fig:h6_event}
\end{figure}

In the second phase, the target structure mechanics was removed and the veto was placed directly in front of the muon system, with the SciFi placed further downstream. 
A third DS station was also added. It was discovered that PCBs on three US stations and two DS stations displayed a number of missing channels, which were then removed from the experimental hall for repair. 
%Because of the silicone epoxy placed on one end, US stations were taken from H6 and  PCBs were removed in a clean room and sent for repair at the CERN electronics workshop. Meanwhile the DS PCBs needing repair were removed directly from the experimental hall. 
The remaining stations were then tested along with the veto and SciFi, with the addition of  grounding cables to the veto and muon system as mentioned before. 
An example of an event display including all electronic sub-systems is shown in Figure~\ref{fig:h6_event}. 
Preliminary analysis of the data also indicated missing channels in two PCBs of the veto system, which were sent for repair at the end of the commissioning phase. 

\subsubsection{SciFi results}
\label{subsubsec:commissioning_scifi_result}
Several runs in different conditions were collected: the T1 and T2 thresholds, described in Section~\ref{subsec:tofpet}, were varied, while the E threshold was not used.
T1 is lower and determines the timestamp of the hit, so its influence on the time resolution was studied, while T2 is higher, and determines whether a hit is collected or not.
It was studied to find the optimal compromise between dark-count rate and efficiency.
In addition, data were collected at three different QDC gain values: \num{1.00}, \num{2.50} and \num{3.65}.

The alignment of the SciFi stations, both their relative position and the inner degrees of freedom within one station, were studied by performing track reconstruction using four of the five stations, extrapolating the reconstructed tracks on the fifth one and minimizing the residuals, i.e. the difference between the track extrapolation and the corresponding cluster position.

\begin{figure}
    \centering
    \includegraphics[width=0.49\textwidth]{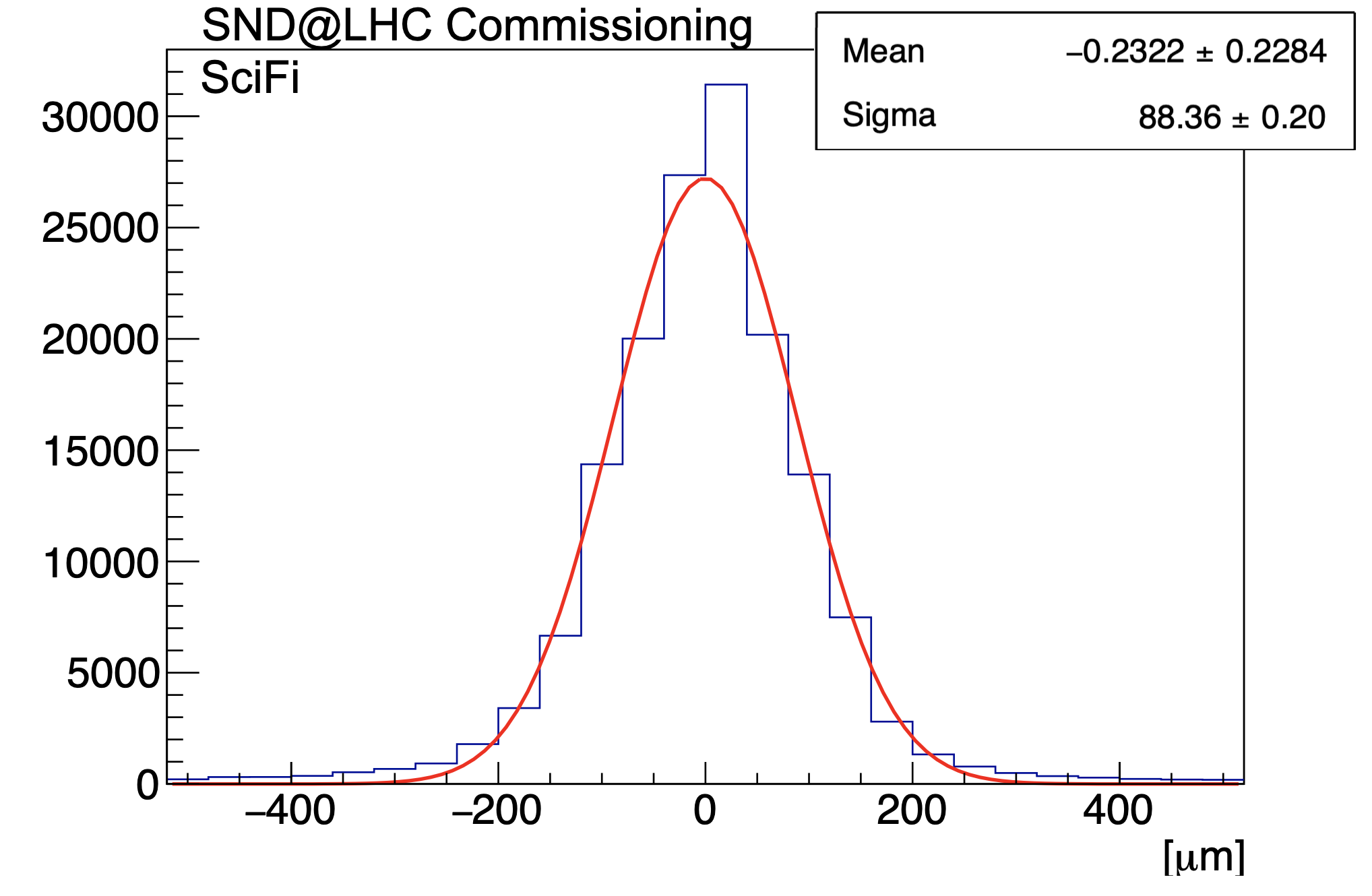}
    \includegraphics[width=0.49\textwidth]{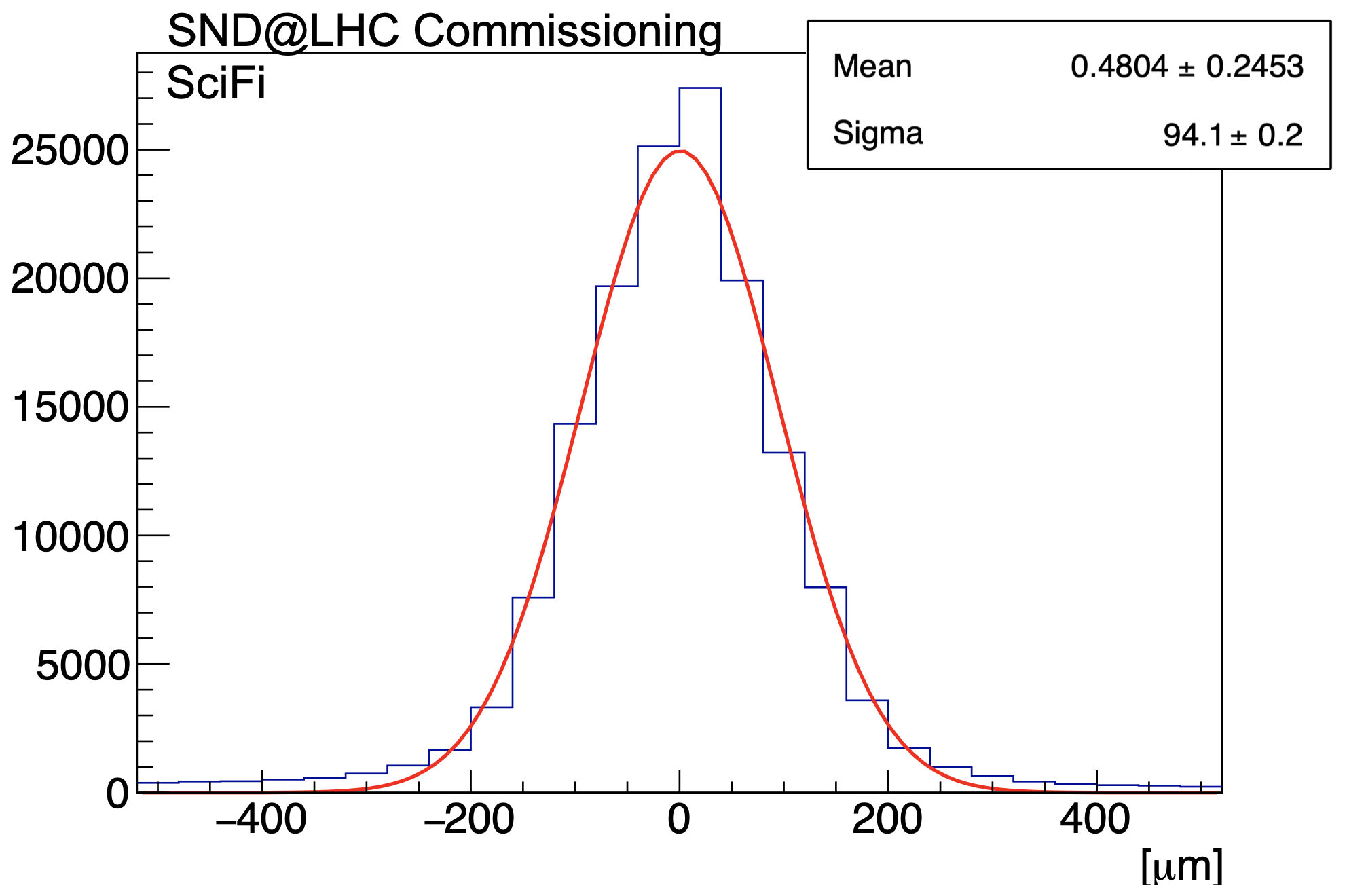}
    \caption{Track-cluster residuals along the $x$-axis (left) and $y$-axis (right). The standard deviation of the fitted gaussian distribution is the spatial resolution of SciFi.}
    \label{fig:residual-scifi}
\end{figure}

Results obtained for one station are shown in Figure~\ref{fig:residual-scifi}.
They show that the alignment procedure works as expected, as the residuals distribution is peaked at 0 and that the spatial resolution of the SciFi system is below \SI{100}{\micro m}.
%They show the  displacement of the station with respect its the nominal position, to be corrected via software, and they give a first hint on the spatial resolution of SciFi, expected to be better that \SI{100}{\micro m} after alignment.

% \begin{figure}[h]
%     \centering
%     \includegraphics[width=0.7\textwidth]{images/commissioning/residual-scifi.png}
%     \caption{Relative displacement of one SciFi station with respect to the nominal position along the $x$-axis (top) and $y$-axis (bottom). The width of the distributions (reported in the Gaussian fit inset) give a first hint on the spatial resolution of SciFi.}
%     \label{fig:residual-scifi}
% \end{figure}

%The relative rotation was also studied, by correlating the residual in one coordinate with the cluster position in the other.
%The result is shown in Figure~\ref{fig:rotation-scifi}.
%This rotation also needs to be corrected in software.

% \begin{figure}[h]
%     \centering
%     \includegraphics[width=0.7\textwidth]{images/commissioning/rotation-scifi.png}
%     \caption{Correlation between the $y$ position of the $x$ residuals, allowing to determine the relative rotation around the $z$ axis of one SciFi station.}
%     \label{fig:rotation-scifi}
% \end{figure}

%Finally, the inner degrees of freedom within one plane were studied, in particular the relative displacement of the three SiPMs assemblies on each side and the relative rotation of  fibre mats.
%Results show that some stations require a correction of the position of up to \SI{300}{\micro m} and a correction of the relative rotation of \SI{\sim 1}{mrad}.
%All the necessary alignment corrections have been implemented in the reconstruction software.

The particle detection efficiency of the SciFi detector was studied at two different T2 thresholds (a higher one, producing \SI{\sim 2}{Hz} of dark rate per channel and a lower one, producing \SI{\sim 20}{Hz}).
The efficiency was studied similarly to the alignment, by reconstructing tracks using four stations, extrapolating them to the fifth one and looking for an associated cluster within a radius of \SI{1}{cm}.
The efficiency is calculated as the ratio between the tracks with associated cluster (in the $x$ plane, $y$ plane or both) and the total number of tracks.

\begin{figure}[h]
    \centering
    \includegraphics[width=\textwidth]{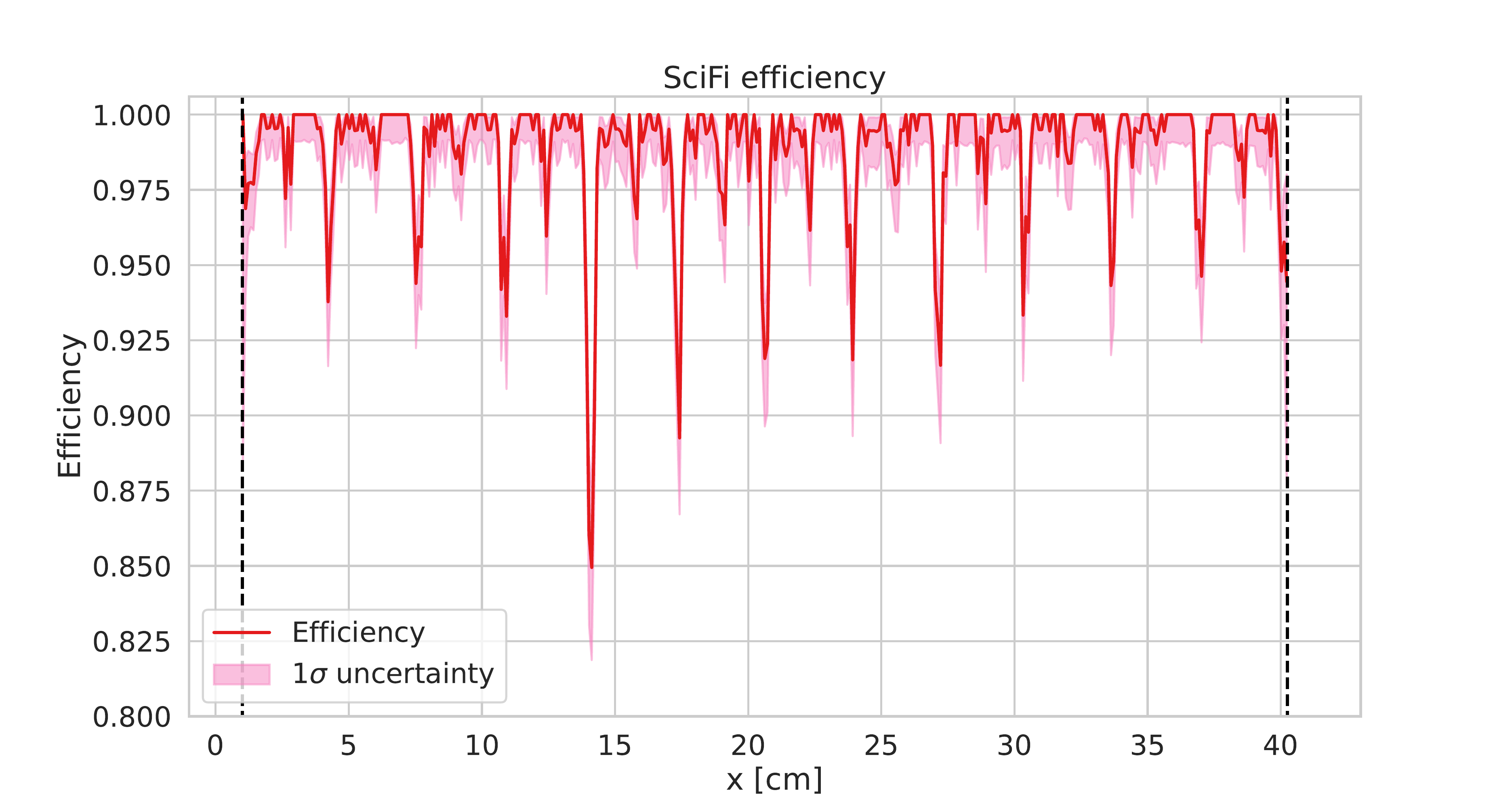}
    \caption{Efficiency of the x layer of the third SciFi module. The dips in efficiency are caused by the gaps between SiPM arrays. The black dashed lines indicate the edges of the layer.}
    \label{fig:scifi-eff}
\end{figure}

%The result for one SciFi station at low threshold, requiring an associated cluster on both the $x$ and $y$ plane, is shown in Figure~\ref{fig:scifi-eff}.
At low threshold, the efficiency is \SI{97}{\percent} over the whole station, while it drops down to \SI{\sim 65}{\percent} at the higher threshold.
This result, in combination with the maximum hit rate allowed by the DAQ server (see Section~\ref{subsec:daq-events}) lead to the decision of running the detector an even lower T2 threshold, producing \SI{\sim 25}{Hz} of dark rate per channel.
The efficiency of a representative SciFi layer at this threshold is shown in Figure~\ref{fig:scifi-eff}.
All layers show a consistent efficiency of \SI{\sim 98}{\percent}, which raises to \SI{\sim 99}{\percent} if the gaps between SiPMs are excluded from the computation.
% Using a lower threshold significantly improves the efficiency: the minimum is \SI{\sim 60}{\percent} in the high threshold run (Figure~\ref{fig:scifi-eff-high}) and \SI{\sim 92}{\percent} in the low threshold one (Figure~\ref{fig:scifi-eff-low}).
%The beam region shows a lower efficiency, while in the rest of the detector it is close to \SI{100}{\percent}.
%This inefficiency has been investigated and its cause has not been found yet, but a data loss and a too high particle rate have been excluded.

\begin{figure}[h]
    \centering
    \includegraphics[width=\textwidth]{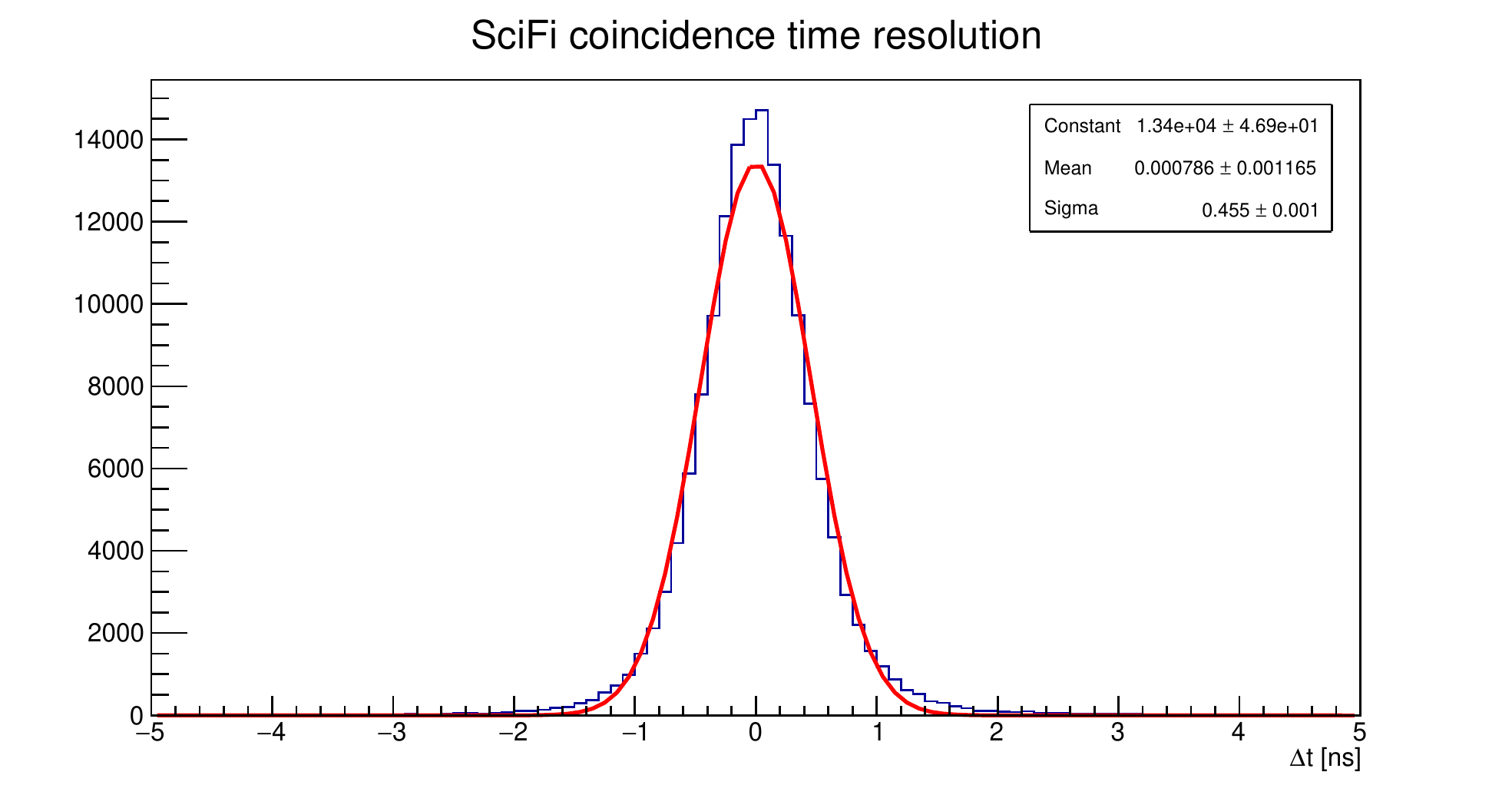}
    \caption{Distribution of the timestamp difference recorded between two layers of the second SciFi module. The CTR is obtained by fitting the data with a Gaussian curve. The single layer time resolution is obtained dividing the standard deviation by $\sqrt{2}$, in the assumption that the timestamps recorded by each layer are independent.}
    \label{fig:scifi-timeres}
\end{figure}

The time resolution of the SciFi tracker is limited by the number of detected photons and the scintillator decay time.
It has been measured by calculating the coincidence time resolution (CTR) between two fibre layers, correcting for the light propagation delay in the fibres.
The results are shown in Figure~\ref{fig:scifi-timeres}: each fibre layer has a time resolution of \SI{\sim 330}{ps}, which translates in \SI{\sim 230}{ps} per plane or \SI{\sim 100}{ps} for the whole target tracker 
%For multiple tracks or showers, the total number of photons is significantly larger and the time resolution better.

%The study of energy resolution is currently ongoing.

\subsubsection{Commissioning of the target structure}

\begin{figure}[h]
    \centering
    \includegraphics[width=1.0\textwidth]{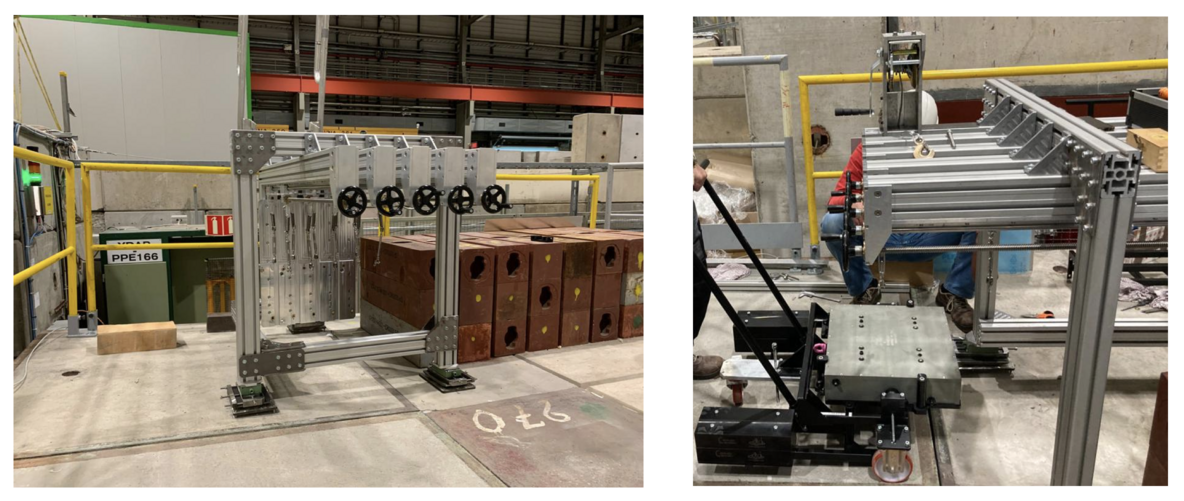}
    \caption{Commissioning of the target mechanical structure in H6.}
    \label{fig:structure_comm}
\end{figure}

In order to perform the commissioning of the target mechanical structure, the upstream section of the floor in H6 was inclined by \SI{4}{\degree} to reproduce the floor inclination in the TI18 tunnel. The whole structure was assembled and installed on the three alignment feet, as shown in Figure~\ref{fig:structure_comm}.

A test of the transportation along the slope of the wall box with the trolley, of the insertion and extraction of the wall box inside the structure, as well as of the fixation of the SciFi plane was successfully performed.

\subsection{Target wall commissioning}

The commissioning of the target wall was performed in November 2021 at the Emulsion Facility at CERN.
A test with a first batch of \SI[product-units=power]{192 x 192}{mm} emulsion films was conducted in order to test the chemical compatibility of tungsten plates with emulsions, the light tightness of the wall box, the uniformity of track reconstruction in different bricks and in different positions within the brick.

\begin{figure}[h]
    \centering
    \includegraphics[width=1.0\textwidth]{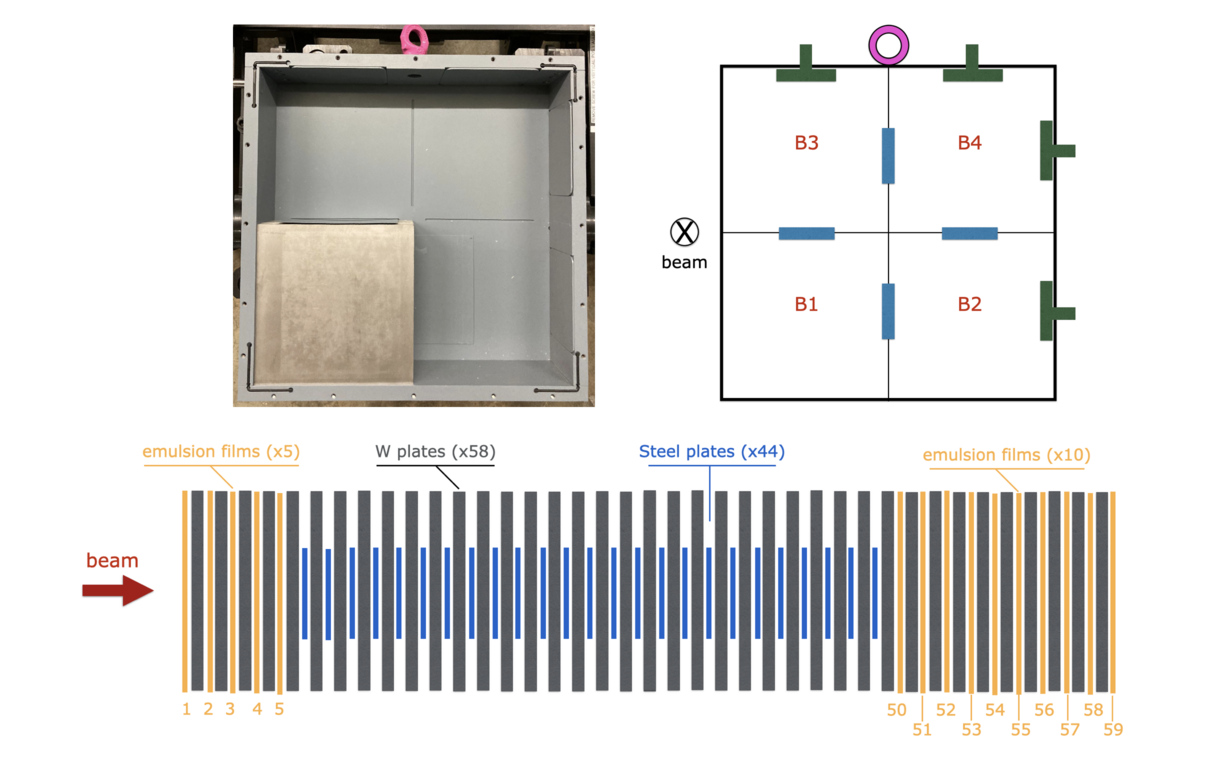}
    \caption{Top: picture taken during the wall assembly and schematic drawing of the wall structure. Bottom: schematic drawing of the position of emulsion films within in the bricks used for the commissioning.}
    \label{fig:wall_comm}
\end{figure}

A full wall made of four bricks, each consisting of 58 tungsten plates, was assembled in dark room conditions.
A stack of 30 emulsion films were used for the test, disposed in two bricks (B1 and B4) as reported in the schematic drawings in Figure~\ref{fig:wall_comm}. Steel plates with a surface of \SI[product-units=power]{10x10}{cm} and a thickness of \SI{300}{\micro m} were used to replace emulsion films in the remaining part of the walls.

After the assembly, the wall box was exposed to cosmic radiation for \SI{48}{h} without any dedicated shielding. 
Then emulsion films were developed and scanned with automated optical microscopes in one of the emulsion scanning laboratories of the Collaboration. During the scanning, aligned grains in adjacent emulsion layers are recorded by a camera and stored as digital pixels.  
After the scanning, an image processor recognized aligned clusters, formed by groups of pixels. 
These clusters need to be separated from a background of thermally excited grains, which get developed even if not exposed to radiation.
This background is usually referred to as \textit{fog}, and its density was measured by counting the number of grains per unit volume in both emulsion layers. 
An average grain density of $4.5 \pm 0.2$  per \SI{1000}{\cubic \um} was measured, compatible with that of reference ($i.e.$ not exposed) emulsion films, showing that contact of the films with neither the tungsten plates nor the internal coating of the wall had chemically contaminated the emulsion. 
The grain density was measured in different points of the emulsion surface and for different positions of emulsion films within the brick, demonstrating the light tightness of the wall box.

\begin{figure}[h]
    \centering
    \includegraphics[width=0.45\textwidth]{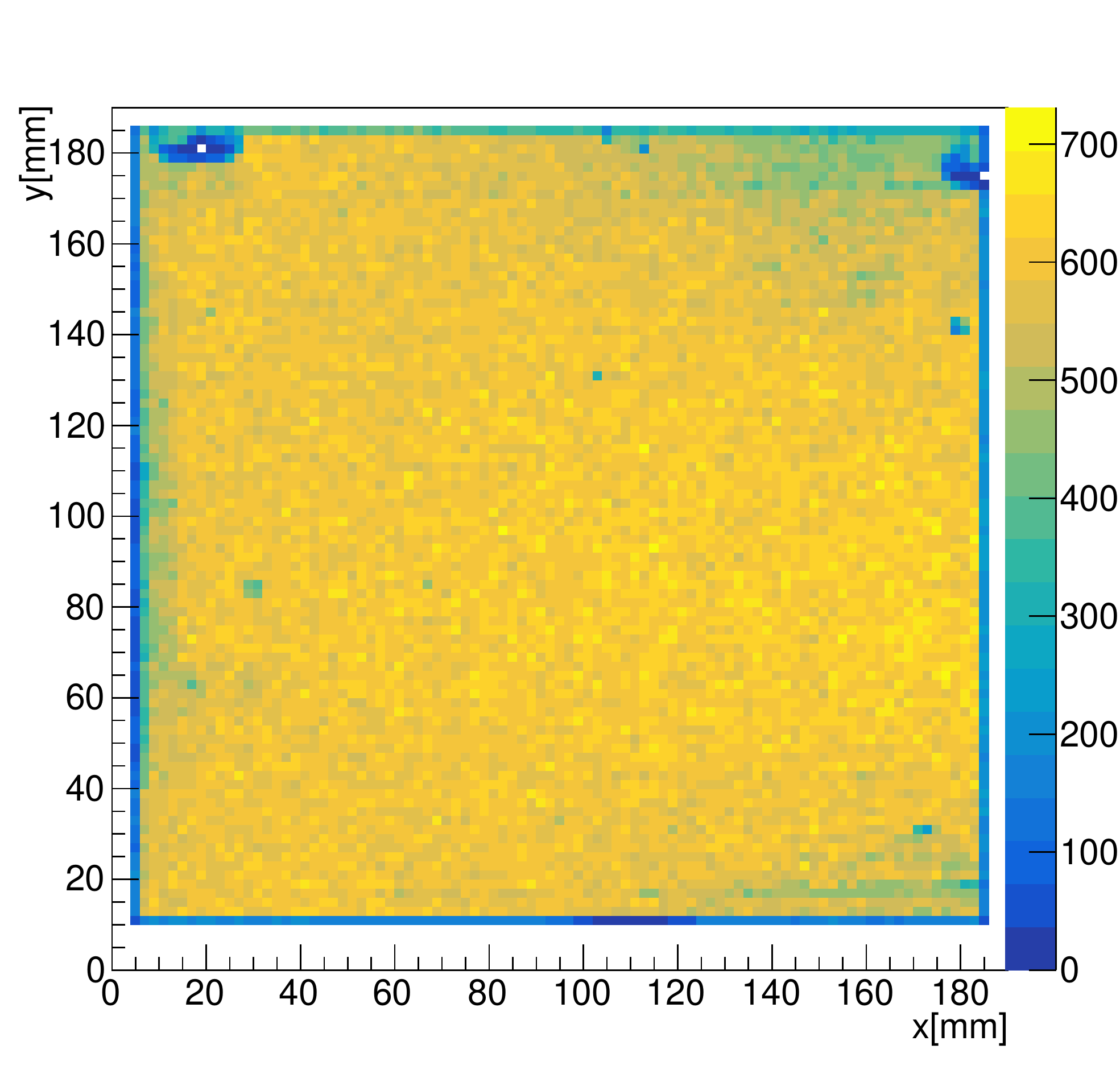}
    \includegraphics[width=0.45\textwidth]{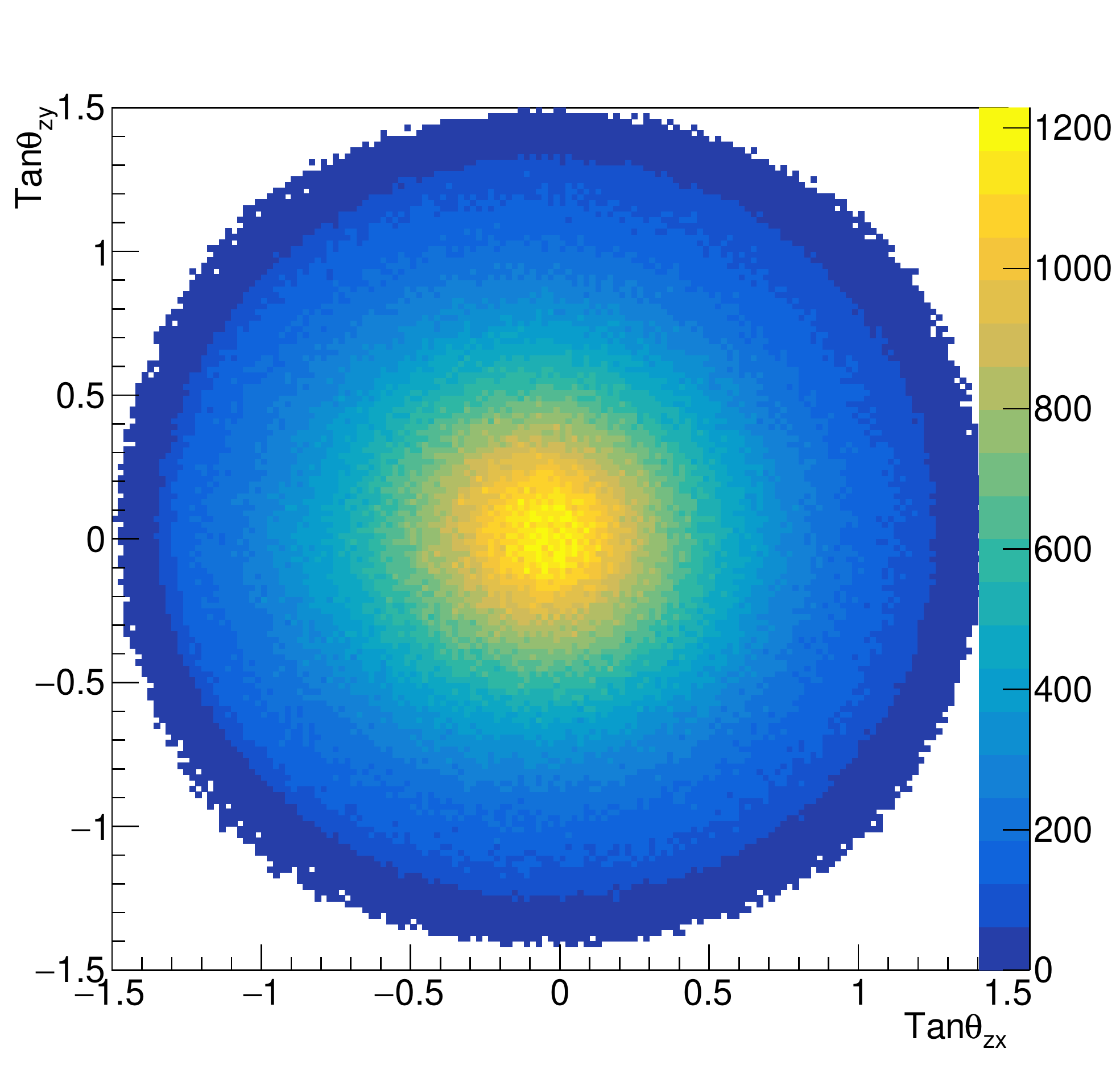}
    \caption{ Reconstructed segments on a single emulsion film used in the target wall commissioning. Left: segments position in the transverse plane; the color palette represents the number of reconstructed segments per mm$^2$. Right: Distribution of the slopes of reconstructed segments in the $zx$ plane and $zy$ plane.}
    \label{fig:emu_distributions}
\end{figure}

After the scanning, the reconstruction process was performed, as described in Section~\ref{sec:offline}. 
The position and angular distributions of reconstructed base-tracks in an emulsion film are shown in Figure~\ref{fig:emu_distributions}.
A good alignment between consecutive films was obtained, proving that the distortion of the emulsion films is negligible.
The tracks are distributed uniformly on the surface, as expected. 
Since the target was placed horizontally during the exposure, the cosmic rays crossed the emulsion films perpendicularly to their surface, leading to a peak of reconstructed tracks at low angles.

\begin{figure}
    \centering
    \includegraphics[width=0.49\textwidth]{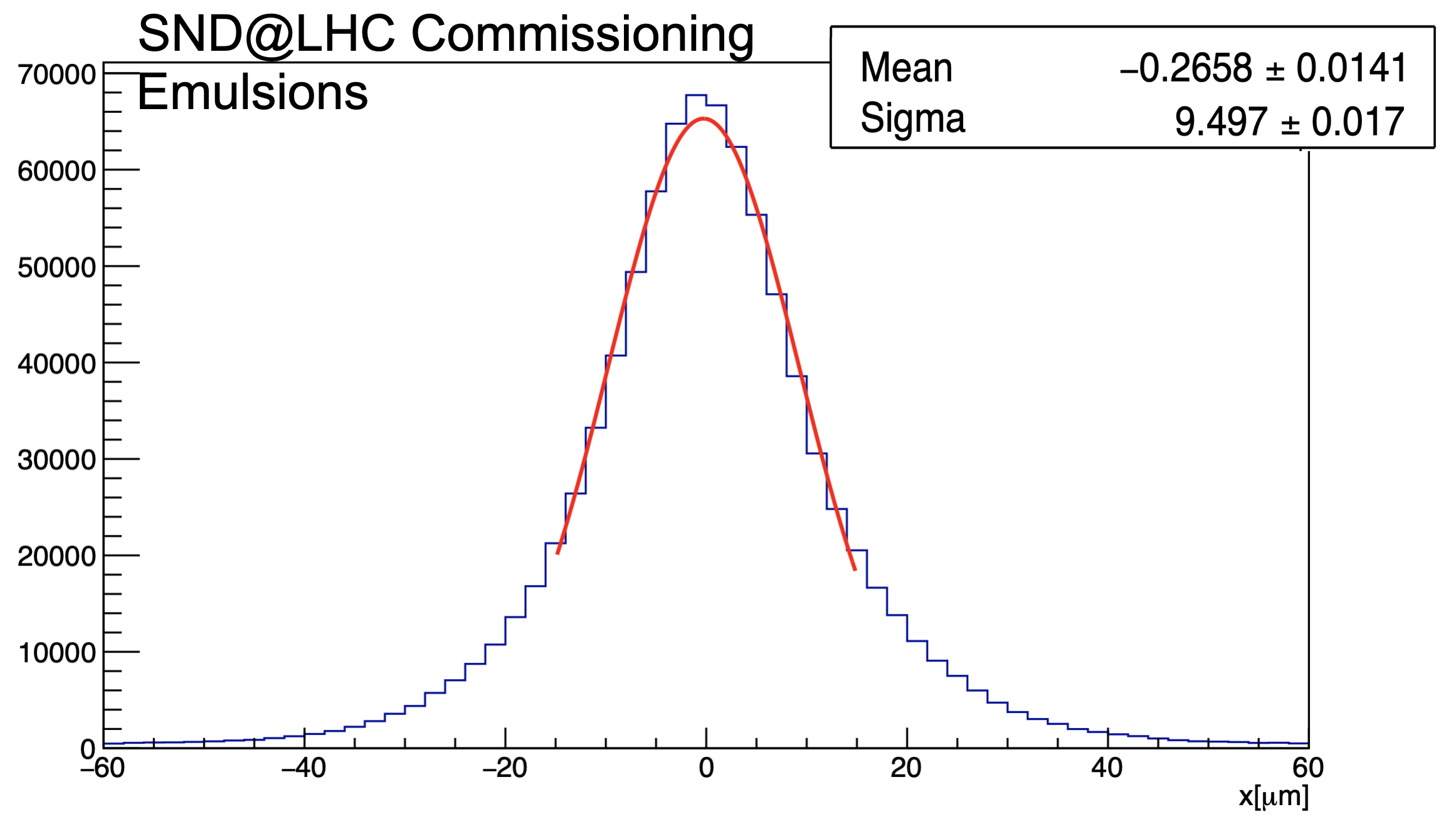}
    \includegraphics[width=0.49\textwidth]{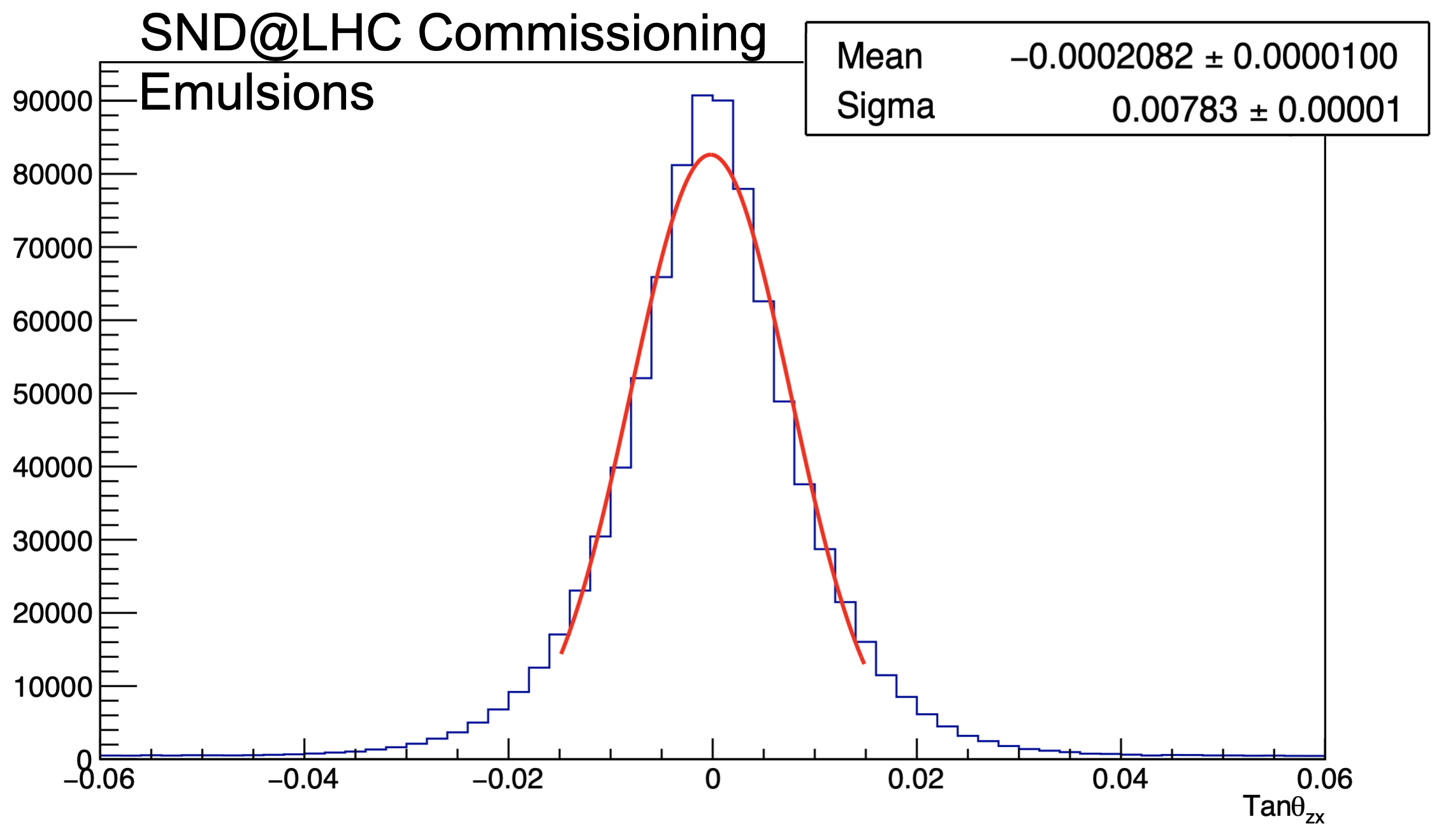}
    \caption{Tracking resolution in $x$ position (left) and slope as the tangent of track angle in the $zx$ plane ($T_X \equiv Tan(\theta_{zx})$)  (right).}
    \label{fig:emutrack_resolution}
\end{figure}

 Finally, track reconstruction in the whole target is performed, with a \textit{Kalman filter} seeded on the base-tracks recorded in the single emulsion films. Comparing the position and angle of each base-track with a linear fit on the $xz$ and $yz$ planes leads to an estimation of the tracking resolution. The results are reported in Figure~\ref{fig:emutrack_resolution}, with Gaussian fits leading to $\sigma_\text{X} \sim \SI{9}{\micro m}$ for the position and $\sigma_\text{TX}\sim \SI{8}{mrad}$ for the slope as the tangent of track angle in the $zx$ plane ($T_X \equiv \tan(\theta_{zx}$).

The average surface density amounts to $(1.5 \pm 0.1) \times 10^3$ tracks/cm$^2$, reconstructed in the angular range from 0 to 1 rad. Over a total exposure time of 48 hours, this density corresponds to a flux of $(0.52 \pm 0.03)$  muons/\SI{}{\square \cm}/min. The expected cosmic-ray in the same angular acceptance amounts to about 0.73  muons/\SI{}{\square \cm}/min. The discrepancy can be attributed to the energy threshold for reconstructed tracks, which are required to pass through the Tungsten layers.

\begin{figure}[h]
 \centering
 \includegraphics[width=0.6\textwidth]{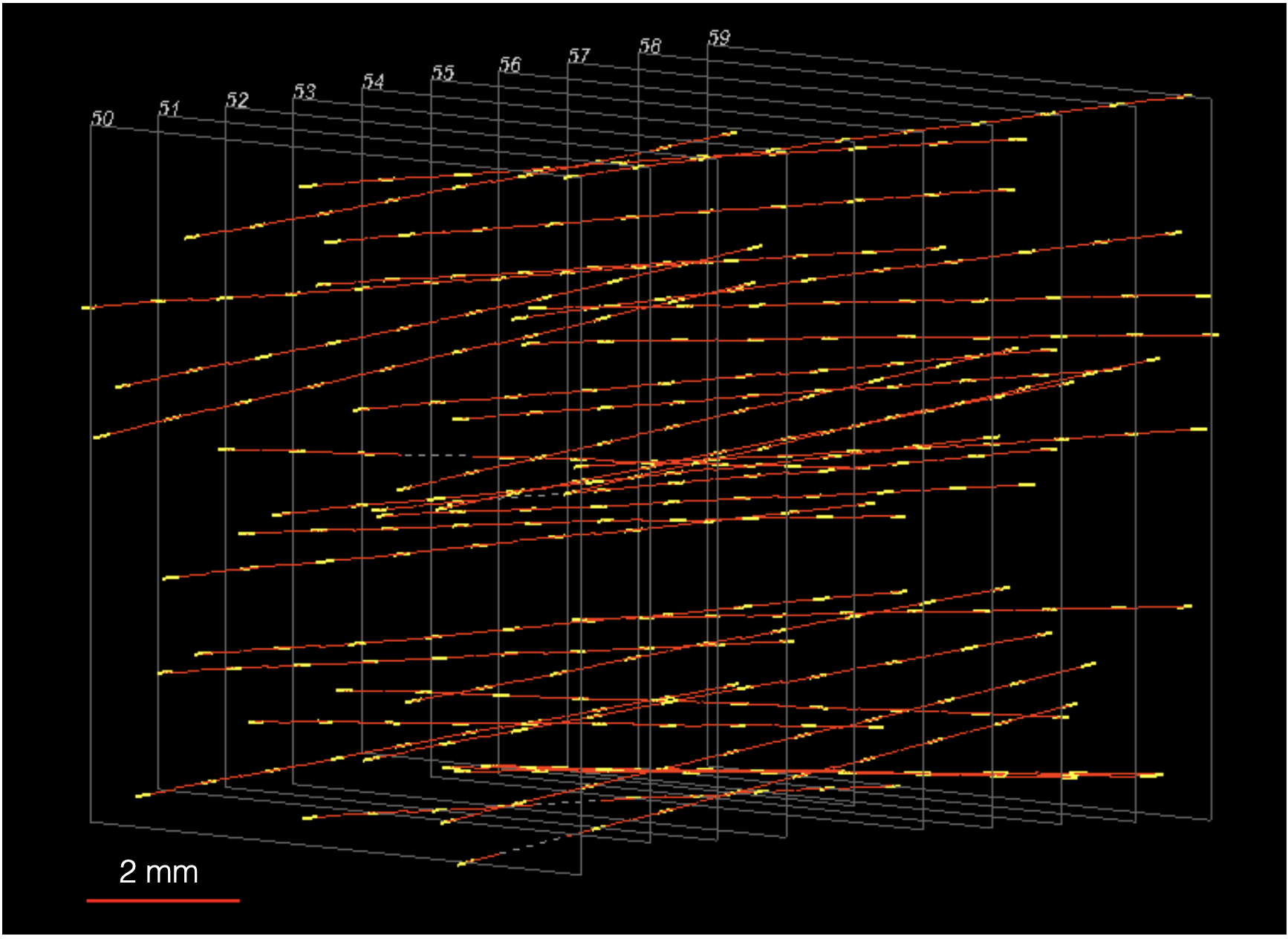}
 \caption{Display of reconstructed tracks (red lines) in ten consecutive films in one brick used for the target wall commissioning. Yellow segments represent base-tracks reconstructed in emulsion films.}\label{fig:emutrack_display}
 \end{figure}

Track reconstruction was performed on both the upstream section (five emulsion films) and downstream section (ten emulsion films) of the two exposed ECC bricks. As an example, a display of the reconstructed tracks in the downstream section of one brick is reported in Figure~\ref{fig:emutrack_display}.

\section{Infrastructure and detector installation in TI18}
\label{sec:installation}

\begin{figure}[h]
    \centering
    \includegraphics[width=\textwidth]{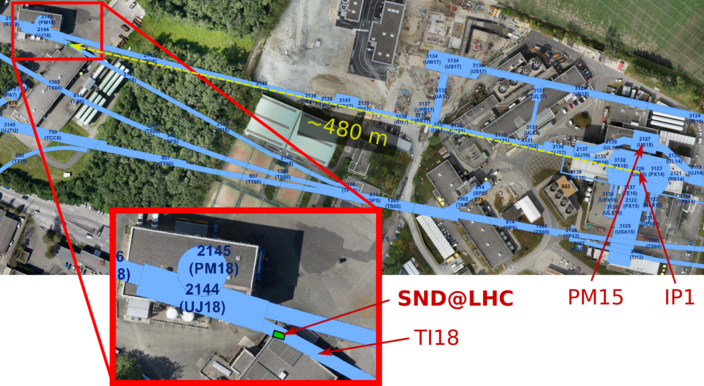}
    \caption{The location of the SND@LHC detector. SND@LHC is located in the TI18 tunnel, accessible from the LHC one via UJ18, \SI{480}{m} from the ATLAS interaction point. Access can be performed via the PM15 and PM18 shafts.}
    \label{fig:map_snd}
\end{figure}

The TI18 tunnel, shown in Figure~\ref{fig:map_snd}, was initially constructed for injection of positrons from the SPS to the LEP accelerator.
It is \SI{280}{m} long and has mostly a steep slope of about \SI{15}{\percent}, but levels out as it enters the LHC ring via the junction cavern UJ18 in the LHC Sector 12, about \SI{480}{m} from the ATLAS interaction point.
The LEP machine elements in TI18 were removed during the preparatory works for the LHC but the tunnel was left unused. All but the last short section of about \SI{20}{m} before entering UJ18 has been closed off. 
At the level of the floor, this short section crosses the collision axis of IP1, making the location particularly suitable for the high pseudo-rapidity region sought by the experiment.
At this location, the tunnel has a slope of \SI{3.6}{\percent} and a \SI{2.9}{m}-wide floor.

Detailed integration studies showed that the detector could be constructed on the floor without any modification to the tunnel structure.
%, by giving up on the magnet for the $\nu-\overline{\nu}$ separation, which will be re-considered for a possible upgrade of the detector for Run 4.   
Yet, the use of TI18 presented a number of challenges.
Firstly, TI18 is on the outside of the LHC ring while the \SI{450}{m} transport path from the access shaft PM15 at IP1 to UJ18 is on the inside, requiring preparation of dedicated transport paths above and under the machine for the infrastructure and detector components.
The transport path had to be made compatible with the machine cryogenics under helium pressure.
Secondly, TI18 was lacking all services in terms of ambient lighting, power, cooling and safety as required by the experiment. 

\begin{figure}[h]
    \centering
    \includegraphics[width=0.8\textwidth]{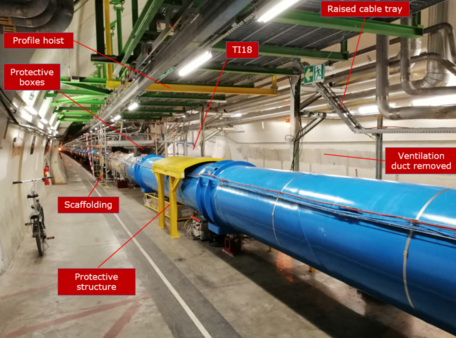}
    \caption{Principal modifications in UJ18 integration.}
    \label{fig:UJ18}
\end{figure}

Figure~\ref{fig:UJ18} shows the main modifications in UJ18.
The transport path over the machine is ensured by an added rail fixed to the UJ18 ceiling and carrying a manual hoist with a \SI{500}{\kilogram} capacity.
A protective table, capable of resisting against a fall of an object of up to \SI{1.3}{\tonne}, was produced and installed under the hoist and over the cryostat.
A transport volume of \SI[product-units=power]{75 x 90 x 170}{cm} was opened up by modifying the location of the existing cable trays.
Space below the machine was also freed to guarantee a path for transporting smaller objects with the help of low-profile trolleys.
The passage will also be used to pass the trolley for the exchange of the emulsion walls during the run (Figure~\ref{fig:wall}b).

Space for storage of detector components and assembly was freed by removing obsolete ventilation ducts in UJ18.
This allowed for detector components and infrastructures items to be brought in batches to avoid transport bottlenecks in the LHC access system.
The required detector electrical power of \SI{11}{kW} could be provided from the existing electrical grid in UJ18.
A dedicated circuit with an electrical box and associated emergency stop buttons were installed in TI18. 

\begin{figure}
\begin{subfigure}{.5\textwidth}
  \centering
  \vspace{0.5cm}
  \includegraphics[width=.99\linewidth]{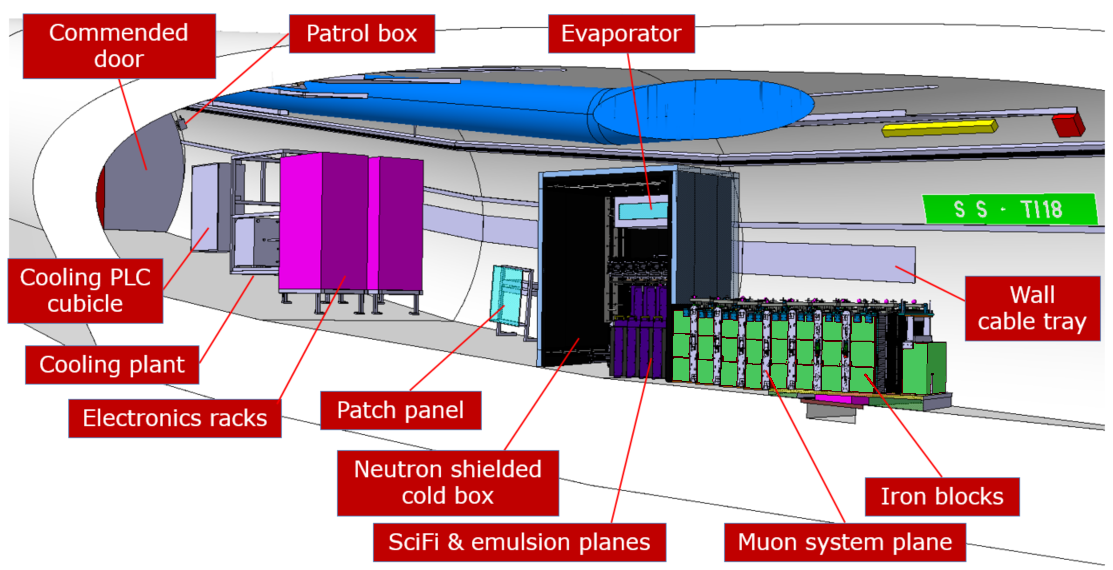}
  \vspace{0.5cm}
  \caption{ Overview of the SND detector integration in TI18}
\end{subfigure}
\begin{subfigure}{.5\textwidth}
  \centering
  \includegraphics[width=.99\linewidth]{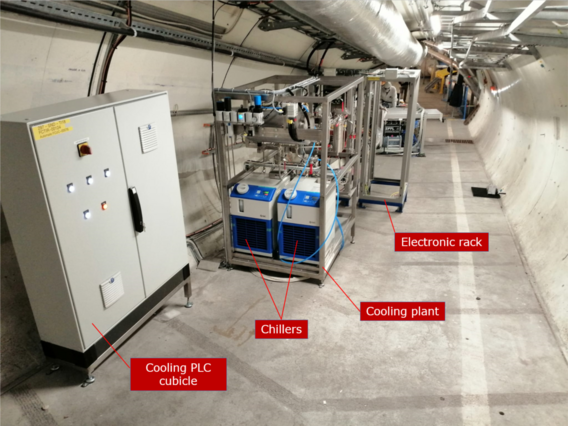}
  \caption{Detector services upstream in TI18}
\end{subfigure}
\begin{subfigure}{.5\textwidth}
  \centering
  \includegraphics[width=.99\linewidth]{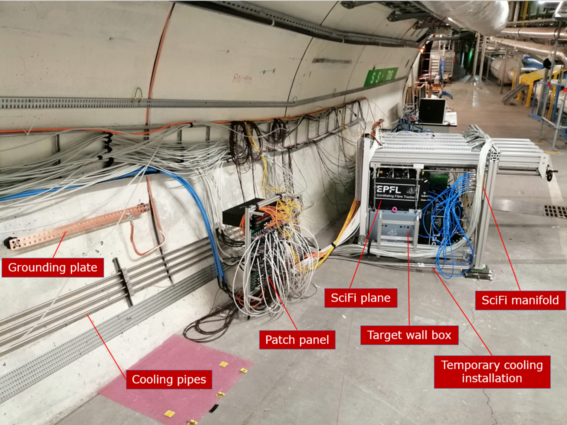}
  \caption{Front view of the detector}
\end{subfigure}
\begin{subfigure}{.5\textwidth}
  \centering
  \includegraphics[width=.99\linewidth]{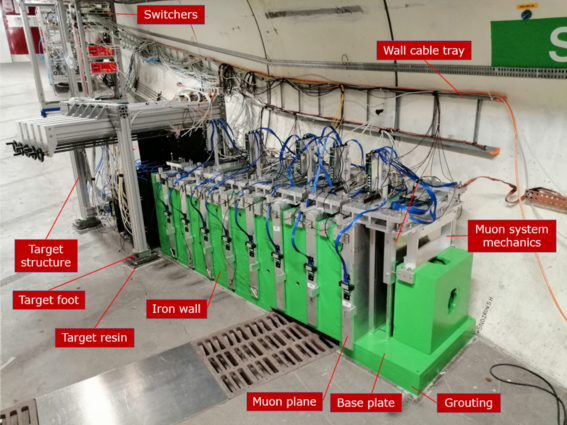}
  \caption{Rear view of the detector}
\end{subfigure}
\caption{Views of the SND@LHC experimental area in TI18, before the installation of the neutron-shielded box.}
\label{fig:TI18}
\end{figure}

Figure~\ref{fig:TI18} shows an overview of the service and detector integration in TI18, together with detailed images of the experimental area.
To free additional space for the detector installation in TI18, a \SI{20}{\meter}-long and obsolete ventilation duct was removed.

\begin{figure}[h]
    \centering
    \includegraphics[width=0.7\textwidth]{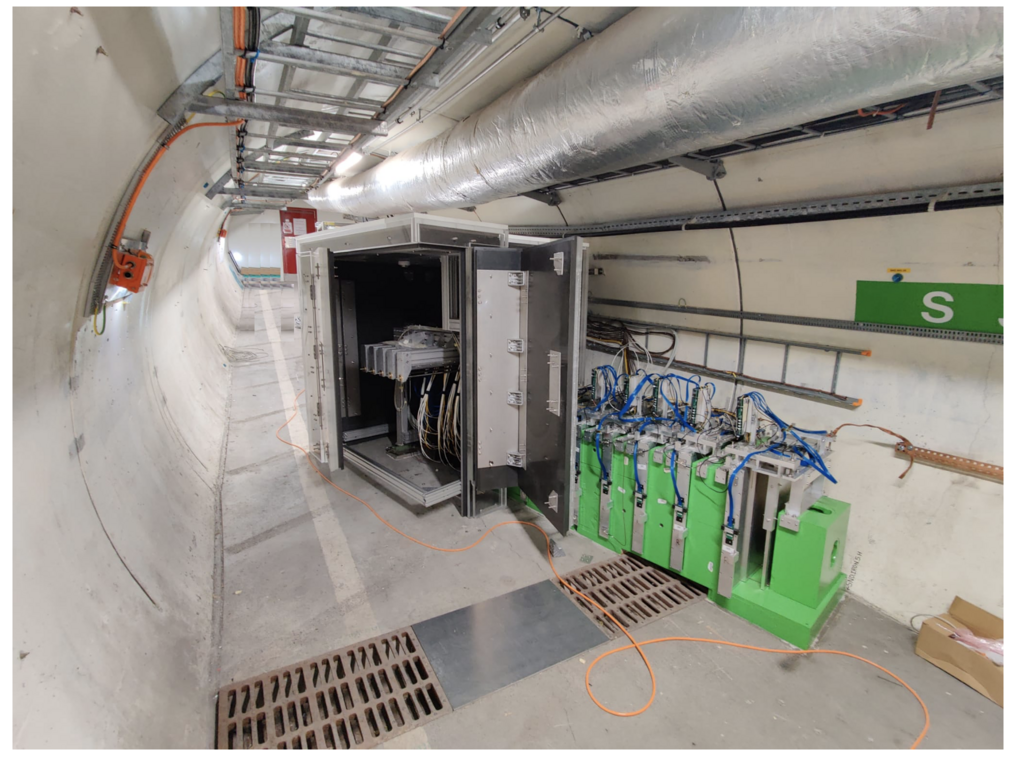}
    \caption{Global view of the detector in TI18 after the installation of the neutron shield.}
    \label{fig:det_all}
\end{figure}

The neutron-shielded box that surrounds the target region has dimensions of \SI[product-units=power]{2.19 x 1.76 x 1.86}{m} and is shown in Figure~\ref{fig:det_all}. 
In order to provide the required shielding, the walls of the box are made of acrylic and \SI{30}{\percent} borated polyethylene panels, having a thickness of \SI{50}{mm} and \SI{40}{mm}, respectively. 
The whole structure is sustained by a skeleton of aluminium profiles. 
Doors on the upstream side and the corridor side of the detector provide easy access for maintenance and for emulsion wall replacements.
The neutron-shielded box is equipped with a closed circuit cooling system that guarantees a stable ambient relative humidity and temperature of \SI{45}{\percent} and \SI{15}{\celsius}, respectively, as required in order to prevent fading of the emulsion films.

Two racks were installed in TI18 to house the detector power supplies and readout system, and  dedicated optical fibre tubes were installed over \SI{600}{m} in the LHC tunnel in order to connect with existing fibres up to the surface rack that is hosting the timing system, and the DAQ and control computer servers.

The eight iron walls of the muon system, each with dimension
\SI[product-units=power]{80 x 60 x 20}{cm} and a weight of \SI{750}{kg}, rest on horizontal steel base plates which were positioned at an accuracy of \SI{0.5}{mm} and grouted against the floor to compensate for the tunnel slope.
Together with a smaller iron block at the end, the walls are by themselves providing the support for the mechanical structure holding the eight muon detector planes.
The emulsion walls and the SciFi planes are carried by the target system structure that is grouted to the existing floor with custom-made wedges.
All detector components were aligned with an accuracy of \SI{0.5}{mm}.
 
The goal of being ready for data taking at the start of the LHC Run~3 in 2022 limited the entire schedule for the infrastructure, detector installation and commissioning to nine months.
The final phase of the LHC Long Shutdown~2 and the preparation of the machine for startup in 2022 set additional strict constraints on the planning.
A large part of the works had to be done with the LHC dipoles cooled to \SI{4.5}{K}, requiring further attention on the procedures.
The main infrastructure modifications in UJ18 and TI18 were performed between the end of June and September 2021.
September and October were dedicated to detector assembly and beam tests on the surface in the North Area Hall EHN1, while LHC was closed for the pilot run.

The detector installation, including the iron blocks, cooling plants and the related electronics, was successfully carried out in November and December, allowing the start of global commissioning by the end of December 2021.
The neutron shield surrounding the target region was constructed in January and February 2022 and completed underground by March 15$^{th}$. A picture of the full detector installed in TI18 is shown in Figure~\ref{fig:det_all}.

On April 7$^{\text{th}}$, one-fifth of the target region was partially instrumented with emulsion films, together with a few independent small emulsion bricks to check machine-induced background during the LHC commissioning, as the very final step of the detector installation. On May 24$^{\text{th}}$, SND@LHC registered the first muons from {\it pp} collisions and at the beginning of July the first ever events at a centre-of-mass record energy of \SI{13.6}{TeV} were recorded.
The first target fully equipped with emulsions was installed on July 26$^{\text{th}}$ and emulsion were replaced three times during the 2022 run, integrating a total of \SI{\sim 40}{fb^{-1}}.

%The detector has been continuously taking cosmic ray data since the beginning of the year. 

\section{Ideas for an HL-LHC upgrade}
\label{sec:outlook}

An advanced version of the SND@LHC detector is envisaged for the HL-LHC. It will consist of two detectors: one placed in the same $\eta$ region as  SND@LHC, $i.e.$~$7.2 < \eta < 8.4$ and the other one in the region $4 < \eta < 5$. The first apparatus will have a similar angular acceptance as for the SND@LHC and will perform the charm production measurement and lepton flavour universality tests with neutrinos at the percent level; the second detector will benefit from the overlap with LHCb to reduce systematic uncertainties and will perform neutrino cross-section measurements. In order to increase the azimuth angle coverage of the second detector, the idea is to search for a location in existing caverns, closer to the interaction point. We consider this second module as a near detector meant for systematic uncertainty reduction.

Each detector will be made of three elements. The upstream one is the target region for  the vertex reconstruction and the electromagnetic energy measurement with a calorimetric approach. It will be followed downstream by a muon identification and hadronic calorimeter system. The third and most downstream element will be a magnet for the muon charge and momentum measurement, thus allowing for neutrino/anti-neutrino separation for $\nu_\mu$ and for $\nu_\tau$ in the muonic decay channel of the $\tau$ lepton. 

%The target will be made of thin sensitive layers interleaved with tungsten plates, for a total mass of a few tons. Giving that the use of nuclear emulsion at the HL-LHC may result to be incompatible with technical stops, the Collaboration is investigating the use of compact electronic trackers with high spatial resolution  fulfilling both tasks of  vertex reconstruction with micrometric accuracy and electromagnetic energy measurement. The hadronic calorimeter and the muon identification system will also be optimised.

The target will be made of thin sensitive layers interleaved with tungsten plates, for a total mass of $\sim$ 5 tons. The use of nuclear emulsion at the HL-LHC is prohibitive due to the very high intensity that would make the replacement rate of the the target incompatible with technical stops. The Collaboration is investigating the use of compact electronic trackers with high spatial resolution  fulfilling both tasks of  vertex reconstruction with micrometric accuracy and electromagnetic energy measurement. The hadronic calorimeter and the muon identification system will be larger than 10~$\lambda$ which will bring the average length of the hadronic calorimeter above 11.5~$\lambda$, thus improving the muon identification efficiency and energy resolution.
The magnetic field strength is assumed to be about 1 T  over a $\sim$2 m length. 

The configuration of the detectors allows efficiently distinguishing between all three neutrino flavours and measure their energy. The SND@LHC upgrade will open a unique opportunity to probe physics of heavy flavour production at the LHC in a region   inaccessible to other experiments.
\section{Acknowledgements}

The support from the ANID--Millennium Program--ICN2019\_044 is acknowledged.
The support from the Marie Sklodowska-Curie Innovative Training Network Fellow-ship of the European Commissions Horizon 2020 Programme under contract number 765710 INSIGHTS is acknowledged. This research was
carried out in the frame of Programme STAR Plus, financially supported
by UniNA and Compagnia di San Paolo.
%The support from the National Research Foundation of Korea with grant numbers of \linebreak 2018R1A2B2007757, 2018R1D1A3B07050649, 2018\-R1D\-1A\-1B07050701, 2017R1D1A1B03036042, 2017\-R1A6A3\-A01075752, 2016R1A2B4012302, and 2016R1A6A3A1\-1\-930680 is acknowledged.
%The support from the Russian Foundation for Basic Research, grant 17-02-00607, and the support from the TAEK of Turkey are acknowledged.
%The work was carried out with financial support from the Ministry of Education and Science of the Russian Federation in the framework of the Competitiveness Enhancement Program of NUST ‘‘MISIS”, implemented by a governmental decree dated 16th of March 2013, No 2.  

\bibliographystyle{JHEP} %
\bibliography{references}

%\begin{thebibliography}{99}
%\bibitem{a}
%Author, \emph{Title}, \emph{J. Abbrev.} {\bf vol} (year) pg.
%\end{thebibliography}

\newpage

%\input{authorlist_5may2022}
%\input{authorlist_3july2022}
%\documentclass{article}
%\usepackage[utf8]{inputenc}

%\begin{document}

\centerline{\Large\bf The SND@LHC Collaboration}
\vspace*{1mm}

%\input{authorlist_february2021}

%\documentclass{article}
%\usepackage[utf8]{inputenc}

%\begin{document}

%
%\centerline{\Large\bf The SHiP Collaboration}
\vspace*{1mm}
\begin{flushleft}
%-- 
%-- SND@LHC Authorlist as of 1 March 2021
%-- 

G.~Acampora$^{9,c}$, %NAPOLI
C.~Ahdida$^{23}$, %CERN
R.~Albanese$^{9,c,g}$, %NAPOLI
C.~Albrecht$^{25}$, %ZURICH
A.~Alexandrov$^{9,18,20,c}$, %NAPOLI
M.~Andreini$^{23}$, %CERN
A.~Anokhina$^{21,i}$, %SINP
T.~Asada$^{9,c}$, %NAPOLI
N.~Auberson$^{24}$, %EPFL
C.~Baldanza$^{7}$
% ...
\vspace*{1cm}

{\footnotesize \it

$ ^{1}$Faculty of Physics, Sofia University, Sofia, Bulgaria\\
$ ^{2}$Niels Bohr Institute, University of Copenhagen, Copenhagen, Denmark\\
$ ^{3}$Humboldt-Universit\"{a}t zu Berlin, Berlin, Germany\\
$ ^{4}$Department Physik, Universit\"{a}t Siegen, Siegen, Germany\\
$ ^{5}$Institut f\"{u}r Physik and PRISMA Cluster of Excellence, Johannes Gutenberg Universit\"{a}t Mainz, Mainz, Germany\\
$ ^{6}$Sezione INFN di Bari, Bari, Italy\\
$ ^{7}$Sezione INFN di Bologna, Bologna, Italy\\
$ ^{8}$Sezione INFN di Cagliari, Cagliari, Italy\\
$ ^{9}$Sezione INFN di Napoli, Napoli, Italy\\
$ ^{10}$Laboratori Nazionali dell'INFN di Gran Sasso, L'Aquila, Italy\\
$ ^{11}$Nagoya University, Nagoya, Japan\\
$ ^{12}$Toho University, Funabashi, Chiba, Japan\\
$ ^{13}$Department of Physics Education and RINS, Gyeongsang National University, Jinju, Korea\\
$ ^{14}$Gwangju National University of Education, Gwangju, Korea\\
$ ^{15}$Korea University, Seoul, Korea\\
$ ^{16}$Sungkyunkwan University, Suwon-si, Gyeong Gi-do, Korea\\
$ ^{17}$University of Leiden, Leiden, The Netherlands\\
$ ^{18}$P.N.~Lebedev Physical Institute (LPI RAS), Moscow, Russia\\
$ ^{19}$National Research Centre 'Kurchatov Institute', Moscow, Russia\\
$ ^{20}$National University of Science and Technology 'MISiS', Moscow, Russia\\
$ ^{21}$Skobeltsyn Institute of Nuclear Physics of Moscow State University (SINP MSU), Moscow, Russia\\
$ ^{22}$National Research University Higher School of Economics, Moscow, Russia \\
%Yandex School of Data Analysis, Moscow, Russia\\
$ ^{23}$European Organization for Nuclear Research (CERN), Geneva, Switzerland\\
$ ^{24}$Institute of Physics, Ecole Polytechnique F\'{e}d\'{e}rale de Lausanne (EPFL), Lausanne, Switzerland\\
$ ^{25}$Physik-Institut, Universit\"{a}t Z\"{u}rich, Z\"{u}rich, Switzerland\\
$ ^{26}$Middle East Technical University (METU), Ankara, Turkey\\
$ ^{27}$Imperial College London, London, United Kingdom\\
$ ^{28}$University College London, London, United Kingdom\\
$ ^{29}$Laboratory of Instrumentation and Experimental Particle Physics (LIP), Lisbon, Portugal\\
$ ^{30}$Millennium Institute for Subatomic physics at high energy frontier-SAPHIR, Fernandez Concha 700, Santiago, Chile \\
$ ^{31}$Center for Theoretical and Experimental Particle Physics, Facultad de Ciencias Exactas, Universidad Andrés Bello, Fernandez Concha 700, Santiago, Chile$ ^{j}$ \\
$ ^{32}$Pontificia Universidad Católica de Chile, Santiago, Chile$ ^{j}$\\
$ ^{33}$Departamento de Física, Facultad de Ciencias, Universidad de La Serena, Avenida Cisternas 1200, La Serena, Chile$ ^{j}$\\
$ ^{34}$International Institute of Physics at the Federal University of Rio Grande do Norte, Rio Grande do Norte, Brazil\\
$ ^{a}$Universit\`{a} di Bari, Bari, Italy\\
$ ^{b}$Universit\`{a} di Bologna, Bologna, Italy\\
$ ^{c}$Universit\`{a} di Napoli ``Federico II'', Napoli, Italy\\
$ ^{d}$Taras Shevchenko National University of Kyiv, Kyiv, Ukraine\\
$ ^{e}$National Research Nuclear University (MEPhI), Moscow, Russia\\
$ ^{f}$Ankara University, Ankara, Turkey\\
$ ^{g}$Consorzio CREATE, Napoli, Italy\\
$ ^{h}$Present address: Faculty of Engineering, Kanagawa University, Yokohama, Japan\\
$ ^{i}$Also at: Lomonosov Moscow State University, Faculty of Physics, Moscow, Russia\\
$ ^{j}$Associated to: Millennium Institute for Subatomic physics at high energy frontier-SAPHIR, Fernandez Concha 700, Santiago, Chile \\
%$ ^{e}$Associated to Gyeongsang National University, Jinju, Korea\\
%$ ^{f}$Associated to Petersburg Nuclear Physics Institute (PNPI), Gatchina, Russia\\
%$ ^{g}$Also at Moscow Institute of Physics and Technology (MIPT),  Moscow Region, Russia\\
%$ ^{i}$Universit\`{a} della Basilicata, Potenza, Italy\\
$ ^{k}$ENEA Research Centre E. Clementel, Bologna, Italy\\
$ ^{l}$Universit\`{a} di Napoli Parthenope, Napoli, Italy\\
$ ^{m}$Universit\`{a} del Sannio, Benevento, Italy\\
}
\end{flushleft}

\end{document}